\documentclass{aa}  

\usepackage{graphicx,txfonts}

\usepackage{mathtools,amsmath,amssymb,CJK,lineno,hyperref,epigraph,textcomp,gensymb,color,siunitx,graphicx,adjustbox,txfonts,color}
\include{journal_abbrv.tex}

\modulolinenumbers[2]
\newcommand{\Lsol}{$L_\odot$}

\begin{document} 


\title{Activity of the Eta-Aquariid and Orionid meteor showers}

\author{A. Egal
    \inst{1, 2, 3}\fnmsep\thanks{\email{aegal@uwo.ca}}
    \and
    P. G. Brown \inst{1, 2}
    \and
    J. Rendtel \inst{4}
    \and
    M. Campbell-Brown \inst{1, 2}
    \and
    P. Wiegert \inst{1, 2}
    }

\institute{
    Department of Physics and Astronomy, The University of Western Ontario, London, Ontario N6A 3K7, Canada
    \and
    Institute for Earth and Space Exploration (IESX), The University of Western Ontario, London, Ontario N6A 3K7, Canada
    \and 
    IMCCE, Observatoire de Paris, PSL Research University, CNRS, Sorbonne Universit\'{e}s, UPMC Univ. Paris 06, Univ. Lille, France
    \and 
    Leibniz-Institut f. Astrophysik Potsdam, An der Sternwarte 16, 14482 Potsdam, Germany, and International Meteor Organization, Eschenweg 16, 14476 Potsdam, Germany
    }

\date{Received XYZ; accepted XYZ}

 
  \abstract
   {}
   {We present a multi-instrumental, multidecadal analysis of the activity of the Eta-Aquariid and Orionid meteor showers for the purpose of constraining models of 1P/Halley's meteoroid streams. }
   {The interannual variability of the showers' peak activity and period of duration is investigated through the compilation of published visual and radar observations prior to 1985 and more recent measurements reported in the International Meteor Organization (IMO) Visual Meteor DataBase, by the IMO Video Meteor Network and by the Canadian Meteor Orbit Radar (CMOR). These techniques probe the range of meteoroid masses from submilligrams to grams. The $\eta$-Aquariids and Orionids activity duration, shape, maximum zenithal hourly rates (ZHR) values, and the solar longitude of annual peaks since 1985 are analyzed. When available, annual activity profiles recorded by each detection network were measured and are compared. }
  {Observations from the three detection methods show generally good agreement in the showers' shape, activity levels, and annual intensity variations. Both showers display several activity peaks of variable location and strength with time. The $\eta$-Aquariids are usually two to three times stronger than the Orionids, but the two showers display occasional outbursts with peaks two to four times their usual activity level. CMOR observations since 2002 seem to support the existence of an $\sim$12 year cycle in Orionids activity variations; however, additional and longer term radar and optical observations of the shower are required to confirm such periodicity.}
   {}


   \maketitle
%
\section{Introduction}

Comet 1P/Halley is known to produce two annual meteor showers, the $\eta$-Aquariids in early May and the Orionids in late October. These two related showers are often collectively termed the Halleyids. Both showers exhibit a total duration of about 30\si{\degree} in solar longitude (\Lsol) and a complex fine structure characterized by several subpeaks of variable location and intensity. The average activity levels of the Orionids and the $\eta$-Aquariids are similar. The activity of the showers is frequently characterized by the zenithal hourly rate (ZHR), which represents the number of meteors an observer would observe per hour under standard reference conditions.

The maximum activity of the Orionids usually varies between 15 and 30 meteors per hour near 209\si{\degree} of solar longitude (\Lsol, ecliptic J2000), while the $\eta$-Aquariids maximum rates are two to three times higher, near 45.5\si{\degree} solar longitude. Both showers are frequently cited as being rich in small (masses $\approx$ 10$^{-8}$ kg) particles \citep{Jenniskens2006,CB2015,Schult2018}.

Despite moderate average annual activity, the two showers are known to produce occasional outbursts. Recent examples include the Orionids apparitions of 2006 and 2007, when the peak ZHR exceeded 60 \citep{Rendtel2007,Arlt2008}, and the 2013 $\eta$-Aquariids outburst that reached an activity level of 135 meteors per hour \citep{Cooper2013}. Mean motion resonances of Halley's meteoroids with Jupiter were identified as responsible for these periods of enhanced activity \citep{Rendtel2007,Sato2007,Sato2014,Sekhar2014}. 

The Orionids and $\eta$-Aquariids are of particular interest because of their connection to comet 1P/Halley. Meteor observations can be used to constrain the nature of the meteoroid trails that are ejected by the comet (meteoroids size, density variations along the stream, etc.) and characterize its past activity. Constraining meteoroid stream models using meteor observations could therefore provide insights into 1P/Halley's past orbital evolution \citep{Sekhar2014,Kinsman2017}, which is mostly unknown prior to 1000 BCE. 

The two showers are also of concern from a spacecraft safety perspective. The Orionids and especially the $\eta$-Aquariids are among the more significant impact hazards out of all the major showers throughout the year. This is a consequence of several characteristics of the showers including their long duration, comparatively high flux, the occurrence of occasional outbursts and their high velocity ($\sim$66 km/s). In spite of their significance, few predictions of future Halleyid activity are currently available in the literature. Moreover, no comprehensive numerical models of the streams have been published which utilize the full suite of modern observational data, in part because no compilation of such data have become available. This is a major goal of the current work.  

Most of Halley's meteoroid stream models were developed in the 1980s, when the expected return of the comet in 1986 renewed the interest in the study of these showers. Several characteristics of the Halleyids (similar duration and average intensity, complex structure) were successfully explained by the long-term evolution of meteoroids under the influence of planetary perturbations (mainly induced by Jupiter, see \citealt{McIntosh1983} and \citealt{McIntosh1988}). 

However, almost no modeling effort had been made to explain the annual activity variations of the Orionids and $\eta$-Aquariids noticed by \cite{Hajduk1970} and \cite{Hajduk1973}, until the observation of the recent Orionids outbursts. By investigating the role of resonances highlighted in previous works, numerical simulations of \cite{Sekhar2014} successfully reproduced the apparition dates of recent and ancient Orionids outbursts, and predicted a future outburst of the shower in 2070. Unfortunately, no similar analysis of the $\eta$-Aquariids has been conducted by those authors, in part because of the small number of reliable observations of this shower compared to the Orionids \citep{Sekhar2014}.

In this work, we investigate the long-term activity of the Orionids and $\eta$-Aquariids as measured by visual, video and radar observations. Published and original observations of both showers are compiled and analyzed to examine the structure, the duration, and the annual variation of their activity profiles.
Following the work of \cite{CB2015} which focused solely on the $\eta$-Aquariids up to 2014, we provide recent measurements of the $\eta$-Aquariids and a complete set of Orionids recorded by the Canadian Meteor Orbit Radar (CMOR) between 2002 and 2019, which constitutes the longest consistent observational set of the Halleyids to date. Results from CMOR are compared to the visual observations contained in the IMO Visual Meteor Data Base (VMDB) and measurements of the Video Meteor Network (VMN). 

The specific goal of this paper is to measure characteristics of the Halleyids including their long-term average ZHR-equivalent profile, activity shapes of annual apparitions, maximum activity levels each year, and the location of the  peak using an integrated multi-instrument analysis of the showers. In particular we wish to compare results across the different detection systems to better identify biases.

The structure of this paper is as follows: sections \ref{sec:Halley}, \ref{section:ori} and \ref{section:eta} review the discovery circumstances, observational history and general characteristics of comet 1P/Halley and previously published analyses of the Orionids and the $\eta$-Aquariids meteor showers. Sections \ref{section:Recent_observations} and \ref{section:analysis} present the methodology for selection, processing and analysis of the showers' visual, video. and radar observations since 2002. The conclusion of our analysis is presented in Section \ref{sec:conclusion}. To aid future modeling efforts, individual activity profiles per year including specific numerical measurements for peak activity and location found in this work are available in Appendices A, B, C and D. 

\section{History of 1P/Halley and early observations of the Halleyids} \label{sec:Halley}

 \subsection{1P/Halley}
 
 1P/Halley is a famous comet, evolving in a retrograde orbit with a period of about 76 years. The comet was named after Edmond Halley, who recognized in 1705 the comets of August-September 1682, October 1607 and August 1531 as being the same object. Halley estimated the orbital period of the comet and anticipated a return in 1758. His eponymous comet was recovered in December 1758, a few years after Halley's death, becoming the first comet whose return close to the Sun was successfully predicted \citep{Hughes1987a}. 
 
 Since Edmond Halley's discovery, several ancient observations of comets have been linked to older apparitions of 1P/Halley \citep{Yeomans1981,Hughes1987a}. The first identified observations of 1P/Halley date back to 240 BCE in Chinese records \citep{Kiang1972} and 164 BCE in Babylonian records \citep{Stephenson1985}. With some adjustments to the comet's eccentricity around AD 837, \cite{Yeomans1981} determined a reliable set of orbits for 1P/Halley until 1404 BCE. Because of a close encounter with Earth in 1404 BCE and the lack of older observational constraints on the comet's motion, the orbital elements of 1P/Halley have not been precisely determined prior to this epoch \citep{Yeomans1981}. 
 
 In March 1986, an international fleet of spacecraft called the "Halley Armada" approached the comet to examine its nucleus. The Armada was composed of five main probes, Giotto (European Space Agency), Vega 1 and Vega 2 (Soviet Union) and Suisei and Sakigake (Japan), supported by additional measurements of the international ISEE-3 (ICE) spacecraft and NASA's Pioneer 7 and Pioneer 12 probes. On the ground, observations of the comet and its associated meteoroid streams were collected and archived by the International Halley Watch (IHW) organization, an international agency created to coordinate comet Halley's observations. Reviews of the extensive research conducted during the comet 1986 apparition can be found for example in \cite{Whipple1987} and \cite{Edberg1988}. The next apparition of the comet is expected in 2061.
 
 \subsection{Halley's meteoroid streams}
 
 1P/Halley's meteoroid streams are responsible for two observed meteor showers on Earth, the Orionids (cf. Section \ref{section:ori}) and the $\eta$-Aquariids (cf. Section \ref{section:eta}). The $\eta$-Aquariids occur at the descending node of the comet, while the Orionids are connected to 1P/Halley's ascending node. 

\begin{figure}[!ht]
    \centering
    \includegraphics[width=0.48\textwidth]{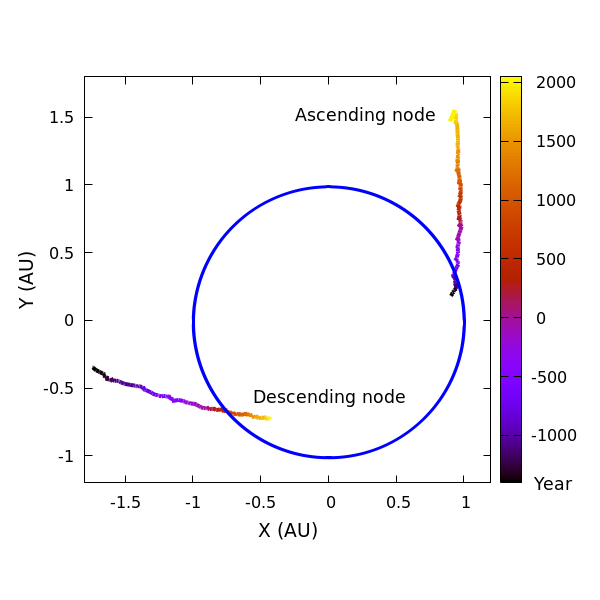}
    \caption{Ecliptic location of 1P/Halley's ascending and descending node as a function of time, in the barycentric frame (ecliptic J2000). The Earth's orbit is represented in blue.}
    \label{fig:Nodes}
\end{figure}

 The evolution of 1P/Halley's nodes as a function of time is presented in Figure \ref{fig:Nodes}. The motion of the comet was integrated from 1404 BCE to AD 2050 using a RADAU15 \citep{Everhart1985} integrator with an external time step of 1 day. Orbital solutions of \cite{Yeomans1981} were used as initial conditions for the comet apparitions before 1910, while the JPL J863/77 solution was used for the 1986 perihelion passage\footnote{https://ssd.jpl.nasa.gov/sbdb.cgi}. Results are presented in the barycentric frame (ecliptic J2000). 
 
As shown in Figure \ref{fig:Nodes}, the orbital precession of 1P/Halley leads to considerable change in the location of its nodes since 1404 BCE. In 1404 BCE, the comet's descending node was located well outside the Earth's orbit, while the ascending node was much closer to Earth's orbit. Currently, the comet crosses the ecliptic plane at an ascending node far outside the Earth's orbit (1.8 AU), and descends below the ecliptic at about 0.85 AU. The current high ascending nodal distance of 1P/Halley from Earth's orbit explains why the link between the Orionids shower and 1P/Halley was difficult to establish (cf. Section \ref{section:ori}). 
 
 Explaining the similar activity levels of the $\eta$-Aquariids and Orionids \citep{Lovell1954,Hajduk1970,Hajduk1973} with such different nodal distances has challenged researchers for many decades. If the showers were produced by a stream centered on the current orbit of the comet, the $\eta$-Aquariids would be considerably stronger than the Orionids and their durations significantly different.
 
 Perhaps the most successful model of the Halleyids is that of the theoretical shell model of \cite{McIntosh1983}. It was later confirmed by \cite{McIntosh1988}'s numerical simulations and offers an explanation for many characteristics of the showers. The model is based on the idea that particles ejected from the comet evolve at different rates, with some remaining on orbits where the comet was a long time ago and others precessing more rapidly than the comet (and eventually reaching its future positions). Each meteoroid stream would therefore evolve into a ribbon-like structure of uniform thickness, producing two meteor showers of similar duration and intensity at the Earth. The superposition of several ribbon-shaped streams, separated through small perturbation relative to the comet orbit, could be responsible for the observed filamentary structure of the $\eta$-Aquariids and Orionids showers (cf. Sections \ref{section:ori} and \ref{section:eta}). 

\begin{figure}[!ht]
    \centering
    \includegraphics[width=0.48\textwidth]{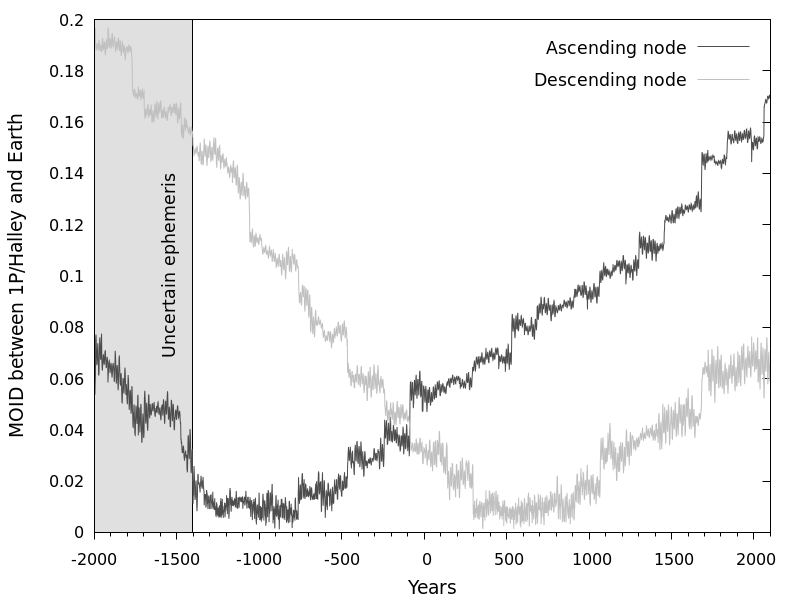}
    \caption{Time variations of the Minimum Orbit Intersection Distance (MOID) of comet 1P/Halley and Earth close to the comet's ascending and descending node.}
    \label{fig:MOID}
\end{figure}

The time evolution of the Minimum Orbit Intersection Distance (MOID) between Earth and the comet around each node is provided in Figure \ref{fig:MOID}. The Earth currently approaches the comet orbit at a minimum distance of 0.154 AU at the time of the Orionids and 0.065 AU at the time of the $\eta$-Aquariids. 1P/Halley's descending node reached its closest distance to Earth's orbit around AD 500, and at about 800 BCE for its ascending node. The proximity of the comet around AD 500 might explain the existence of strong $\eta$-Aquariids outbursts reported in ancient Chinese observations. 

  \subsection{Ancient Chinese, Japanese, and Korean records} \label{sec:chinese}

The long-term (millennium timescale) variation of the strength of the showers is recorded in ancient visual records, though the biases and incompleteness of these sources imply that some strong returns may easily have been missed. The analysis of ancient records of meteor showers reveals that the Orionids and the $\eta$-Aquariids were active a millennium or more ago \citep{Ahn2003,Ahn2004}. Orionids outbursts were identified in Chinese and Japanese observations in AD 585, 930, 1436, 1439, 1465, and 1623 \citep{Imoto1958}. Possible strong displays in AD 288 and 1651 are also mentioned in \cite{Zhuang1977}. 

The first Korean record of an $\eta$-Aquariid outburst could be as old as 687 BCE \citep{Ahn2004}. Chinese observations allowed the identification of several $\eta$-Aquariids outbursts in 74 BCE, and AD 401, 443, 466, 530, 839, 905, 927, 934 \citep{Zhuang1977,Imoto1958}. The comparison of the numerical integration of the comet Halley meteoroid stream with Maya hieroglyphic inscriptions seems to indicate that this civilization also kept track of observed $\eta$-Aquariids outbursts \citep{Kinsman2017}. In that work, the existence of a strong outburst in AD 461, observed but not classified as an $\eta$-Aquariid in \cite{Zhuang1977} or \cite{Imoto1958}, was also discussed.

The Halleyids intensity and year-to-year variations cannot be rigorously estimated from these ancient observations. Indeed, missing records for specific years and missing information about the observing conditions limits the interpretation of the showers annual activity. However, the records are able to highlight the existence of years with particularly strong activity \citep[e.g., "hundreds of meteors scattered in all directions" in AD 585, or "hundreds of large and small meteors" in AD 1439, see][]{Imoto1958}, with maximum dates falling a few days earlier than the current shower peaks \citep{Zhuang1977}.

  \section{Orionids}\label{section:ori}
  
   \subsection{History}
   
  Despite earlier observations of Orionids meteors, the discovery of the meteor shower is independently attributed to E. C. Herrick and Quetelet in 1839 \citep{Lindblad1999,Kronk2014}. The first precise Orionid radiant was determined by A. Herschel in 1864 and 1865 \citep{Denning1899,Herschel1865}, and shortly after \cite{Falb1868} proposed a connection between 1P/Halley, the $\eta$-Aquariid, and the Orionid meteor showers. The similarity of the Orionids with the $\eta$-Aquariids was noticed again by Olivier in 1911.  
  
  However, despite the already established connection between the $\eta$-Aquariids and 1P/Halley, the link between the Orionids and the comet was not immediately accepted because of the large orbital distance of 1P/Halley at the time of the shower. The relation between the Orionids and its parent comet was finally accepted after decades of controversy \citep{Obrubov1993,Zhuang1977,Rendtel2008}.
  
  The complexity of the Orionids activity structure was noticed at the beginning of the 20th century. 
  Considerable but variable rates have been reported in early analyses of visual observations \citep[e.g.,][]{Prentice1931,Prentice1933,Prentice1936}. In 1918, Denning emphasized that the Orionids radiant appeared to be stationary during the entire duration of the shower. The stationary radiant hypothesis was at the center of a great controversy for many years, until proof of a drift was presented by (among others) Olivier \citep[e.g.,][]{Olivier1923,Olivier1925} and Prentice \citep[e.g.,][]{Prentice1939}. The existence of multiple subradiants within the stream, investigated by several observers around the same period, was not confirmed by further observations \citep{Lindblad1999}. 
  
  The first photographic Orionid was captured in 1922 at the Harvard Observatory \citep[King 1923 in][]{Lindblad1999}, but only few double station observations of the shower were obtained during the program \citep{Lindblad1999}. In the following decades, the shower was the subject of several visual \citep{Stohl1981}, telescopic \citep{Znojil1968,Porubcan1973}, and radar studies \citep[e.g.,][]{Hajduk1982b,Jones1983,Cevolani1985}. 
  
  Most of the Orionid data gathered over the period 1900-1967 was collected and analyzed by \cite{Hajduk1970}. This work compiled and processed visual and photographic observations of the shower, as well as an extensive set of radar data from Canada and Czechoslovakia. Despite the heterogeneity of the observational methods and instruments used, \cite{Hajduk1970} managed to estimate the peak  meteor rate and solar longitude of the maximum activity over this time period. The derived visual meteor rates typically ranged between 10 and 30 meteors per hour, with increased rates during some individual returns. 
  
  The shower showed several maxima between \Lsol=201\si{\degree} and \Lsol=216\si{\degree}, varying in location from apparition to apparition (sometimes by more than 5\si{\degree}). No regular periodicity in the maxima locations or the annual activity was identified. The temporal variability of the shower peaks of activity, reflecting density variations along and across the stream, were interpreted as a sign of the strong filamentary nature of the stream. 
  The existence of density variations within the stream was identified in subsequent radar campaigns at Dushanbe and Ondrejov \citep[e.g.,][]{Babadzhanov1977,Babadzhanov1979,Hajduk1981,Jones1983,Hajduk1984,Cevolani1985,Hajduk1987}, as well in CMOR measurements between 2008 and 2010 \citep{Blaauw2011}. 

  \subsection{Activity}
  
  \subsubsection{Average activity profile} \label{section:ori_prev_obs}
  
  Between 1944 and 1950, visual observations carried out at the Skalnaté Pleso Observatory provided the first reliable activity profile of the shower. From this set of data, \cite{Stohl1981} and \cite{Porubcan1984} identified a main peak of activity at \Lsol=208.5\si{\degree} (J2000) and a secondary maxima at \Lsol=210.5\si{\degree}.  At the occasion of the International Halley Watch, multiple visual observations were carried out by amateurs spread all over the world. However, because of the heterogeneity of the resulting data and the little information provided about the observers, the analysis of the observations collected was particularly complex \citep{Porubcan1991}. 
  
  The average activity profile of the Orionids around the 1986 apparition of the comet is presented in Appendix \ref{old_obs}. The shower displayed normal levels of activity, with a maximum ZHR of about 20 meteors per hour and a Full Half Width Maximum (FWHM) of 7 to 8 days. Several subpeaks were detected during the period of activity (extending roughly between \Lsol=198\si{\degree} and \Lsol=220\si{\degree}). The analysis of the 1990 Orionids revealed that subpeaks of activity lasted for about 24h \citep{Koschack1991}. 
  
  From the analysis of 60 photographic Orionids contained in the IAU Meteor Data Center, \cite{Lindblad1999} derived a slightly asymmetric activity profile of the shower, with highest rates attained between \Lsol=208.7\si{\degree} and 210.7\si{\degree} and an estimated maximum around \Lsol=209.6\si{\degree}. \cite{Svoren2017}'s recent analysis of a very similar data set identified two maxima of photographic Orionid activity at \Lsol=208.5\si{\degree} and \Lsol=210.5\si{\degree}, in good agreement with \cite{Stohl1981} and \cite{Porubcan1984}'s results. A central dip was noticed close to the main peak of activity, as previously noted by  \cite{McIntosh1983} and \cite{Lindblad1999} for the Orionids, and by \cite{Hajduk1980} for the $\eta$-Aquariids. 
  
  Radar observations carried out by the CMOR radar between 2002 and 2008 highlighted an activity lasting from October 11 to November 9 \citep[from 198\si{\degree} to 227\si{\degree}, cf.][]{Brown2010}, slightly longer than the duration reported by \cite{Porubcan1991}. The main maximum was identified around 208\si{\degree} of solar longitude. The maximum Orionid rate recorded by the Middle Atmosphere ALOMAR Radar System (MAARSY) between 2013 and 2015 was identified as being around \Lsol=210.5\si{\degree} \citep{Schult2018}.
 
  The long-term variability of the Orionids and the $\eta$-Aquariids was investigated by \cite{Dubietis2003}. Processing the standardized observations of the IMO Visual Meteor Data Base (VMDB), the author derived the population index and an average peak ZHR for every apparition of the Orionids between 1984 and 2001. No reproducible trend in the population index variations was identified. As in previous works, the peak ZHR was found to vary between 10 to 35 meteors per hour. Based on the annual variation of the ZHR within these bounds, a 12 year periodicity was proposed (cf. Section \ref{sec:Lsolmax}). 
  
  A similar analysis was conducted by \cite{Rendtel2008}, who processed visual observations dating back to 1944 along with the data contained in the VMDB since the 1980s. As expected, the peak ZHR and the population index of the shower were found to vary over time. An average maximum ZHR of the shower was estimated to be about 20 to 25 between \Lsol=207\si{\degree} and \Lsol=211\si{\degree}. 
  
  Examples of visual activity profiles of the Orionids between 1985 and 2001 are presented in Appendix \ref{old_obs}. The results of the aforementioned studies were plotted along with the ZHR profiles available in the IMO VMDB website\footnote{https://www.imo.net/members/imo\_live\_shower}. The complex structure of the stream is not always discernible in each individual profile, when the low number of observations restricts the time resolution of the activity profiles \citep{Dubietis2003}. The variability in the number of peaks, in their intensity, and solar longitude from year to year is nonetheless clearly noticeable. 
  
  However, the quality and quantity of visual data in any given year are heavily influenced by the lunar phase around the Orionid maximum. These type of data are further complicated by the often poor weather at the end of October in the Northern Hemisphere, where the bulk of Orionid observations are performed. As a result, interannual variability of the shower is very difficult to conclusively prove based on visual observations alone.
  
  \subsubsection{Orionid outbursts}
  
  The Orionids are known to have produced strong outbursts over the past century, reaching two to four times the usual intensity of the shower \citep{Lovell1954,Hajduk1970}. The analysis of visual and radio measurements of the shower revealed increased meteor rates around 1934-1936, 1946-1948, 1966-1968, and potentially around 1927 as well \citep[][Figure 1]{Hajduk1970}. A small Orionid outburst (ZHR $>$ 30) was reported by European observers in 1993 \citep{Miskotte1993,Rendtel1993,Jenniskens1995}. The low population index at the time of the peak \citep[r=1.8, see][]{Rendtel1993} suggests that the Earth encountered a portion of the stream rich in large particles, a result consistent with the number of bright photographic meteors recorded during the outburst. However, similar ZHR rates were reached in 1995 and 1997 \citep[cf. Figure 2 of][]{Rendtel2008}. 
  
  The strongest activity reported in modern times was for the 2006 and 2007 apparitions of the shower \citep{Rendtel2007,Rendtel2008,TR2007,Spurny2008,Arlt2008}. In 2006, several peaks approached a ZHR of 60 \citep{Rendtel2007}. The highest activity occurred when the population index of the shower was at its lowest value, revealing a different meteoroid population from the shower background \citep{Rendtel2007}. The outburst was accompanied by an exceptional number of fireballs recorded by the European Fireball Network \citep{Spurny2008}. The following year, a maximum ZHR of about 80 around \Lsol=208.45\si{\degree} was reported by \cite{Arlt2008}. \cite{Rendtel2007} and \cite{Arlt2008} both suggested that enhanced activity of the Orionids can occur for a few consecutive years. 
  
  \subsubsection{The role of resonances}
  
  The results of the International Halley Watch campaign demonstrated that, as was the case for the 1910 apparition, the passage of 1P/Halley through perihelion in 1986 yielded no rate enhancement of the $\eta$-Aquariids or the Orionids \citep{Spalding1987,Porubcan1991}. In addition, the 1993 Orionids outburst occurred when the comet was far from its perihelion position \citep{Jenniskens2006}. The existence of Halleyids outbursts is therefore not correlated with recent perihelion passages of the comet, a conclusion that is easy to understand when we consider the significant orbital distance between 1P/Halley and the Earth.  
  
  The influence of Jupiter on the stream evolution, particularly on the spatial density and the meteoroid size distribution, was first investigated by \cite{Hajduk1970} and \cite{McIntosh1983}. After the observation of a strong outburst in 2006, the hypothesis of enhanced Orionid activity caused by meteoroids trapped in resonant orbits was investigated by several authors. Among the possible Mean Motion Resonances (MMR) with Jupiter \citep{EmelYanenko2001}, the 1:6 MMR was identified by \cite{Rendtel2007} as the probable cause of the 2006 outburst and suspected enhanced Orionid activity between 1933 and 1938 \citep{Rendtel2008}. Consistent with that result, \cite{Sato2007} determined that the 2006 outburst was caused by 1:6 resonant meteoroids ejected between 911 BCE and 1266 BCE. In contrast, \cite{Spurny2008} suggested the source of the observed fireballs during this apparition was most likely the 1:5 MMR with Jupiter. 
  
  The resonant behavior of the Orionids was investigated by \cite{Sekhar2014}. By performing numerical simulations, the authors highlighted the influence of the 1:6 and 2:13 MMR with Jupiter on recent and ancient Orionids outbursts. They identified the 2:13 MMR as being responsible for the observed 1993 outburst and of a possible older outburst in 1916. 
  
  Results of their simulations indicated several meteor outbursts due to particles trapped in the 1:6 resonance  between AD 1436 and 1440, when Orionid outbursts were reported in ancient Chinese records (see Section \ref{sec:chinese}). Following a similar approach, \cite{Kinsman2020}'s  numerical simulations pointed toward a strong Orionid outburst in AD 585 caused by the center of the 1:6 resonance. In addition, \cite{Sekhar2014}'s model also identified this resonance as causing the observed enhanced activity of 2006, 2007, 2008 and 2009 \citep{Kero2011}. The authors predicted that a future Orionid outburst in 2070 could be produced by particles currently trapped in the 2:13 MMR with Jupiter.
  
 \subsection{Mass index}
 
  \subsubsection{Definition}
  
  It is typically assumed that the mass distribution of meteor showers follows a power law, that is, $$\text{d}N\propto M^{-s}\text{d}M$$ 
  with $dN$ the number of meteoroids of masses between $M$ and $M+\text{d}M$. The exponent $s$ is called the differential mass index and characterizes the proportion of big and small particles in a shower. Having $s$ values strictly greater than 2 imply a stream mass concentrated in small particles, when values strictly lower than 2 indicate the opposite. The mass index is related to the population index $r$, the ratio of the number of meteors of magnitude M+1 to those of magnitude M, using the modern empirical relation $s\simeq1+2.3\text{ log }r$ \citep{Jenniskens2006}. 
  
  Several estimates of the mass indices of the $\eta$-Aquariids and Orionids are available in the literature. Mass indices  deduced from visual and radar observations between 1953 and 1980 were compiled by \cite{Hughes1987b}. Estimates ranged from 1.4 to 2 for the $\eta$-Aquariids (mass ranges of $[10^{-3},10^{-2}]$g and $[10^{-2},10^{-1}]$g respectively) and 1.85 and 2.51 for the Orionids (masses in $[10^{-3},1]$g and $[10^{-2},1]$g respectively). Though the scatter in reported values for the same magnitude ranges makes firm conclusions difficult \citep{Hughes1987b}, the data hints that the Orionids have a slightly higher $s$ than the $\eta$-Aquariids.   
  
  In-situ mass distribution indices measured by the Giotto, Vega-1, and Vega-2 spacecrafts when approaching 1P/Halley's nucleus are summarized in \cite{Hughes1987b}. Mass distribution indices of the dust changed with the distance between the spacecrafts and the nucleus in an uncertain manner. Vega-1 and Vega-2 measurements led to $s$ estimates between 1.54 and 1.92 (masses between $10^{-6}$ and $10^{-12}$g) with the SP-2 dust detector and between 1.84 and 2.53 (masses between $10^{-10}$ and $10^{-13}$g) with the DUCMA instrument. Giotto/DIDSY results spanned from 1.49 to 2.03 for masses between $10^{-6}$ and $10^{-9}$g. Most of the in-situ $s$ estimates are lower than 2 \citep{Hughes1987b}. 
  
  Confronting the meteoroids mass index measured close to the comet's nucleus and deduced from meteor observations is difficult for several reasons. The observed mass ranges barely overlap, preventing the comparison of the $s$ measured for meteoroids of similar mass. In addition, the variability of in-situ $s$ estimates in function of the distance to the comet prevents a clear estimate of the dust mass index close to the nucleus' surface. In consequence, we can only conclude that mass index estimates of 1P/Halley's meteoroids of different masses and observed at different locations before 1986 range between 1.4 and 2.5.
  
  \subsubsection{Orionids mass index}
  
  Subsequent meteor studies showed better agreement in the mass index measured for specific years, and highlighted the variability of $s$ for both showers. From visual data, \cite{Dubietis2003} measured mass indices range from 1.83 to 2.11 for the Orionids between 1984 and 2001, with an average of 1.87. A particularly low value of 1.83 ($r=2.25$) was estimated during the 1993 apparition of the shower.  
  
  Similarly, \cite{Rendtel2008} measured $s$ values close to the peak of maximum activity varying between 1.46 and 1.96 between 1979 and 2006. The lowest mass index was found during the 2006 outburst, with an average of 1.69 and a peak value of 1.46 \citep{Rendtel2008} or less \citep{TR2007}. As in \cite{Dubietis2003}, the mass index was found to present small variations around an average value of $s=1.87$ ($r=2.4$).
  
  The computation of several meteor showers mass indices from the data of the Canadian Meteor Orbit Radar (CMOR) between 2007 and 2010 confirmed the time-variability of the $\eta$-Aquariids and Orionids $s$ \citep{Blaauw2011}. Mass indices varied between 1.93 and 1.65, reaching their minimum value around the peak of maximum activity. The comparison of $s$ around the peak time between 2007 and 2009 also showed a variability of the mass index from year to year (varying from 1.65 to 1.77). These results are consistent with \cite{Cevolani1996}, who measured $s>1.8$ from forward scatter radio data for both showers. 
   
  \cite{Kero2011} estimated a mass index slightly above 2.0 from Orionids measurements of head echoes by the MU radar. Similarly, head echo measurements performed with the high power large aperture radar MAARSY between 2013 and 2016 resulted in $s=1.95$, appropriate to meteoroid masses between $10^{-12}$ and $10^{-7}$ kg \citep{Schult2018}.

\section{Eta-Aquariids}\label{section:eta}

  \subsection{General features}
  
  The $\eta$-Aquariids were the first meteor shower to be linked to comet 1P/Halley. In 1876, A. S. Herschel calculated the theoretical radiant of meteors associated with several comets, and estimated that 1P/Halley could be responsible for a shower in early May \citep{Kronk2014}. In 1910, the correlation between the $\eta$-Aquariids and Halley was established by Olivier \citep{Olivier1912}.  
  
  The $\eta$-Aquariids is the third strongest annual meteor shower observable at Earth, and one of the most active showers observable from the Southern Hemisphere. Visual observations of the $\eta$-Aquariids are strongly favored for Southern Hemisphere observers compared to those in the north, where a higher proportion of meteor observers are located. The radiant elongation ($\leq$70\si{\degree} from the Sun) restricts the observation window from the Northern Hemisphere to a few hours before the Sun rises, and poor weather conditions in early May frequently hamper visual observations of the shower. 
  
  The smaller number of reliable $\eta$-Aquariids records has limited the modeling of this shower \citep{Sekhar2014}. Fortunately, observational constraints for $\eta$-Aquariids modeling are provided by radar measurements of the shower. In 1947, the $\eta$-Aquariids became one of the first streams to be detected using specular backscattering radio techniques at the Jodrell Bank Experimental Station \citep{Clegg1947}. Subsequently, multiple specular radar observations of the shower were conducted between the 1950s and 1990s \citep[e.g.,][]{Hajduk1981,Hajduk1982,Hajduk1985,Chebotarev1988}. The shower was also the first one to be clearly identified in head echo measurements, in particular by the interferometric 49.92 MHz high-power large aperture radar at the Jicamarca Radio Observatory \citep{Chau2008}. In addition, the $\eta$-Aquariids is the strongest stream detected in specular backscatter by the Advanced Meteor Orbit Radar \citep{Galligan2000}, which has a limiting sensitivity near +13. These observations support the idea that the $\eta$-Aquariids are particularly rich in small meteoroids \citep{Jenniskens2006,CB2015}, with masses below $10^{-8}$kg \citep{Schult2018}.
  
 \subsection{Activity}

 Visual observations of the $\eta$-Aquariids during the 20th century originate from a limited number of sources. As with the Orionids, \cite{Porubcan1991} processed observations gathered during the International Halley Watch campaign to derive an average activity profile of the shower between 1984 and 1987. The main peak of activity was identified as a sharp double maximum at solar longitudes of 45.5\si{\degree} and 46.5\si{\degree}, with an average peak ZHR of 50. The total period of activity exceeded one month, with a FWHM of 7 to 8 days. The existence of a small dip just after the maximum was also identified, with the presence of possible secondary maxima of activity. The profile is presented in Appendix \ref{old_obs} for reference. 

 An analysis of visual observations of the $\eta$-Aquariids from the Southern Hemisphere (South Africa) between 1986 and 1995, and in 1997 and 1998 is presented in \cite{Cooper1996,Cooper1997,Cooper1998}. The author estimated an average peak ZHR of 60-70 between \Lsol=43.5\si{\degree} and \Lsol=44\si{\degree}, with a possible second maximum around \Lsol=46-47\si{\degree}. A secondary maximum around \Lsol=48\si{\degree} was also noticed for the 1997 $\eta$-Aquariids by \cite{Rendtel1997}. Enhanced activity (ZHR higher than 100) was reported by \cite{Cooper1996} for the 1993 and 1995 apparitions.
 
 The comparison of the activity profiles computed by northern and southern observers highlight the importance of the location in the visual observation of the shower. On some occasions, both profiles show good agreement \citep[like in 1993 or in 1997, cf.][]{Cooper1997,Rendtel1997}. For the other apparitions, observers from the Southern Hemisphere were able to provide profiles of higher temporal resolution and longer duration, permitting, for example, the identification of enhanced activity in 1995 that was missed by northern observers. Cooper's activity profiles are presented in Appendix \ref{old_obs}. In the Appendix, Cooper's original activity profiles (computed with a population index $r_1$ of 2.3 and a limiting magnitude LM=5.8) were rescaled to a population index $r_2$ of 2.46 (see Section \ref{sec:selected_mass_index}) with the relation:
 
 \begin{equation}
     ZHR_{r_2} \simeq  ZHR_{r_1}*\left(\frac{r_2}{r_1}\right)^{(6.5-LM)}.
 \end{equation}
 
Following their analysis of the Orionids, \cite{Hajduk1973}, \cite{Hajduk1981}, \cite{Hajduk1982} and \cite{Hajduk1985} analyzed the variations of the $\eta$-Aquariids activity in visual and radar data. Again, the maximum activity was detected as a sharp double maximum around \Lsol=45\si{\degree} and \Lsol=47\si{\degree}, for an average location of about 45.5\si{\degree}. Much as its twin shower, the $\eta$-Aquariids displayed considerable variations in density along the orbit and a possible drift of the main peak's solar longitude was observed (from 44.7\si{\degree} in 1971 to 45\si{\degree} in 1975 and 47\si{\degree} in 1978). \\

 The long-term evolution of the $\eta$-Aquariids was also investigated by \cite{Dubietis2003}. As with the Orionids, the author computed the shower population index $r$ and average peak ZHR from visual observations in the IMO VMDB between 1989 and 2001. The population index of the $\eta$-Aquariids displayed a minimum in 1992-1994, when a particularly low $r$ value was also recorded for the Orionids in 1993 \citep{Rendtel1993}. If the $\eta$-Aquariids annual ZHR seemed to present some periodic trends, the reliability of these variations is reduced by the smaller number of available observations. No regular periodicity in the $\eta$-Aquariids rates was clearly identified by \cite{Hajduk1973} over the period 1900-1967. 

  A detailed analysis of the shower observed by the Canadian Meteor Orbit Radar (CMOR) between 2002 and 2014 is presented in \cite{CB2015}. The variability of the activity profiles is an additional indication of the existence of fine structure within the stream, as already noted by \cite{Blaauw2011}. The main peak was generally localized around \Lsol=45\si{\degree}, with a secondary peak around \Lsol=48\si{\degree}. In some cases however, a secondary peak supplemented the first one. The existence of a third peak around \Lsol=54\si{\degree} was observed in 2002 and 2006. The analysis also revealed the existence of two strong outbursts, in 2004 and 2013. In 2004, the maximum activity occurred close to the full moon and no visual observations can confirm or contradict the existence of an outburst. To our knowledge, CMOR is the only source of $\eta$-Aquariids measurements during the 2004 apparition. The 2013 outburst, predicted by \cite{Sato2014}, was successfully recorded by radio, visual, and video detection networks \citep[e.g.,][]{Molau2013b,Cooper2013,Steyaert2014,CB2015}. 

   \subsection{Mass index}
  
   \cite{Dubietis2003} estimated a mass index of 1.78 to 1.94 from visual observations of the shower between 1989 and 2001. The lowest value of 1.78 ($r=2.18$) was reached in 1992. Like the Orionids, an average mass index of 1.87 was estimated for the $\eta$-Aquariids \citep{Dubietis2003}. However, much lower values of $s$ were measured close to the maximum activity of the shower \citep[for example a $s$ of 1.72 in 1997, cf.][]{Rendtel1997}. 

   The $\eta$-Aquariids recorded by the CMOR radar in 2008 display a mass index varying between 2 (at the beginning and end of the activity) to 1.85 around \Lsol=44.5\si{\degree} \citep{Blaauw2011}. \cite{CB2015} estimated a peak mass index varying from 1.75 to 1.95 between 2012 and 2015. The mass index was found to be below 1.9 at the peak and above these values far from the peak.

 \section{Halleyids observations between 2002 and 2019} \label{section:Recent_observations}
 
 In the following sections, we analyze the long-term activity of the $\eta$-Aquariids and the Orionids, focusing on the period of coverage of the CMOR radar (since 2002). CMOR measurements are of particular importance between 2003 and 2010, when only a few observations of the Orionids (and even fewer for the $\eta$-Aquariids) were published in the literature. When available, CMOR results are compared with visual observations contained in the IMO VMDB and measurements of the VMN network. In this section, details about the available observations and our data processing are provided. The analysis of the resulting activity profiles is presented in Section \ref{section:analysis}. 
 
\subsection{Observations}

\subsubsection{IMO VMDB}

Activity profiles of the $\eta$-Aquariids are available in the IMO Visual Meteor Data Base$^\text{2}$ (VMDB) back to 1989. Because of the difficult observing conditions from the Northern Hemisphere, several apparitions of the shower were missed or only partially recorded. Observations are in particular missing in 1991, and also in 1993, 1996, 2004 and 2015 when the main activity occurred close to the full Moon. Because Northern observers tend to focus their attention on the estimated peak date, the total $\eta$-Aquariids duration is hardly retrievable on the sole basis of these profiles. Exceptions are years 2012, 2013, 2018 and 2019 when complete activity profiles are available on the website. 

Activity profiles of the Orionids are available in the VMDB going back to 1985, and continuously since 1989. When allowed by the Moon phase, visual observations cover the full period of activity of the shower (for example in 2006, 2008, 2010, 2012 or 2019). These visual profiles are particularly valuable when comparing the results of different detection networks. 

\subsubsection{IMO VMN}

The IMO Video Meteor Network (hereafter called VMN) is comprised of about 130 cameras dedicated to meteor observations. The network coverage extends mainly over Europe, with additional cameras located in the United States and Australia. The cameras are capable of recording meteors down to a limiting magnitude of about 3.0$\pm$0.8, and stars with a limiting magnitude of 4.0$\pm$0.9\footnote{http://www.imonet.org/imc13/meteoroids2013\_poster.pdf}. The meteor detection is automatically performed by the MetRec software$^3$, which also provides flux estimates. Monthly reports of the VMN are regularly published in WGN, the Journal of the IMO. 

A complementary web interface allows users to visualize and analyze the flux profile of the showers recorded by the network since 2011\footnote{https://meteorflux.org/}. Several filters (choice of time bin, the cameras used, radiant location, limiting magnitude, single or multiyear analysis, etc.) can be applied when computing the flux profile. A recent modification of the software offers the opportunity to modify the population index of a shower, which is now applied in the flux density calculation (and not only in the transformation of the flux density into a ZHR). 

\subsubsection{CMOR}

The CMOR radar has been providing consistent single-station and orbital, multifrequency observations of the Halleyids since 2002. The equipment consists of three independent radar systems running at frequencies of 17.45 MHz, 29.85 MHz, and 38.15 MHz. A detailed description of the instrument is presented in \cite{Brown2008,Brown2010}. In this work, the flux computation of the $\eta$-Aquariids and Orionids was performed using the 29.85 MHz and 38.15 MHz data, as 17.45 MHz suffers significant terrestrial interference, particularly in the early years of operation. The data processing was performed as described in \cite{CB2015}.

It is important to note that the hardware, experimental setup (pulse repetition frequency, receiver bandwidth, pulse shape, and duration) as well as the software and detection algorithms used by the 38 MHz CMOR system were completely unchanged since 2002. The 29 MHz system underwent a transmitter power upgrade in the summer of 2009, but the receiver, antennas, and software detection algorithms plus experimental setup remained unchanged compared to the pre-2009 period. 

The main change in the systems over time due to hardware aging is the transmit power output which is directly measured and recorded for each system every 30 minutes or less and included as a correction in flux calculations.  As a result, we expect that shower profiles over this time frame can be compared and differences (particularly those showing up in both systems) confidently associated with real flux variability. 

 \subsection{Data processing}
 
  The comparison of visual, video, and radar data is challenging. Each observation method suffers its own biases, related to the observing conditions (atmospheric conditions, radiant elevation, etc.), the meteoroids' characteristics (mass, size, deceleration), and instrumental constraints. In addition, the three systems considered here are not equally sensitive to the same mass range, which could prevent a reliable comparison between the systems. Indeed, differences in the shower activity between systems could simply reflect different data processing assumptions, or the presence of different meteoroid size distributions. 
  
  As a result, in this study, we focus on a comparison of the global characteristics of the Halleyids activity profile (shape, maximum ZHR, and approximate location of the maximum activity). When possible, activity profiles are determined using a consistent population index and a consistent time resolution. Though the recomputation of the activity profiles with consistent parameters may reduce our sensitivity to some activity variations (e.g., rebinning with a longer time interval could obscure short-term variations), it is necessary for reliable comparison between the different data sets. 
  
  \subsubsection{Mass index} \label{sec:selected_mass_index}
  
 Using radar measurements as our baseline, we assume for the rest of this work a constant mass index of 1.9 ($r=2.46$) for the $\eta$-Aquariids \citep{CB2015} and 1.95 ($r=2.59$) for the Orionids \citep{Schult2018}. These values were used for all flux computations and ZHR estimates derived from  VMN and CMOR data; no additional correction was applied to these observations, except an additional normalizing factor described in Section \ref{sec:normalization}. 
 
 In the VMDB, the choice of the population index and the temporal resolution of the profile is not directly offered to the user. As a first approximation, we could attempt to rescale the visual profiles to our selected population index values. However, such a transformation requires knowledge of the limiting magnitude ($LM$) of the observation, which is not accessible and varies substantially with moonlight conditions. In addition, ZHR estimates in the VMDB frequently result from an average of several interval counts, each one processed with different corrective factors. Applying a uniform correction (assuming for example a constant $LM$ value) to the VMDB profiles, recorded in very different observing conditions, is therefore not realistic. However, population indices usually applied by the IMO to compute the activity profiles are $r=2.4$ ($s=1.87$) for the $\eta$-Aquariids and $r=2.5$ ($s=1.92$) for the Orionids, which are close to our selected values. Differences induced by use of these  mass indices as opposed to our values, are lower than the usual uncertainty on the ZHR computation.

  \subsubsection{Time resolution}
  
   A constant bin of 1\si{\degree} in solar longitude was considered for the activity profile computations of the VMN and CMOR data. The selection of a rather large time bin was made to ease the comparison of our results with previous published works \citep{Koschack1991,Dubietis2003}. Because of the limited visibility of the $\eta$-Aquariids radiant from CMOR's latitude of +43, the time resolution for CMOR profiles for this shower are necessarily limited to half a day anyway. 
   
     \begin{table*}[!ht]
      \centering
      \begin{tabular}{ccccc}
        \hline
        \hline
          & VMDB & VMN & CMOR 29 MHz & CMOR 38 MHz  \\
  $\eta$-Aquariids & 1 &  0.6 & 0.45  & 1 \\
          Orionids &  1 & 0.8 & 0.74 & 1.63 \\
        \hline
        \hline
      \end{tabular}
      \caption{Normalization factors applied to the $\eta$-Aquariids and Orionids activity measured by the VMDB, VMN, and CMOR networks. Each ZHR estimate presented in Section \ref{section:analysis} results from the multiplication of the original ZHR measured by each system with the coefficients listed above. }
      \label{table:normalizing_factors}
  \end{table*}
   
   Much as is the case for the shower population index, the resolution of the profiles available in the VMDB cannot be selected directly by the user. A variable time bin (related to the quantity and frequency of the reported observations) is usually applied to different portions of the profile to increase the reliability of the ZHR rates presented. In this work, no smoothing or interpolation of the available visual profile as a function of solar longitude was applied. The individual activity profiles were plotted against the video and the radar data without further correction. As a result, the analysis of the main peak time and intensity requires a caution as is described in later sections. 
   
\subsubsection{Normalization} \label{sec:normalization}

When we started comparing the original activity profiles derived for each system, an initial divergence of the ZHR-equivalent levels recorded was immediately evident.  The main peak activity measured from CMOR (29 MHz) was systematically 1 to 1.5 times higher than the VMN measurements, and 1 to 2.5 times higher than the VMDB records. These differences were observed for both showers and each year considered. They are therefore likely issues of the limiting sensitivity being systematically under or over estimated, calibration differences or observational biases, rather than caused by differences in the observed meteoroids population with mass range which would produce more random scatter. 

Since this work focuses on the relative long-term variability of the Orionids and $\eta$-Aquariids meteor showers, we decided to scale the activity levels measured by each system to obtain similar ZHR rates over a long period of time ($\sim$1985 to 2019). To ease comparison of recent measurements with older visual observations (published or contained in the VMDB), VMN and CMOR profiles were scaled to the VMDB activity curves. Normalizing factors for each system were determined by adjusting the maximum ZHR rates measured by the system and the VMDB over the observation period. 

In this work, each VMN profile has been multiplied by 0.8 for the Orionids and by 0.6 for the $\eta$-Aquariids. These different correction factors for the two showers might be due to the reduced number of visual observations available for the $\eta$-Aquariids. CMOR 29 MHz profiles were multiplied by 0.74 for the Orionids and 0.45 for the $\eta$-Aquariids.

As mentioned in \cite{CB2015}, the flux calculated from the 38 MHz data is about two to three times lower than the flux determined from the 29 MHz system (probably because of uncertainties in the mass index correction factors and/or initial radius bias). As a consequence, the Orionids profiles computed from the 38 MHz data needed to be increased by a factor of 1.63, while no modification was required for the $\eta$-Aquariids profiles. The normalizing factors for each system are summarized in Table \ref{table:normalizing_factors}. 
 
 Each curve and ZHR estimates presented in Section \ref{section:analysis} and Appendices B \& C were scaled by the factors of Table \ref{table:normalizing_factors}, and are noted ZHR$_\text{v}$ in the text. Therefore, results and figures of the following sections should not be interpreted to represent absolute ZHR estimates of the Halleyids meteor showers. Scaling the video and radar profiles to visual observations results from an arbitrary choice to make the longer time base of visual data the standard, and not from a conviction that visual records better match the real activity of the showers. 
 The purpose of this work is instead to provide consistent measurements of the activity profiles (duration, shape) and year-to-year variations in activity between different detection networks. 
 
  \section{Results and analysis} \label{section:analysis}
  
  CMOR activity profiles of the $\eta$-Aquariids and Orionids between 2002 and 2019 are presented in Appendix \ref{new_obs}, Figures \ref{fig:new_etas} and \ref{fig:new_oris}. Results of the 29 MHz (blue) and 38 MHz (green) systems are plotted along with the VMDB profiles (in black) and the VMN observations (beginning in 2011, in red).

    \subsection{Individual activity profiles} \label{sec:individual_profiles}
 
    \subsubsection{$\eta$-Aquariids}
    
  The $\eta$-Aquariids are generally active between \Lsol=35\si{\degree} and \Lsol=60\si{\degree}, with highest rates recorded between \Lsol=40\si{\degree} and \Lsol=55\si{\degree}. ZHR$_\text{v}$ estimates usually reach a maximum value of 60 to 80 meteors per hour, except for two years of enhanced activity: the 2004 outburst recorded by CMOR and already reported by \cite{CB2015}, and the 2013 outburst predicted by \cite{Sato2014} and well covered by visual, video, and radar observations. The location of the main peak varies between \Lsol=44\si{\degree} and 47\si{\degree}, with the existence of several peaks of lower intensity after the main maximum. Most of the visual profiles peak around 45\si{\degree} to 45.75\si{\degree} (2001, 2005, second maxima in 2006, 2007, 2009-2014 and 2017), with some occurring earlier than this time (2003, first maxima in 2006, 2015 and 2016) and a few later (2002, 2008, 2018 and 2019). The changing activity profiles of the shower from return to return suggests that there is structure in the stream, as such large changes in peak times are not common among the major showers  \citep[cf.][]{Rendtel2008b}.
  
  Good agreement in the duration and peak time in a given year is found between the different techniques. VMN measurements are comparable to the VMDB, at least when the visual observations provide a complete activity profile with good resolution (like from 2017 to 2019). In some years, the difference between the two networks is larger (e.g., in 2011 or 2014). CMOR profiles typically follow the shape defined by optical measurements, but present more variations in the main peak location and suggest the existence of subpeaks. 
  
  For example, the 2012, 2013, 2017 and 2019 apparitions are very consistent between the networks, but the radar main peak location diverges from the optical data in 2005, 2015, 2016 and 2018. Profiles obtained using the 38 MHz and 29 MHz data also differ slightly from one another, but there is no evidence that one frequency better reproduces the optical observations for every apparition of the shower. Sudden gaps in the radar profiles indicate periods for which the records are missing (because of instrumental issues or lack of reliable measurements). 
  
  Unfortunately, radar observations were not available around the estimated peak time in 2002, 2003, 2006 and potentially 2010. In addition, the 2009 radar profile largely diverges from the VMDB observations around the peak time, in this instance because of a large scale equipment change to the 29 MHz CMOR system which occurred in this time frame. As a result, only the 38 MHz data is available to characterize this specific apparition.

      \begin{figure*}[!ht]
        \centering
        \includegraphics[width=\textwidth]{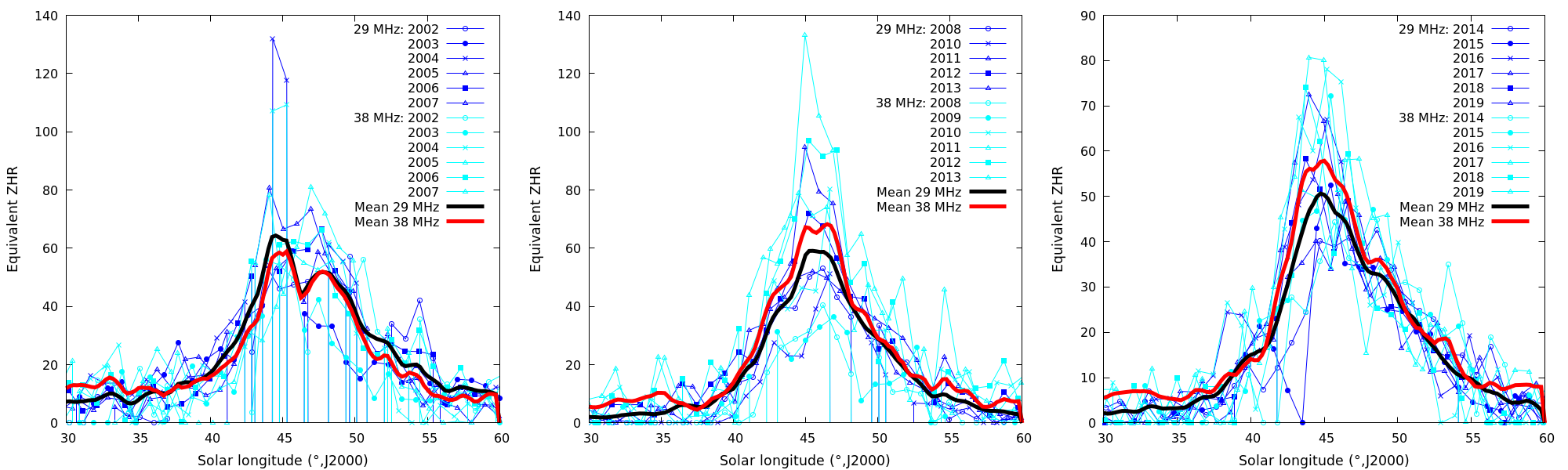}
        \includegraphics[width=\textwidth]{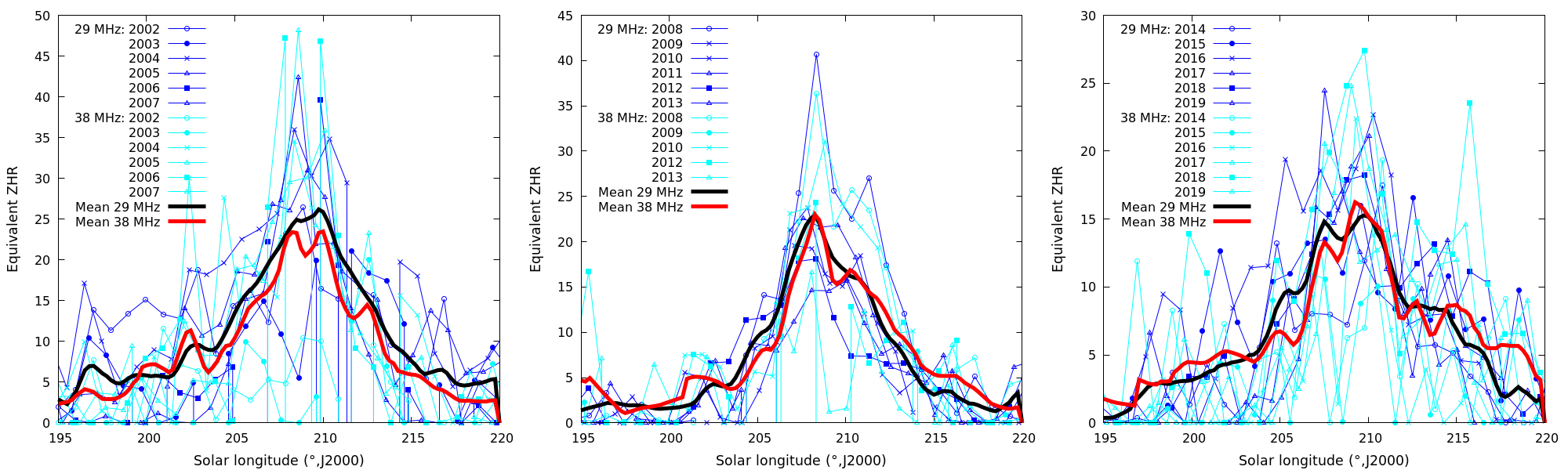}
        \caption{Individual and average activity profiles (ZHR$_\text{v}$) of the $\eta$-Aquariids (top) and Orionids (bottom) recorded by the radar CMOR between 2002 and 2007 (left), 2008 and 2013 (center), and 2014 and 2019 (right).}
        \label{fig:CMOR_averages}
    \end{figure*}

    \subsubsection{Orionids}
    
   In our data, the Orionids display noticeable activity between \Lsol=195\si{\degree} and \Lsol=220\si{\degree}. The peak ZHR$_\text{v}$ rates vary between 20 and 40 meteors per hour up to 2003 and after 2012, and between 40 and 80 around the 2006 and 2007 resonant return years. Years of enhanced activity tend to present a sharper and less scattered profile than apparitions of moderate intensity. As noticed by many observers (cf. Section \ref{section:ori}), the Orionids present a broad maximum between 206\si{\degree} and 211\si{\degree}, with several subpeaks of variable intensity and location. The main peak of activity is therefore difficult to assess for several Orionids apparitions. On some occasions, the highest ZHR$_\text{v}$ rates do not coincide with the center of the broad maximum (e.g., in 2018, 2019) or the presence of multiple subpeaks prevents the identification of the main peak (e.g., in 2009 or 2017). In 2017 for example, the visual profile displays two maxima of equivalent strength, clearly separated by a dip already observed in the past (see Section \ref{section:ori}). 
   
    For the Orionids, the observations by different detection methods are less consistent than for the $\eta$-Aquariids. This is in part a consequence of the lower number statistics for the Orionids compared to the $\eta$-Aquariids. From figure \ref{fig:new_oris}, it is clear that the shower does not show a stable activity profile from return to return.  The overall duration and location of the broad maximum of activity are similar between the systems, but the fine structure of the profiles differs. Clearly, the lower activity level of this shower increases the apparent discrepancies between our different data sets due to small number statistics. 
    
    Relatively good agreement is found between the VMN and VMDB profiles in 2014, 2017, and 2019, but not for earlier apparitions (e.g., 2011, 2012). Observations from the 38 MHz system appears to be particularly sensitive to the low-level activity of the shower, leading to very broad profiles. CMOR activity profiles are similar to visual observations in 2005, 2007, 2008, and 2010. A lack of observations around the peak in 2006 does not allow us to clearly define the main peak location and intensity, but high ZHR$_\text{v}$ estimates ($>$70) were recorded at \Lsol=207\si{\degree} and \Lsol=209\si{\degree}. The overall shape of the CMOR and VMN profiles are consistent in 2011, 2012, and 2013, but differ for other apparitions of the shower, especially in 2016 and 2018. In 2018, an early peak of activity is noticeable in the VMDB data (ZHR$_\text{v}$ of 54$\pm$7 around \Lsol=198.54\si{\degree}) that might be detected by the 38 MHz system too, but does not appear in the VMN activity curve.

  \subsection{Average activity} \label{sec:average}

   \subsubsection{2002-2019}
 
  Figure \ref{fig:CMOR_averages} presents the average activity profiles of the $\eta$-Aquariids and Orionids recorded by CMOR over the periods 2002-2007, 2008-2013, and 2014-2019. The average profiles display significant variations as a function of the period considered. Early CMOR measurements match published estimates of $\eta$-Aquariids activity (see Section \ref{section:eta}). The average $\eta$-Aquariids profiles of both frequencies from 2002-2007 show an initial maximum at \Lsol=44.5-45\si{\degree}, followed by a broad secondary peak at \Lsol=47.5-48\si{\degree}. In the average profiles between 2008 and 2019, the presence of these two clear peaks of activity becomes less apparent. The average activity profile over the period 2014 to 2019 shows a broad maximum around \Lsol=45\si{\degree}, with secondary bumps around \Lsol=48.5\si{\degree} and \Lsol=53\si{\degree}. 
  
  For the Orionids, the presence of subpeaks in the average profile varies as a function of the period considered, but two main activity peaks (around \Lsol=208.5\si{\degree} and \Lsol=210.5\si{\degree}) remain at a similar location over time. The two peaks are separated by a dip of varying depth, and the relative intensity between the two maxima also changes in the three average profiles. 
  
\begin{figure}[!ht]
        \centering
        \includegraphics[width=0.48\textwidth]{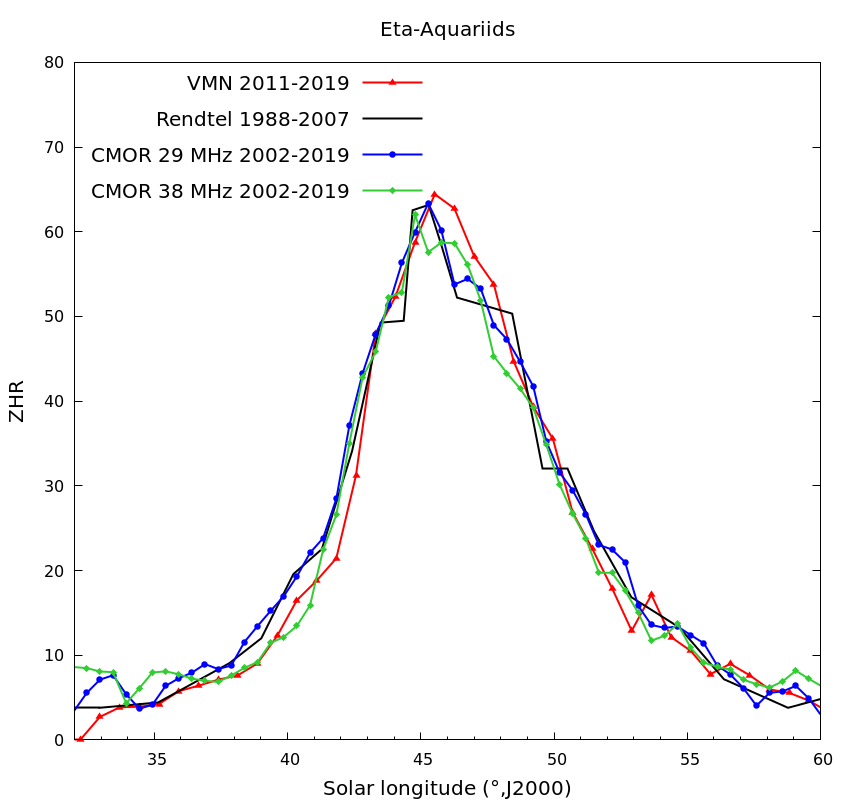}
        \caption{Average activity (ZHR$_\text{v}$) of the Eta-Aquariid meteor shower, as measured with the CMOR 29 MHz system (blue) and the 38 MHz system (green) between 2002 and 2019, the VMN (red) between 2011 and 2019, and visual observations from the IMO from 1988 to 2007 \citep[][in black]{Rendtel2008b}.}
        \label{fig:Eta_average}
    \end{figure}

\begin{figure}[!ht]
        \centering
        \includegraphics[width=0.48\textwidth]{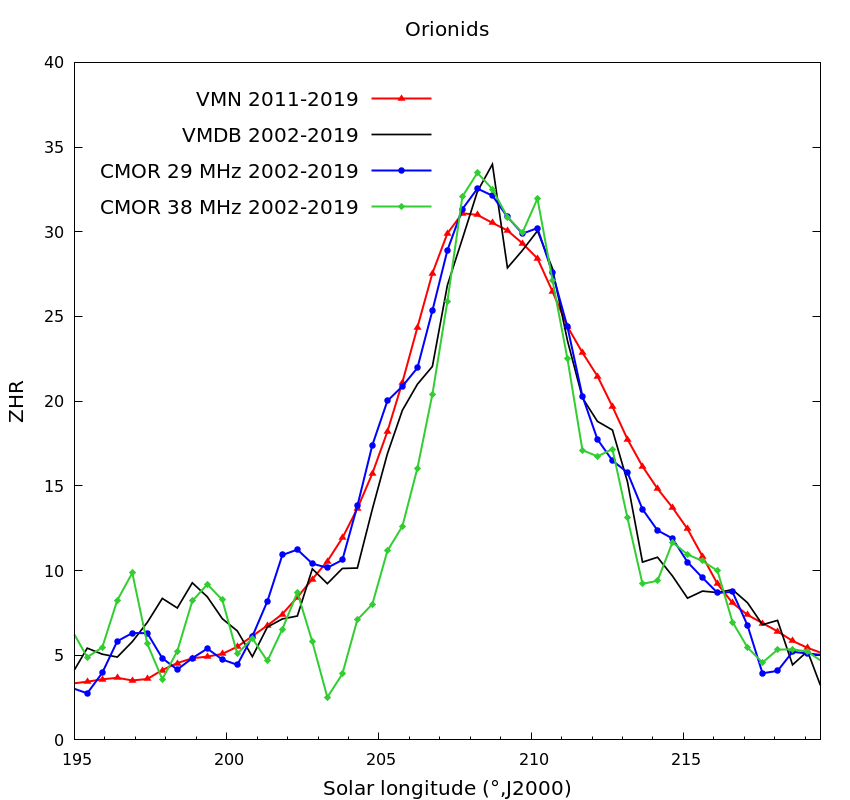}
        \caption{Average activity (ZHR$_\text{v}$) of the Orionid meteor shower, as measured with the CMOR 29 MHz system (blue) and the 38 MHz system (green) between 2002 and 2019, the VMN (red) between 2011 and 2019, and visual observations from the VMDB between 2002 and 2019 (black). Differences in the 38 MHz profile are increased by the lower number of available observations. More information is given in the text.}
        \label{fig:Ori_average}
    \end{figure}

  The overall average activity profiles as measured by the 29 MHz (in blue) and the 38 MHz (in green) radar systems between 2002 and 2019 for the $\eta$-Aquariids and Orionids are presented in Figure \ref{fig:Eta_average} and \ref{fig:Ori_average}. The average profile computed from the VMN  data (in red) is also shown for comparison. The average VMDB profile is shown for the Orionids but because of the limited $\eta$-Aquariids data  available on the VMDB website, Figure \ref{fig:Eta_average} instead includes the average IMO observations from 1988 to 2007 determined by \cite{Rendtel2008b}. 
  
 Averaging the showers' activity over long time periods smoothes the year-to-year profiles presented in Appendices \ref{old_obs} and \ref{new_obs}. This is equivalent to applying a low-pass filter to the time series, removing most of the information about the fine structure in the stream. The average $\eta$-Aquariids profiles measured by the different systems show remarkable consistency. The main peaks are located at similar solar longitudes and the profiles display similar rising and falling slopes of activity. The consistency between measurements of different systems supports the idea that there is no significant size sorting of the particles within the $\eta$-Aquariids, in agreement with earlier studies by \cite{Jenniskens2006} or \cite{CB2015}.

 The average Orionids profiles are also broadly similar between the detection systems. The CMOR and VMDB determined average profiles display a double peaked maximum around \Lsol=208.5\si{\degree} and \Lsol=210.5\si{\degree}, a feature noticed in several previous works (see Section \ref{section:ori_prev_obs}). The VMN average presents a more flat broad maximum between \Lsol=207\si{\degree} and \Lsol=211\si{\degree}. 
 
 VMN, VMDB, and the CMOR 29 MHz profiles present a similar rising slope of activity. The CMOR 38 MHz average profile differs noticeably from these others with lower activity measured between 203\si{\degree} to 206\si{\degree} in solar longitude. Low ZHR$_v$ rates between \Lsol=203\si{\degree} and \Lsol=205\si{\degree} were measured by the system during 7 returns of the shower. However, observations at these solar longitudes are missing for 9 Orionid apparitions, decreasing the reliability of the ZHR$_v$ gap at \Lsol=203.5\si{\degree} in CMOR 38 MHz data.
 
 If real, several possible reasons could explain this difference. It could be due to an increase in the shower mass index in this interval, which would attenuate detection numbers through both a larger initial radius effect and a smaller number of echoes detectable by 38 MHz. It may also be a function of the lower sensitivity of 38 MHz to the Orionids compared to the sporadic background in general, an effect noticeable in the larger activity scatter between \Lsol=195-202\si{\degree}.
 
 The falling slope of the average profiles is similar for all four data sets, except for a slight divergence of the VMN profile between \Lsol=211-215\si{\degree}.   The ZHR$_\text{v}$ rates decrease more rapidly in visual and radar data than in the video records. However, it is unclear if these differences are related to the stream characteristics or to biases in the shower observations and the data processing. 

  \subsubsection{General activity profile shape}
 
 Following \cite{Jenniskens1994}, the shape of a meteor shower ZHR profile can be approximated by a double-exponential curve, which can be expressed as: 
 
      \begin{equation} \label{eq:ZHR}
     \left\{
           \begin{aligned}
      ZHR&=ZHR_{m}*10^{B(L_\odot-L_{\odot_{m}})} \\
      B&=+B_p \text{ if } L_\odot \leq L_{\odot_{m}} \\
      B&=-B_m \text{ if } L_\odot > L_{\odot_{m}}, \\
             \end{aligned}
             \right.
     \end{equation}
    where the maximum ZHR ($ZHR_{m}$) at solar longitude $L_{\odot_{m}}$ and the slope coefficients $B_p$ and $B_m$ are fit to the observed profile. 
    
    With the presence of a significant plateau in the Orionids activity profile, we found it preferable to replace $L_{\odot_{m}}$  with two solar longitudes $L_{\odot_{m,1}}$ and $L_{\odot_{m,2}}$ delimiting the plateau location. With this adaptation, the coefficients  $B_{p2}$ and $B_{m2}$ characterize the slope of the ascending (\Lsol$\leq$ $L_{\odot_{m,1}}$) and descending (\Lsol $>L_{\odot_{m,1}}$) branch of the activity profile. The plateau region is modeled by a linear function $ZHR=\alpha x+\beta$, where $\alpha$ and $\beta$ are determined from the estimates of $\{L_{\odot_{m,1}},ZHR( L_{\odot_{m,1}})\}$ and  $\{L_{\odot_{m,2}},ZHR( L_{\odot_{m,2}})\}$ to ensure the continuity of the modeled ZHR profile.
    
    A standard least-squares fit of each average activity of Figures \ref{fig:Eta_average} and \ref{fig:Ori_average} was performed to determine the slope parameters $B_p$, $B_m$, $B_{p2}$ and $B_{m2}$. Results with the associated formal uncertainties of the fit are summarized in Table \ref{table:Bvalues}. 
     
  \begin{table*}[!ht]
  \centering
   $\eta$-Aquariids\\[0.2cm]
   \begin{tabular}{ccccccccc}
   \hline
   \hline
     System & $B_p$ & $B_m$  & $B_{p2}$ & $B_{m2}$  & $ZHR_{m}$ & $ZHR_{m2}$ & $L_{\odot_{m,1}}$ (\si{\degree}) & $L_{\odot_{m,2}}$ (\si{\degree})  \\
     \hline
     29 MHz &  0.111 $\pm$ 0.005 & 0.073 $\pm$ 0.003 & 0.122 $\pm$ 0.008 & 0.094 $\pm$ 0.007 & 76 & 66 & 44.5 & 46.8 \\
     38 MHz &  0.126 $\pm$ 0.012 & 0.074 $\pm$ 0.006 & 0.164 $\pm$ 0.010 & 0.108 $\pm$ 0.006 & 51 &  49 & 44.7 & 45.3 \\
        VMN &  0.129 $\pm$ 0.010 & 0.083 $\pm$ 0.006 & 0.180 $\pm$ 0.012 & 0.105 $\pm$ 0.006 & 70 &  60 & 44.3 & 47.3 \\
       VMDB &  0.104 $\pm$ 0.014 & 0.065 $\pm$ 0.007 & 0.121 $\pm$ 0.011 & 0.106 $\pm$ 0.014 & 70 & 64 & 44.4 & 46.4 \\
    \hline
    {[}1{]} & 0.135 $\pm$ 0.003 & 0.078 $\pm$ 0.003 & - & - & - & - & - & -\\
     \hline 
     \hline
   \end{tabular} \\
   \vspace{0.2cm}\mbox{ }
    Orionids\\[0.2cm]
   \begin{tabular}{ccccccccc}
   \hline
   \hline
     System & $B_p$ & $B_m$  & $B_{p2}$ & $B_{m2}$  & $ZHR_{m}$ & $ZHR_{m2}$ & $L_{\odot_{m,1}}$ (\si{\degree}) & $L_{\odot_{m,2}}$ (\si{\degree})  \\
     \hline
    29 MHz & 0.078 $\pm$ 0.009 & 0.069 $\pm$ 0.007 & 0.106 $\pm$ 0.008 & 0.103 $\pm$ 0.009 & 33 & 31 & 207.3 & 209.7 \\
    38 MHz & 0.215 $\pm$ 0.047 & 0.118 $\pm$ 0.024 & 0.349 $\pm$ 0.043 & 0.233 $\pm$ 0.032 & 35 & 32 &  208.2 & 210.3  \\
       VMN & 0.073 $\pm$ 0.008 & 0.054 $\pm$ 0.006 & 0.128 $\pm$ 0.011 & 0.095 $\pm$ 0.009 & 35 & 32 & 207.7 & 210.0 \\
      VMDB & 0.073 $\pm$ 0.022 & 0.080 $\pm$ 0.023 & 0.120 $\pm$ 0.027 & 0.168 $\pm$ 0.044 & 40 &  33 & 207.7 & 210.5 \\
    \hline
    {[}2{]} & 0.122 & 0.098  & - & - & - & - & - & -\\
    {[}3{]} & 0.140 & 0.140  & - & - & - & - & - & -\\
   \hline
   \hline
   \end{tabular}
   \caption{Fit coefficients of the rising and falling slopes of $\eta$-Aquariids and Orionids average profiles presented in Figures \ref{fig:Eta_average} and \ref{fig:Ori_average}. Each estimate is followed by the fit formal uncertainty. The parameters $B_{p2}$ and $B_{m2}$ are adaptations of Equation \ref{eq:ZHR}'s $B_{p}$ and $B_{m}$ coefficients for profiles containing a dip or a plateau. The modeled maximum ZHR ($ZHR_{m},ZHR_{m2}$) and the solar longitude endpoints of the plateau ($L_{\odot_{m,1}},L_{\odot_{m,2}}$) are provided for information. [1]: $\eta$-Aquariids coefficients provided by \cite{CB2015}. [2,3]: Orionids coefficients determined by \cite{Jenniskens1994} from Northern [2] and Southern [3] Hemisphere observations. \label{table:Bvalues}}
   
  \end{table*}
  
  As mentioned in \cite{Jenniskens2006} and \cite{CB2015}, the $\eta$-Aquariids rise of activity is more sudden than the post-maximum decrease in intensity. Depending on the detection system, we find the slope coefficient of the ascending branch to vary between 0.104 and 0.180, and for the descending branch to change between 0.065 to 0.108. These estimates are consistent with \cite{CB2015} results (cf. {[}1{]} in Table \ref{table:Bvalues}).
  
  As expected, the Orionids display a much more symmetric activity profile, especially in the radar measurements. Our estimates of the ascending branch slope for each system vary between 0.073 and 0.349, for a descending coefficient of 0.054 to 0.233. These value span \cite{Jenniskens1994}'s result derived from visual observations carried out in the Northern hemisphere, and are a bit lower than the values deduced from Southern Hemisphere data (see {[}2{]} \& {[}3{]} in Table \ref{table:Bvalues}). 
 
    \begin{figure*}[!ht]
     \centering
     \includegraphics[width=.99\textwidth]{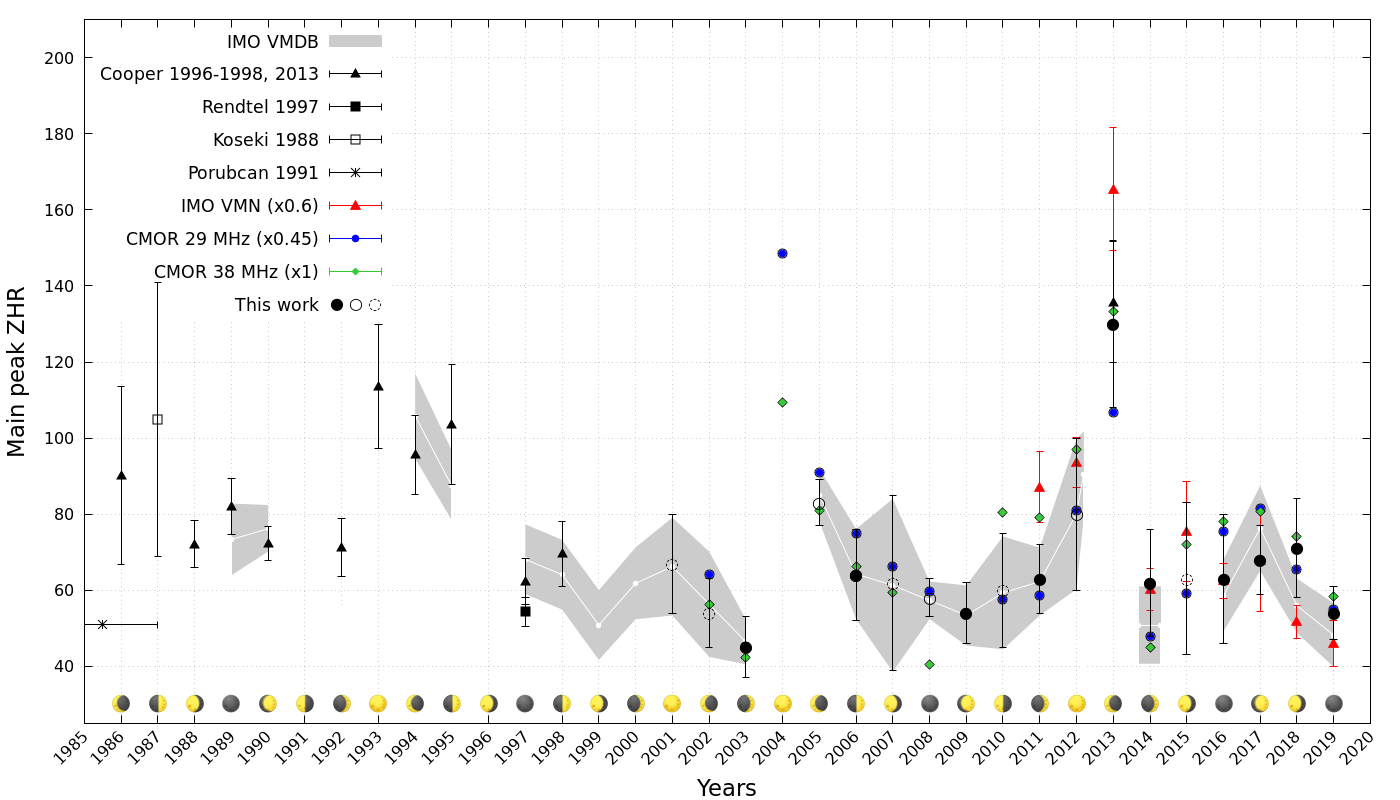}
     \caption{Annual variation of the $\eta$-Aquariids main peak (ZHR$_\text{v}$) between 1985 and 2019. The maximum rates recorded by CMOR 29 MHz (blue) and 38 MHz (green) are plotted along with the results of the VMN system (in red). Gray, black, and empty symbols refer to the maximum ZHR$_\text{v}$ deduced from visual observations. The new analysis of the IMO VMDB observations since 2001 is characterized by circles of different colors depending on the reliability of the peak observations (filled: reliable estimate, empty: uncertain value, empty and dashed: very uncertain estimate). The gray filled curve represents the approximate ZHR$_\text{v}$ maximum and uncertainty from the directly available online VMDB profiles (with no rigorous processing of the data). Additional measurements (especially before 2001) refer to previous publications. Horizontal error bars imply an average estimate of the maximum (ZHR$_\text{v}$) over the indicated period. The bottom symbols represent the phase of the moon around the estimated main peak date. More information is given in the text.      \label{fig:annualZHR_ETA}}
   \end{figure*}
   
      \begin{figure*}[!ht]
     \centering
     \includegraphics[width=.99\textwidth]{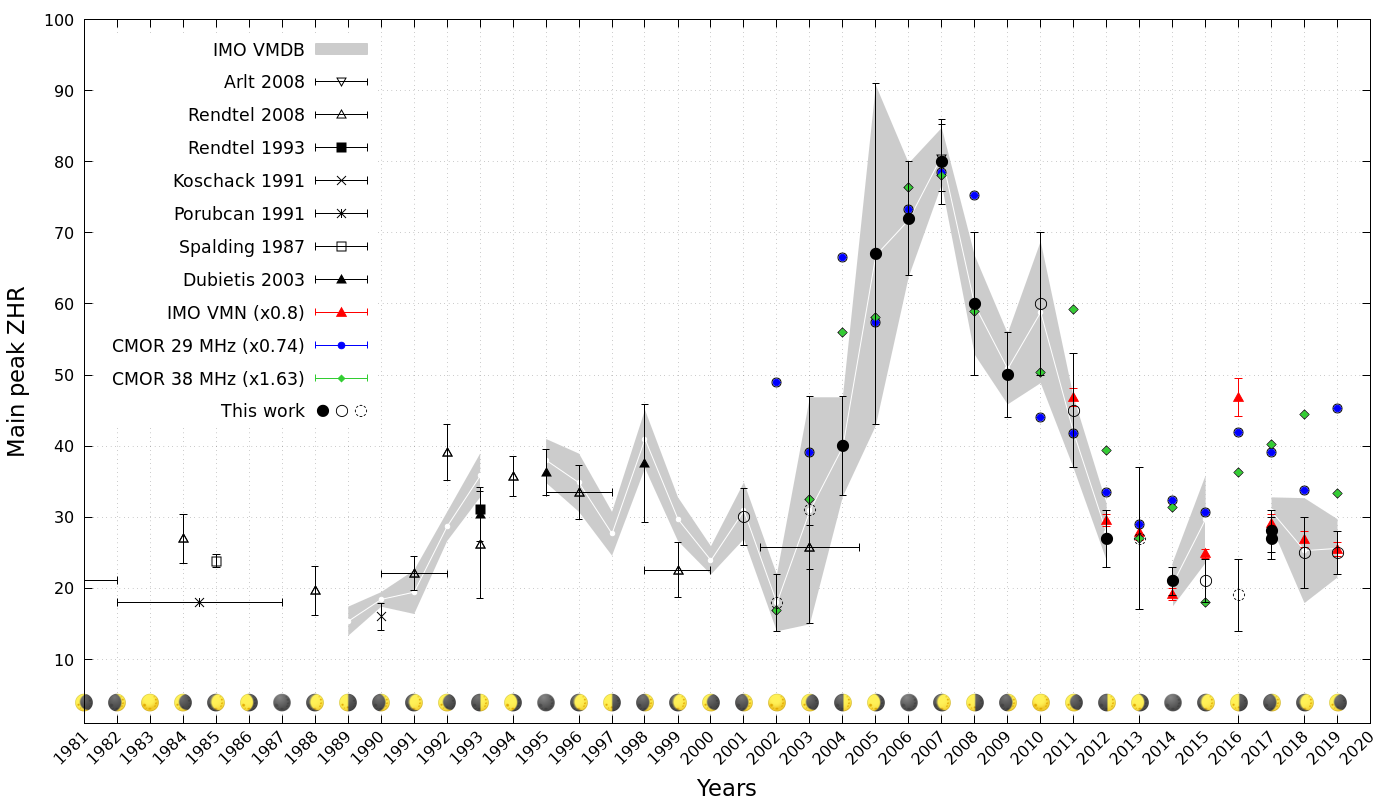}
     \caption{Annual variation of the Orionids main peak (ZHR$_\text{v}$) between 1985 and 2019. Same notations are used as in Figure \ref{fig:annualZHR_ETA}.      \label{fig:annualZHR_ORI}}
   \end{figure*}

  \subsection{Annual variation}
  
  Following \cite{Dubietis2003} and \cite{Rendtel2008}, we investigate in this section the long-term evolution of the Halleyids main peak. The variation of the showers' strength and structure (peak location) since 1985 is analyzed. We remind the reader that the results presented in this section reflect the general behavior of the showers' activity (ZHR$_\text{v}$), and not absolute measurements of the apparitions' ZHRs. Assuming a constant population index with time influences the general shape of the profiles computed, as well as the main peak intensity and location. In addition, the selection of a large solar longitude binning (\Lsol=1\si{\degree}) smears out the presence of short-lived structures of higher (ZHR$_\text{v}$) rates and can modify the location of the maximum intensity for each apparition of the showers. 
  
  \subsubsection{Maximum activity value} \label{sec:ZHRmax}
  
  Figures \ref{fig:annualZHR_ETA} and \ref{fig:annualZHR_ORI} illustrate the evolution of the magnitude of the $\eta$-Aquariids and Orionids main peak of activity per year. Maximum (ZHR$_\text{v}$) rates measured by the VMN (red), CMOR 29 MHz (blue), and CMOR 38 MHz (green) are plotted along with the result of visual observations (black, gray, and empty symbols). 
  
  Though the visual activity profiles provided by the IMO VMDB user interface could be used without additional processing in sections \ref{sec:individual_profiles} and \ref{sec:average}, the estimate of the main peak intensity for each apparition of the shower requires a more cautious analysis of the data. One of us (JR) re-processed the observations available in the IMO VMDB with our preferred constant study population indices of 2.46 for the $\eta$-Aquariids and 2.59 for the Orionids. 
  
  The time binning of several parts of the profile was modified to increase the reliability of the main peak location, making sure that the related ZHR estimate is not based on too few contributing intervals. In this re-analysis, the probable main peak intensity, location, and profile FWHM for every apparition of the $\eta$-Aquariids and Orionids since 2001 was determined. 
  
  The results are plotted as circles in Figures \ref{fig:annualZHR_ETA}  and \ref{fig:annualZHR_ORI}. The colors of the symbols reflect the reliability of the derived maximum ZHR$_\text{v}$ (filled: reliable estimate, empty: uncertain value, empty and dashed: very uncertain estimate), depending on the structure of the profile, and the number and quality of the available observations. Since these measurements are not directly retrievable from the VMDB website, numerical estimates are  provided in Appendix \ref{sec:maximum_rates}.
  
  For comparison, the interval of maximum intensity deduced from the VMDB profiles of Figures \ref{fig:new_etas} and \ref{fig:new_oris} is represented by the gray filled curve in Figures \ref{fig:annualZHR_ETA}  and \ref{fig:annualZHR_ORI}. The identification of the main peak was performed using a combination of the full activity profile and using observations conducted close to the estimated peak time ("Peak" tab in the live ZHR website). The maximum peak ZHR$_\text{v}$ was retrieved without any smoothing, fitting or extrapolation of the original data. When no observations are available around the supposed maximum of activity, no approximated ZHR$_\text{v}$ was computed; this is the reason for the large gaps for some years in the gray curve. 
  
  Additional measurements of the showers' activity (especially before 2001) published by several authors \citep{Spalding1987,Koseki1988,Porubcan1991,Koschack1991,Rendtel1993,Cooper1996,Cooper1997,Rendtel1997,Cooper1998,Dubietis2003,Rendtel2008,Arlt2008,Cooper2013} were also added in Figures \ref{fig:annualZHR_ETA}  and \ref{fig:annualZHR_ORI}  (filled or empty triangles, squares, etc.) for comparison.
  
  From Figure \ref{fig:annualZHR_ETA}, we notice that the annual activity variations of the $\eta$-Aquariids determined by the different detection networks is remarkably similar. Results from the 29 MHz system are in better agreement with optical measurements than for 38 MHz (especially in 2008). However, data obtained with both frequencies show a similar evolution with time, following the general trend found in the optical data. Since 1997, annual ZHR$_\text{v}$ maximum rates vary between an average of 65-70 meteors per hour, with the notable exception of two outbursts in 2004 and 2013 (ZHR$_\text{v}\ge$100). The possible existence of outbursts in 1993 and 1995 has been suggested by \cite{Cooper1996}. 
  
  Higher ZHR$_\text{v}$ rates tend to have been observed in the past. This may be due to a real enhancement of the shower average level activity or simply reflect changes in coverage between Northern and Southern observations. No clear periodicity in outburst years is apparent from this figure. A potential 12 year periodicity of the shower minimum of activity could be imagined if years 1990-1991, 2002-2003, and 2014-2015 corresponded to the lowest $\eta$-Aquariids rates measured since 1985. However, we find no conclusive evidence of such periodicity from the observations. 
  
  The Orionids annual peak activity evolution per year also shows relatively good agreement between the different techniques (cf. Figure \ref{fig:annualZHR_ORI}). Before 1990, published peak ZHR rates are close to 20 meteors per hour on average. A small systematic increase in the peak of the average activity level is noticeable over the period 1990 to 2001, reaching rates of about 40 meteors per hour in 1993 and 1997. A potential minimum of activity is observed in 2002, but is based on uncertain visual observations under poor lunar conditions and a very broad profile from the CMOR 38 MHz system. All techniques agree in an enhancement of Orionids activity between 2002 and 2013, with the highest rates of more than 70 meteors per hour reached for the 2006 and 2007 resonant years. The existence of an additional increase of activity in 2016 is observed in the VMN and CMOR data, but visual observations of the VMDB neither support or contradict this conclusion. 
  
  The greatest difference between the four data sets relates to the most recent apparition of the shower (2019). In this year, CMOR 29 MHz rates were systematically about twice that found from the optical techniques. The variations in the 38 MHz data are less reliable than the 29 MHz profile for the Orionids because of the lower sensitivity of that system and its correspondingly poorer noise statistics. 

   The existence of a periodicity to Orionid activity, caused by Jovian perturbations of the meteoroid stream, has been proposed by several authors \citep[e.g.,][]{McIntosh1983,Dubietis2003,Rendtel2008,Sekhar2014}. Visual and radar measurements suggested a periodic return of low meteor rates, raising the question of a 12 year variation of the Orionid minimum of activity \citep{McIntosh1983}. No periodic variation of the apparition of enhanced meteor rates has been clearly identified for the shower. However, increased meteor activity around 1984-1985 and 1993-1998 led \cite{Dubietis2003} to propose that a similar periodicity is linked to years of maximum Orionid activity. 
   
  The variance in the magnitude of peak activity between the three techniques for recent apparitions of the shower preclude any strong statement about the 12 year periodicity in the Orionids maximum meteor rates that seems to be supported  by Figure \ref{fig:annualZHR_ORI}.   
  In particular, when examining years of low peak intensity, we find a clear minimum in 1990, also reported by \cite{Koschack1991}, and another potential minimum in activity in 2002.  A 12 year periodicity would lead to a subsequent minimum in 2014, which is clearly observed in the VMN and VMDB data sets, and to a lesser extent in CMOR measurements. The existence of a periodicity in Orionid peak activity is therefore not rejected by our analysis. However, because of the uncertainty of the 2002 minimum, no strong conclusion can be drawn - additional observations of several Orionid returns covering another full 12 year cycle are required. 
    
    \begin{figure*}[!ht]
      \centering
      \includegraphics[width=.49\textwidth]{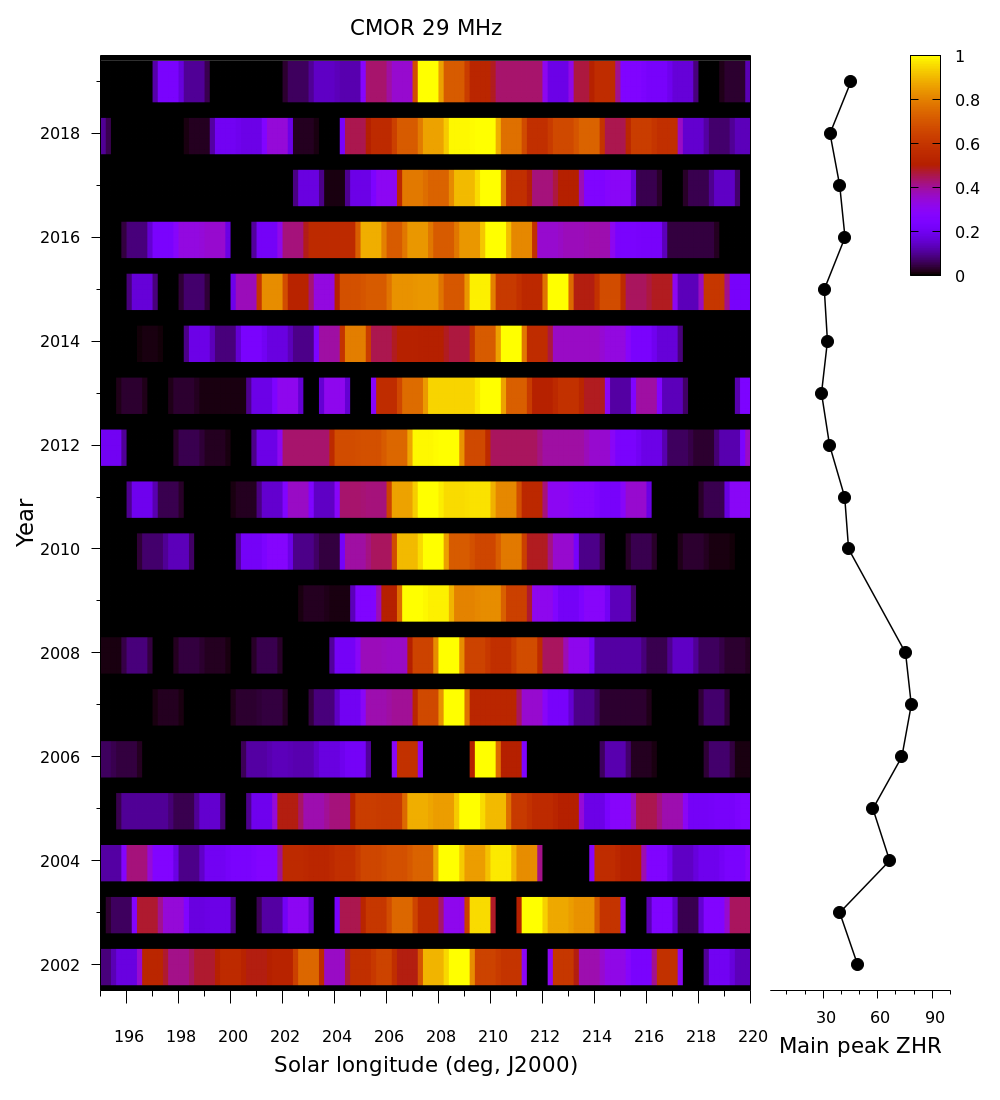}
      \includegraphics[width=.49\textwidth]{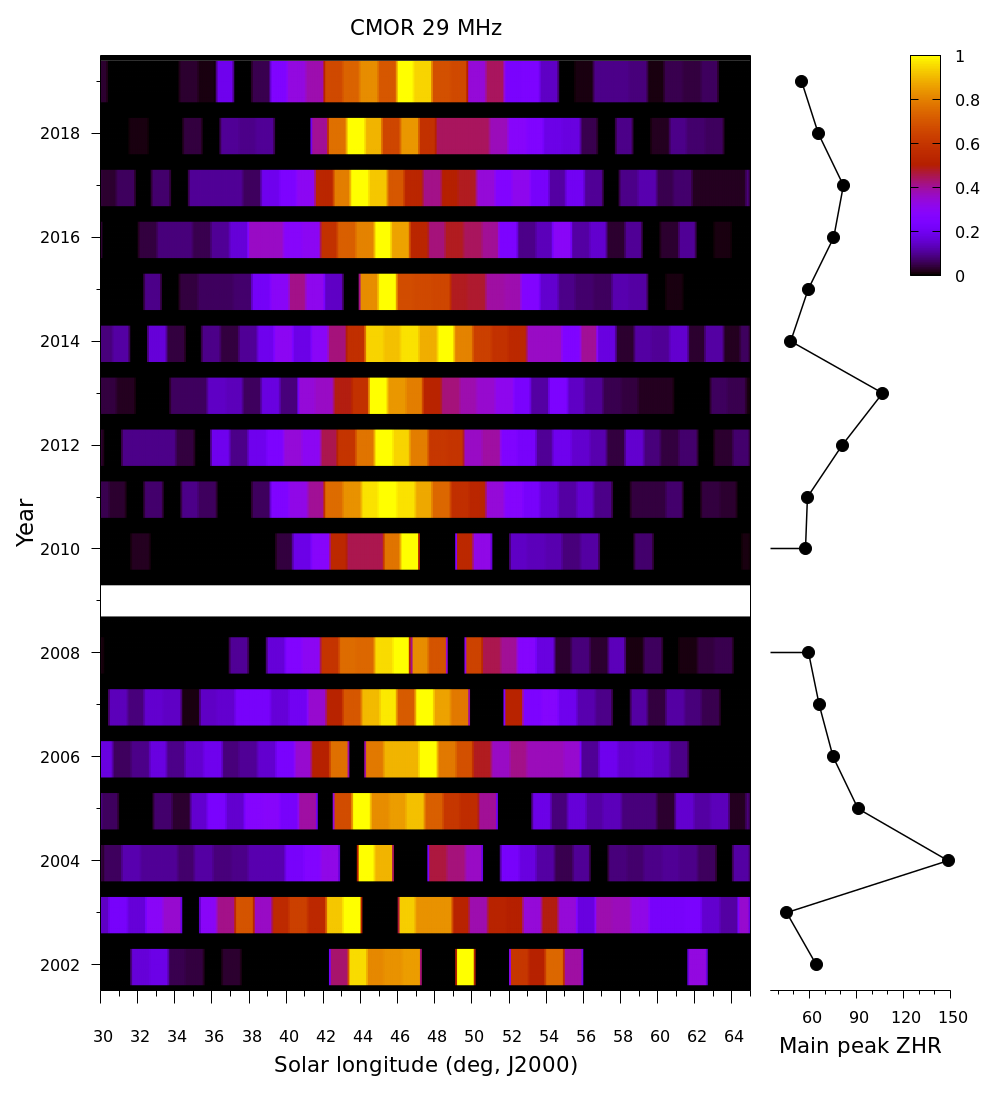}
      \caption{Intensity maps of the normalized Orionids (left) and $\eta$-Aquariids (right) activity as measured by the CMOR 29 MHz system since 2002. For each apparition of the shower, the activity profiles of Figures \ref{fig:new_oris} and \ref{fig:new_etas} are normalized by the maximum ZHR$_\text{v}$ value determined in the previous section. This is represented by the black curve to the right of each map. Timing of maximum activity are highlighted by light colors (yellow to orange), while low meteor rates (or the  absence of measurements) are represented by dark colors (purple to black).}
      \label{fig:CMOR29_maps}
  \end{figure*}
  
 \begin{figure*}[!ht]
      \centering
      \includegraphics[width=.495\textwidth]{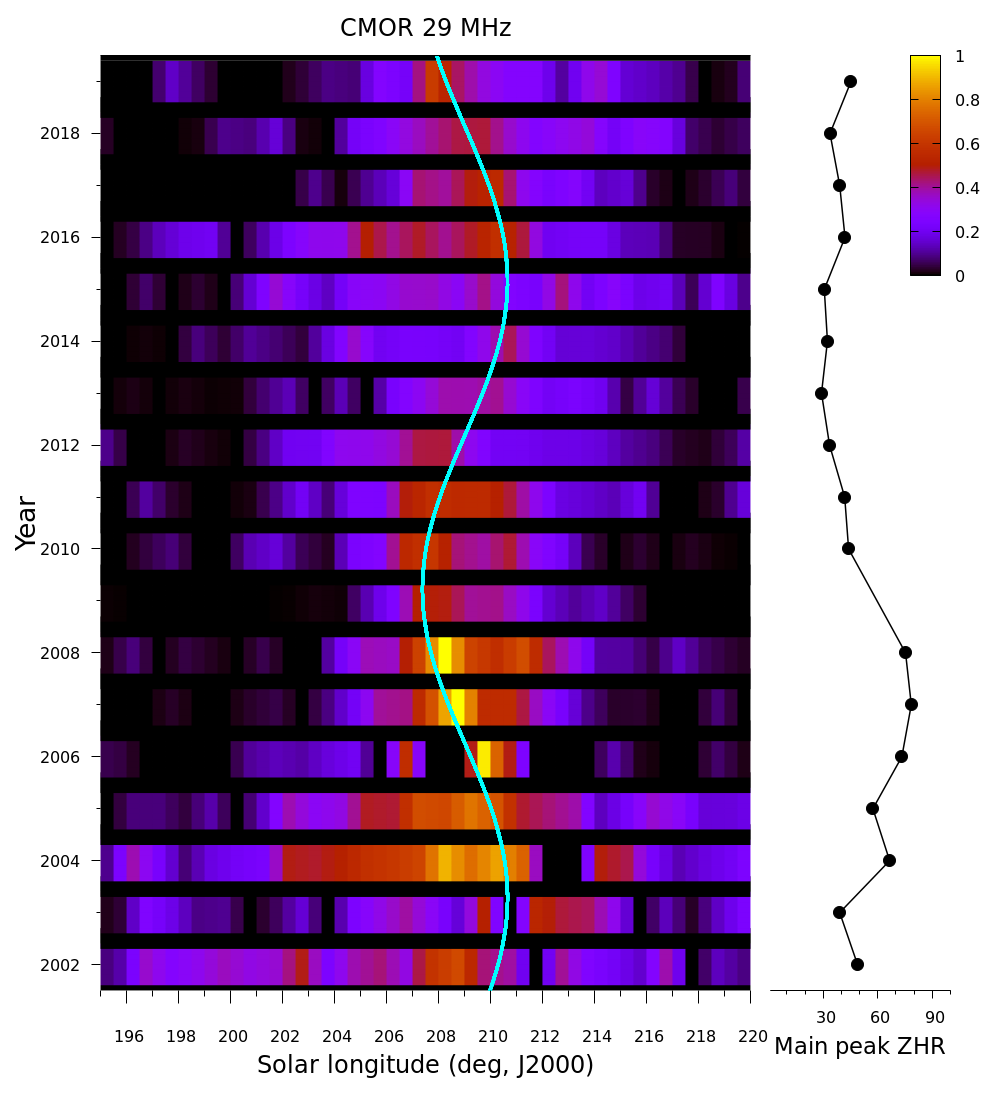}
      \includegraphics[width=.495\textwidth]{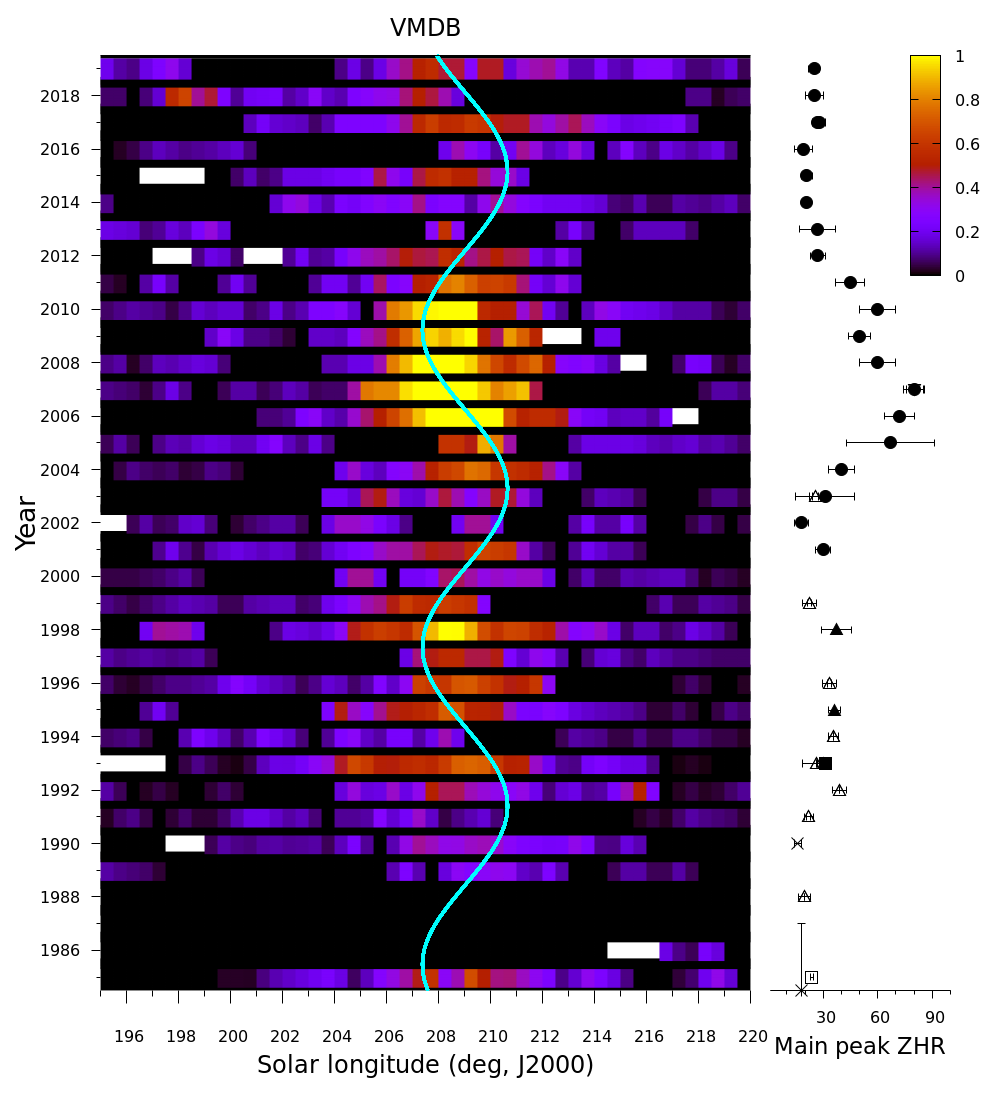}\\
      \includegraphics[width=.98\textwidth]{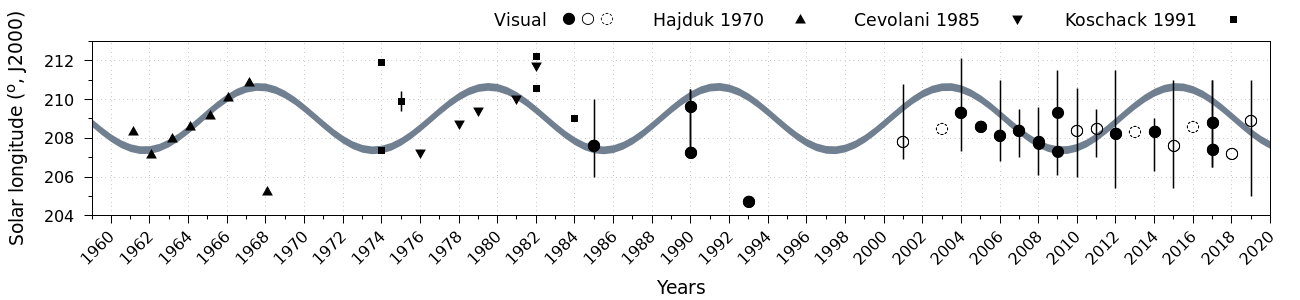}
      \caption{Top: Intensity maps of the Orionids as measured by CMOR 29 MHz (left) and from visual observations (right). Here individual profiles were normalized by the maximum ZHR$_\text{v}$ ever recorded by the network, and not by the maximum ZHR$_\text{v}$ of a particular apparition. The modeled sinusoidal curve, best fit of Figure \ref{fig:CMOR29_maps}, is presented in cyan. Bottom plot: comparison of the same modeled solar longitude variation (gray) as shown in the top plot, but now compared to the visual observations of Table \ref{tab:visualZHR_ori} (empty and filled circles) and including radar observations of \cite{Hajduk1970} (triangles, since 1960), \cite{Hajduk1984} and \cite{Cevolani1985} (inverse trangles), and complementary observations summarized in \cite{Koschack1991} (squares, before 1985). }
      \label{fig:Mapfit}
  \end{figure*}

  \subsubsection{Peak location} \label{sec:Lsolmax}
  
  Trends in the location of the Orionids and $\eta$-Aquariids activity peaks were investigated by \cite{Hajduk1970} and \cite{Hajduk1973}. From his analysis of visual and radar observations, \cite{Hajduk1970} identified a clear variability of the Orionids peak location for each return of the shower, including a shift by as much as 5\si{\degree} in solar longitude in some years. When long observational series were obtained with the radars (early visual observations were considered to be doubtful), the author noticed a gradual displacement of the activity peaks during successive apparitions. If a similar trend was suspected for the $\eta$-Aquariids (cf. Section \ref{section:eta}), no evidence of it was found in the observed changes in the shower's main peak of activity \citep{Hajduk1973}. 
  
  Since the time of those earlier works, the location of the Orionids and $\eta$-Aquariids main peak of activity has been re-examined by several authors \citep[e.g., ][]{McIntosh1983,Hajduk1980,Hajduk1984,Cevolani1985,Cevolani1987,Koschack1991}. Covering periods of 5 to 10 years, displacements in solar longitude of 0.3 to 0.75 degrees per year and 0.42 to 0.92 degrees per year have been claimed for the $\eta$-Aquariids and Orionids respectively \citep{McIntosh1983}. 
  
  As noted earlier, the identification of the showers' main peak of activity is highly dependent on the time resolution of the activity profiles, the selection of the population index (that may vary with time), and the level of continuity in the monitoring of the showers' activity. As a consequence of these factors, in this work we choose to examine the evolution of maximum activity regions (sometimes comprising several peaks) instead of the main peak location as these are likely to produce more robust results. 
  
  To aid in the visualization of the location of the solar longitude \Lsol$_{,max}$ of maximum activity over time, the individual activity profiles of Appendix \ref{new_obs} were converted into the intensity maps as shown in Figures \ref{fig:CMOR29_maps}, \ref{fig:Mapfit}, and \ref{fig:Optical_maps}. In these figures, the ZHR$_\text{v}$ of each apparition of the shower (rows) as a function of the solar longitude (columns) is colored as a function of the shower intensity (yellow: maximum meteor rates recorded, black: low meteor rates or no measurements available). To examine the variability in the activity peak locations, the profiles were normalized by the maximum ZHR$_\text{v}$ recorded for each year of the shower (computed in Section \ref{sec:ZHRmax} in these figures, but the reader is reminded of the peak ZHR values by a black curve adjacent to the map).  
  
  Since CMOR records provide the longest consistent observational set for the Halleyids in our study, we consider the measurements of the 29 MHz system as a reference for the rest of this section. Despite the ability of the 38 MHz data to reproduce the absolute intensity variations of the showers (cf. Section \ref{sec:ZHRmax}), the 29 MHz system was selected because of the lower statistical noise in the Orionids profile. 
  
  Figure \ref{fig:CMOR29_maps} highlights the evolution in \Lsol{}  of the peak activity regions recorded by CMOR for the Orionids (left panel) and $\eta$-Aquariids (right panel). The maximum activity of the $\eta$-Aquariids shows a moderate variation between \Lsol=44\si{\degree} and \Lsol=50\si{\degree}, with no specific pattern in the maximum activity location noticeable. In contrast, the Orionids maximum location seems to oscillate around \Lsol=209\si{\degree} with an amplitude of about 1 to 2\si{\degree}. However, no such trend is immediately obvious in the Orionids intensity maps derived for the other detection systems (cf. Appendix \ref{maps}). 
   To determine if the apparent oscillation in the location of the Orionids maximum activity in the 29 MHZ measurements is an artifact, a sinusoidal signal was fit to the yellow regions of Figure \ref{fig:CMOR29_maps} using the Particle Swarm Optimization method \citep[PSO, cf.][]{PSO}. The best fit solution was for a mean solar longitude of 209\si{\degree}, an amplitude of 1.64\si{\degree} and an angular frequency of 0.529 yr$^{-1}$. The resultant frequency fit corresponds to a period of 11.88 years, very close to Jupiter's orbital period of 11.86 years. 
  
  Figure \ref{fig:Mapfit} compares the resulting sinusoidal curve to the Orionids intensity maps as measured by CMOR (left) and the VMDB (right). For clarity, the activity maps of Figure \ref{fig:Mapfit} were normalized by the highest ZHR measured during the entire period of observation for the technique considered. 
  
  In the VMDB map, global periods of enhanced activity (e.g., around 1998 and 2007) match with times that the sine model is at lowest solar longitudes, while higher modeled solar longitudes are close to the reported minima of activity. On the other hand, the correlation between the yearly maximum activity location and the sinusoidal fit is less clear in these observations. However, because of the low signal-to-noise ratio for the Orionids and the presence of multiple gaps in the VMDB profiles, no robust conclusions can be drawn from the intensity map of Figure \ref{fig:Mapfit}. 
  
  To investigate the veracity of a periodic displacement in shower activity in more detail, we present in Figure \ref{fig:Mapfit} (bottom panel) a comparison of our sinusoidal model (gray curve) with reported past maxima locations of the shower (symbols). The estimated maximum peak location and FWHM of the profile determined from the rigorous analysis of the Orionids visual data in Section \ref{sec:ZHRmax} are represented by circles (empty or filled) and vertical lines in the figure. Additional published observations since 1985 were added to the visual data set and summarized in Table \ref{tab:visualZHR_ori}. 
  
  Between 1960 and 1970, our model is compared to the solar longitude of the maximum activity recorded by radar observations as reported by \cite{Hajduk1970}. Results of the simultaneous measurements of the Budrio and Ondrejov radars, discussed in \cite{Hajduk1984} and \cite{Cevolani1985}, are presented as inverted triangles over the period 1976-1982. Complementary visual observations, listed by \cite{Koschack1991}, were added to Figure \ref{fig:Mapfit} when a single maximum of activity (or double-peak maximum) could be identified from the observations. 
  
  From this figure, it is clear that the sinusoidal fit does not reproduce all the peak locations reported for the Orionids since 1960.  However, the modeled curve falls within the estimated FWHM ranges from visual observations (except for the 1993 and 2014 apparitions). This  could be interpreted as potentially indicating the location of secondary maxima not always captured in the existing data. The $\sim$12 year periodicity determined from CMOR measurements reproduces very well \cite{Hajduk1970} radar observations between 1961 and 1967, but deviates more from the estimates of \cite{Cevolani1985} and \cite{Koschack1991} made between 1975 and 1983. A secondary fit solution, leading to a period of around 11.1 years, offers a correct match of CMOR and the 1975-1983 radar observations, but does not reproduce the 1961-1967 results. Without accurate estimates of the shower FWHM and measurement uncertainties before 1985, the agreement between the sinusoidal model and these older radar observations cannot be clearly established.
  
  From earlier studies and recent visual observations of the shower, there is no evidence of a $\sim$12 year periodicity of the Orionids main peak location (right hand side of the bottom panel of Figure~\ref{fig:Mapfit}). However, the variations presented in Figure \ref{fig:Mapfit} do not exclude the existence of such periodicity either, especially when we consider the difficulty of identifying the main peak of activity. The period of the modeled fit together with the fact that such a trend is most noticeable in the longest and most consistent set of observations so far (CMOR), lends support to the existence of periodic oscillations over many years in the timing of the Orionids maximum activity. We therefore suggest that our measurements point toward a cyclical variation of the location of the Orionids most active period, but emphasize that this trend needs to be confirmed by future radar and optical observations.  

\section{Conclusions} \label{sec:conclusion}

 In this study, we have measured long-term trends in the activity of the $\eta$-Aquariids and Orionids meteor showers as observed with visual, video, and radar techniques. Results from the IMO VMDB and CMOR databases were compared for each annual apparition of the showers since 2002, along with VMN observations since 2011. Despite the different biases inherent to each detection method, observations from the three data sets show good agreement in the general shape, activity level, and annual intensity variations of both showers. This consistency among systems sensitive to different size meteoroids suggests that there is no significant size sorting of the particles within the meteoroid stream. 
 
 The analysis of the Halleyids activity since 2002 is generally consistent with previous published observations of the showers. The main characteristics of the $\eta$-Aquariids and Orionids as derived from our analysis are:
 
 \begin{enumerate}
     \item The $\eta$-Aquariids are generally active between \Lsol=35\si{\degree} and \Lsol=60\si{\degree}, with highest meteor rates recorded between 44\si{\degree} and 50\si{\degree}.
     \item The Orionids are active between \Lsol=195\si{\degree} and \Lsol=220\si{\degree}, and present a broad maximum between 206\si{\degree} and 211\si{\degree}.
     \item Both showers display several subpeaks of variable location and strength with time.
     \item Typical maximum ZHR$_\text{v}$ rates vary around an average of 65 to 70 meteors per hour for the $\eta$-Aquariids, and between 20 to 40 meteors per hour for the Orionids.
     \item Several outbursts, caused by meteoroids trapped in resonant orbits with Jupiter, were observed for the $\eta$-Aquariids (in 2004 and 2013) and the Orionids (in 2006-2007). ZHR maximum rates then reached two to four times the usual activity level of the showers.
     \item The average profile of the $\eta$-Aquariids is asymmetric, with a rise of activity more sudden than the decreasing activity. The profile of the Orionids is more symmetric around the broad maximum of activity. The general shape of the average activity profiles can differ as a function of the period considered. 
 \end{enumerate}
 
No clear periodicity in the annual activity level or the main peak location of the $\eta$-Aquariids can be inferred from our analysis. Consistent radar observations of the Orionids since 2002 support the existence of a periodic displacement in the location of the solar longitude of the shower's peak. A period of $\sim$11.88 years, leading to an angular drift of about $\Delta$\Lsol\mbox{ } of about 0.53\si{\degree}.yr$^{-1}$ was estimated from CMOR 29 MHz data. The existence of such periodicity cannot be established from existing visual observations, and needs to be confirmed by future radar and optical observations of the shower. 

\begin{acknowledgements}
  This work was supported in part by NASA Meteoroid Environment Office under cooperative agreement 80NSSC18M0046 and contract 80MSFC18C0011, by the Natural Sciences and Engineering Research Council of Canada (Grants no. RGPIN-2016-04433 \& RGPIN-2018-05659), and by the Canada Research Chairs Program. We are
  	thankful to the referee for his careful review that helped improve the manuscript.
\end{acknowledgements}

 
  \bibliographystyle{model2-names.bst}
  \bibliography{References}

%

\begin{appendix} 

\onecolumn

  \section{Example of Orionids and $\eta$-Aquariids activity profiles between 1985 and 2001} \label{old_obs}
  
 \mbox{ } \vspace{-0.5cm} \mbox{ }

  \begin{figure*}[!ht]
  \centering
      \includegraphics[width=.29\textwidth]{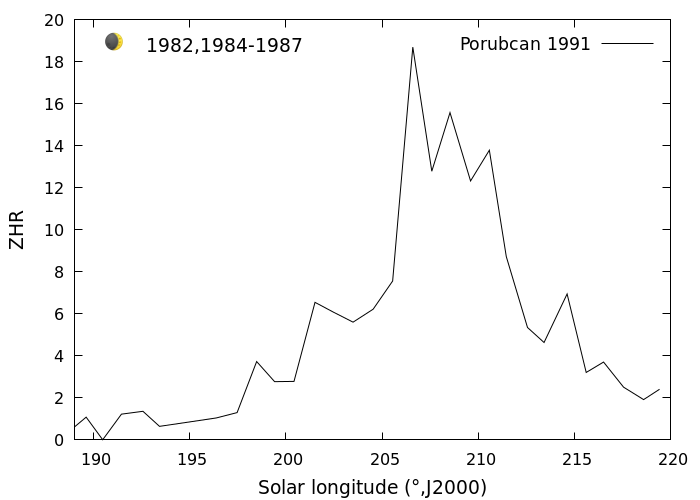}
      \includegraphics[width=.29\textwidth]{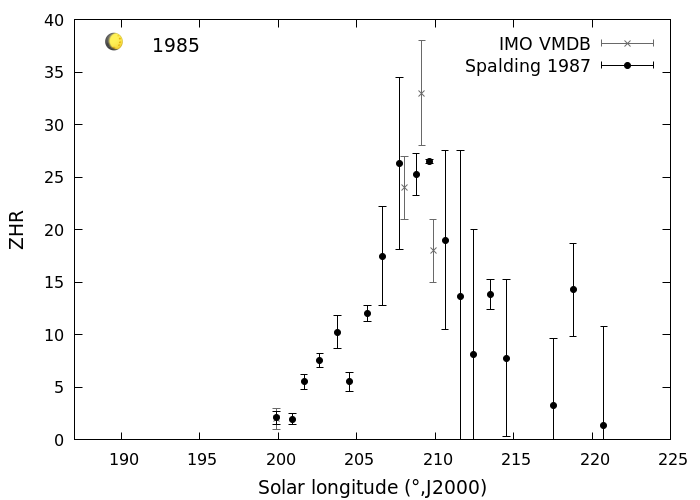}
      \includegraphics[width=.29\textwidth]{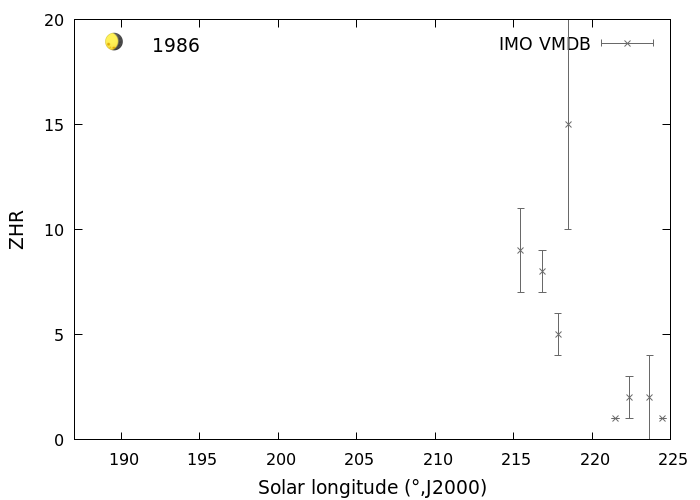}\\
      \includegraphics[width=.29\textwidth]{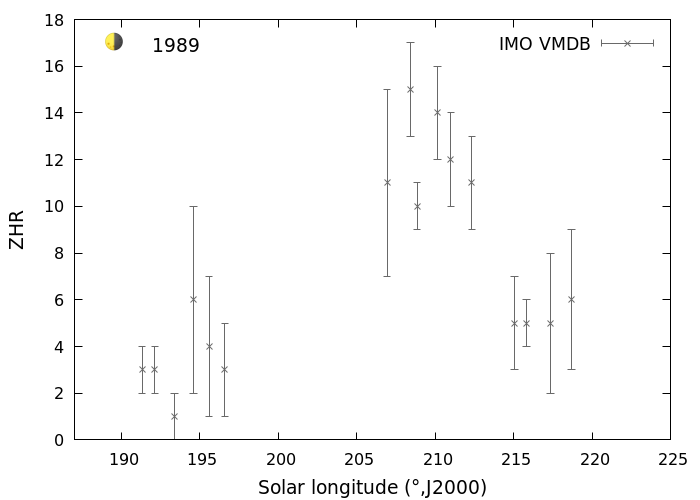}
      \includegraphics[width=.29\textwidth]{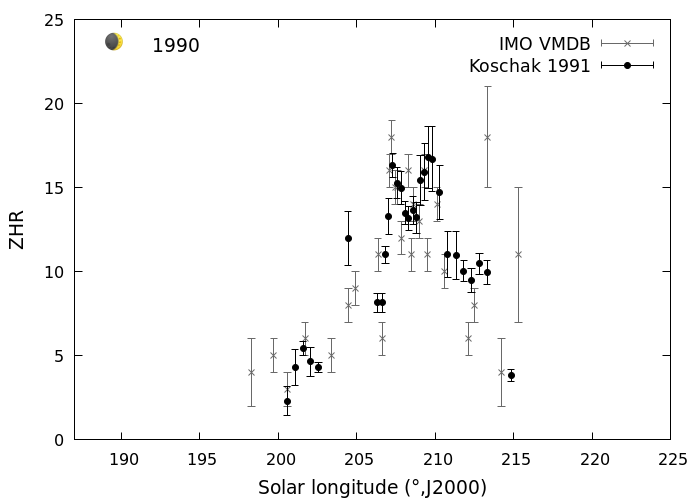}
      \includegraphics[width=.29\textwidth]{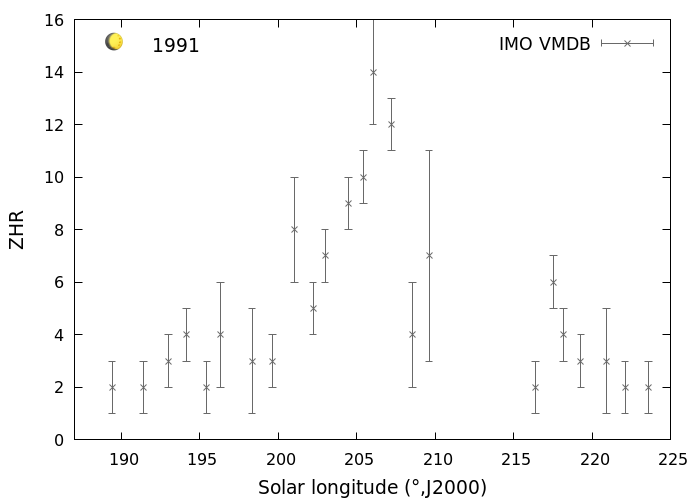}\\
      \includegraphics[width=.29\textwidth]{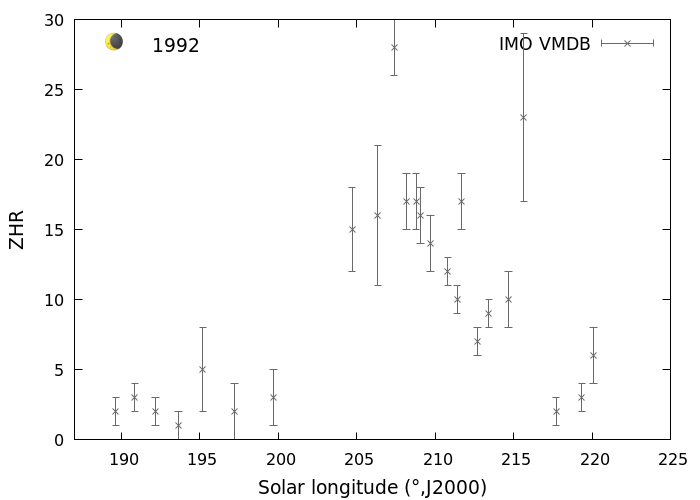}
      \includegraphics[width=.29\textwidth]{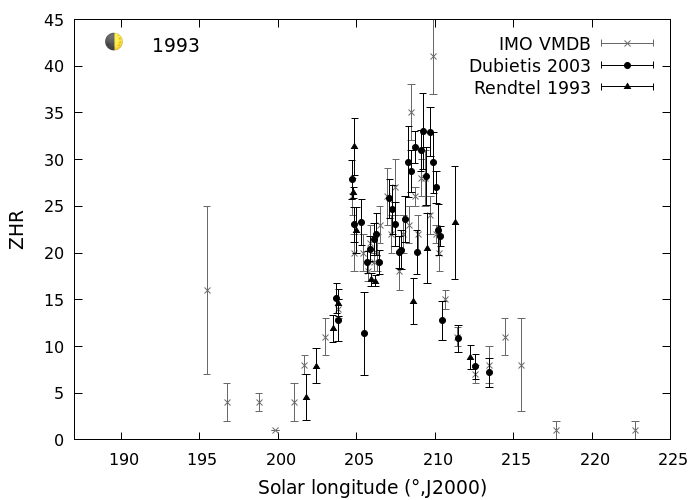}
      \includegraphics[width=.29\textwidth]{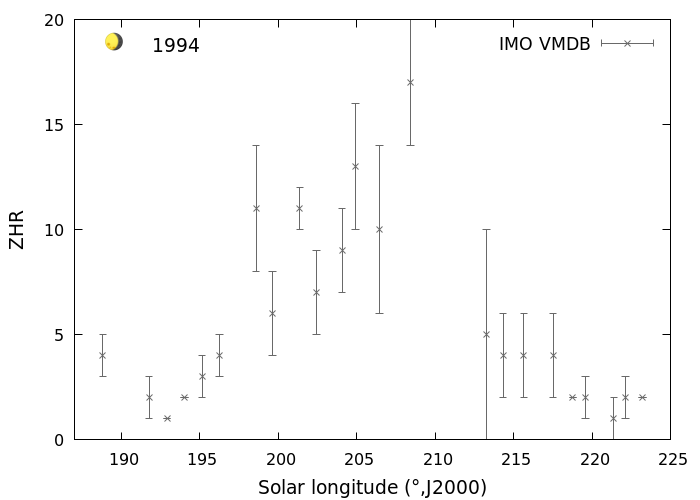}\\
      \includegraphics[width=.29\textwidth]{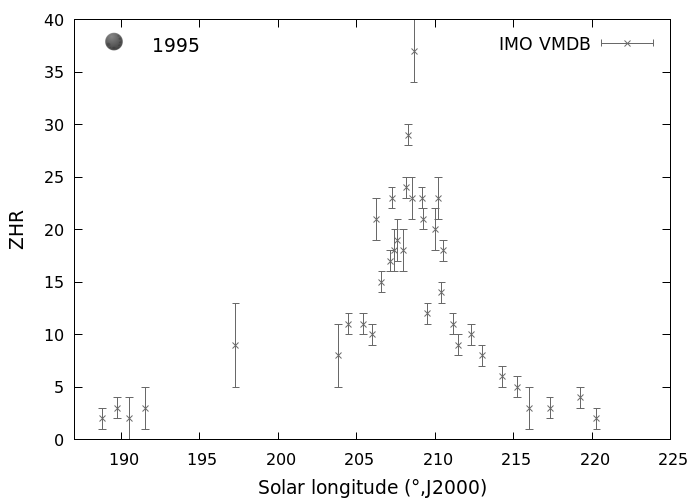}
      \includegraphics[width=.29\textwidth]{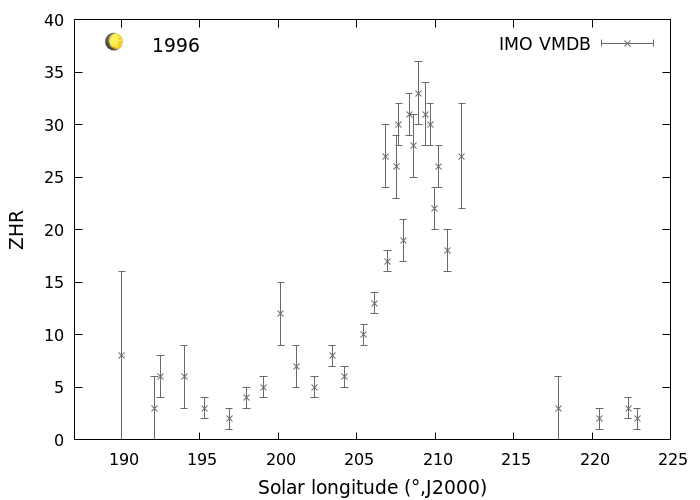}
      \includegraphics[width=.29\textwidth]{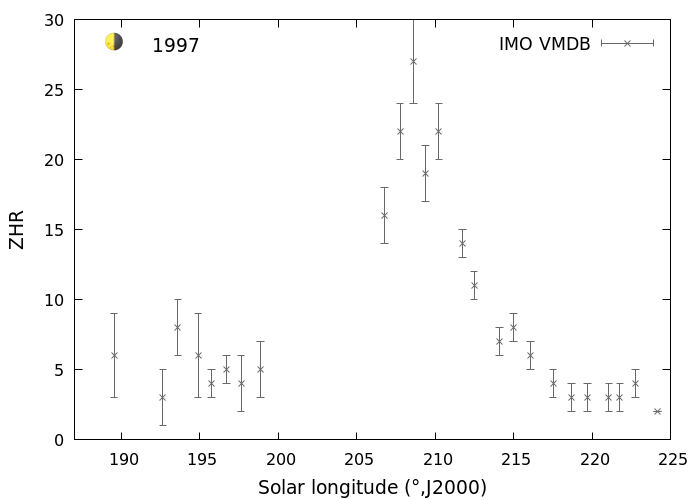}\\
      \includegraphics[width=.29\textwidth]{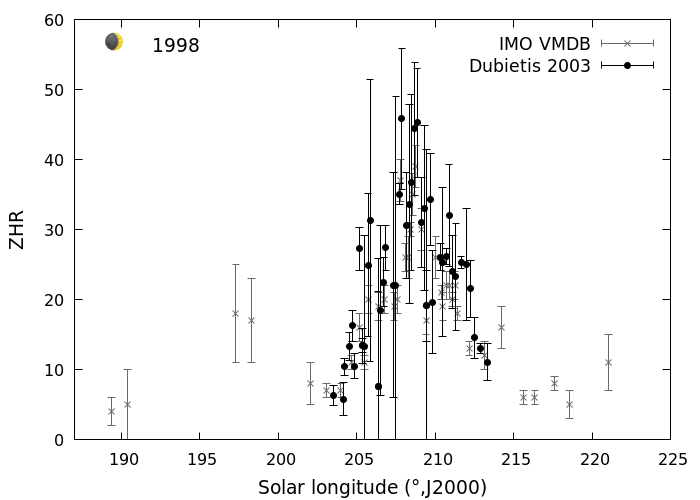}
      \includegraphics[width=.29\textwidth]{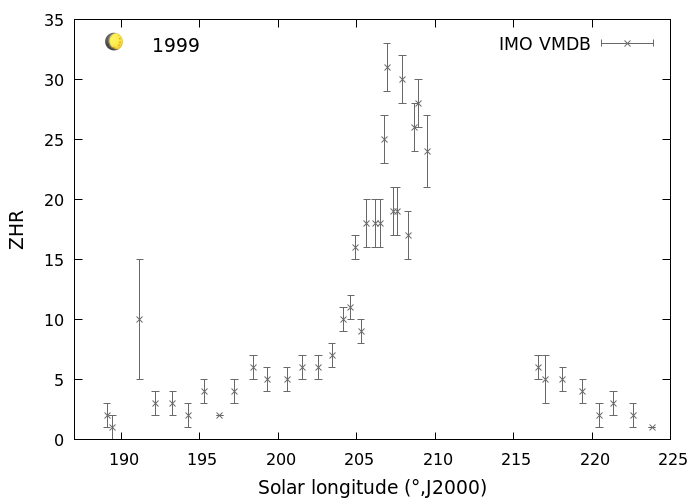}
      \includegraphics[width=.29\textwidth]{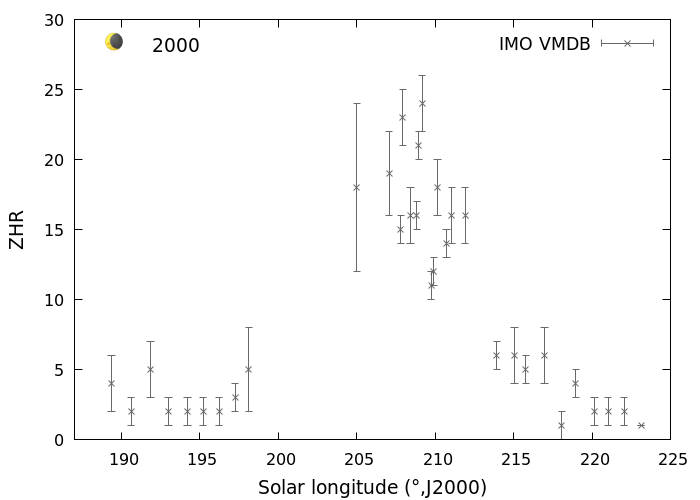}\\
      \includegraphics[width=.29\textwidth]{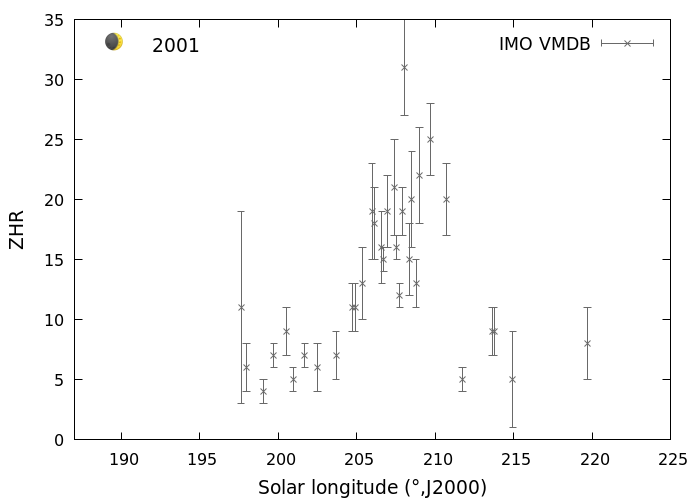}
      \caption{Available activity profiles of the Orionids between 1985 and 2001. Original ZHR were rescaled to a common population index of 2.59.}
      \label{fig:old_oris}
  \end{figure*}
  
  \begin{figure*}[!ht]
  \centering
      \includegraphics[width=.29\textwidth]{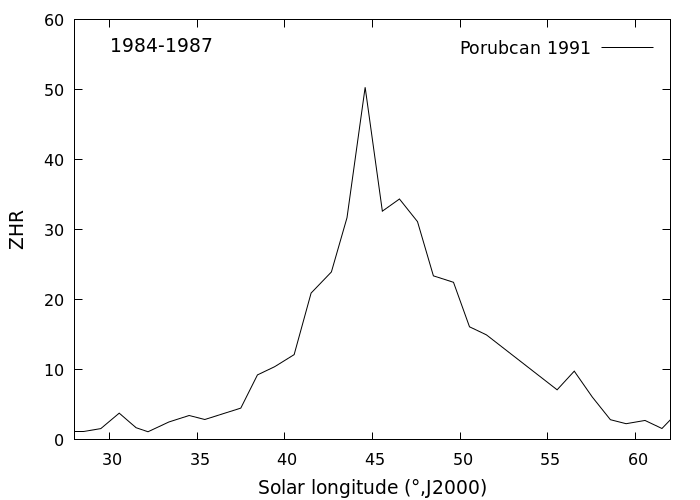}
      \includegraphics[width=.29\textwidth]{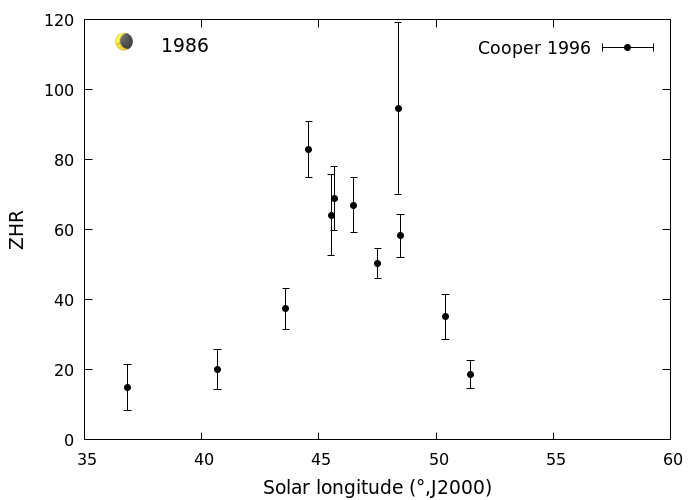}
      \includegraphics[width=.29\textwidth]{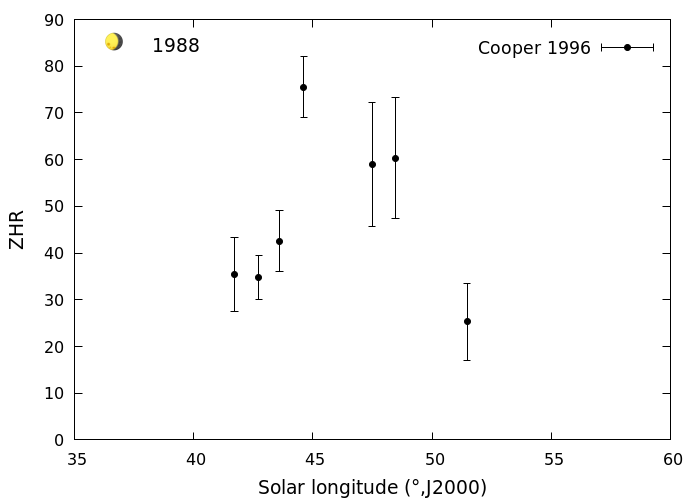}\\
      \includegraphics[width=.29\textwidth]{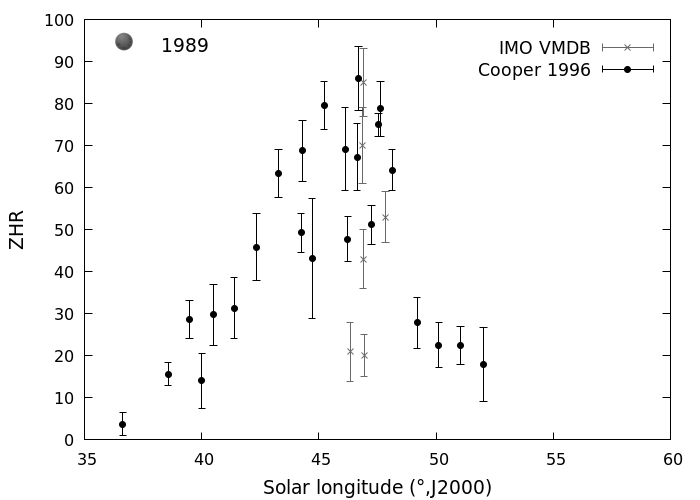}
      \includegraphics[width=.29\textwidth]{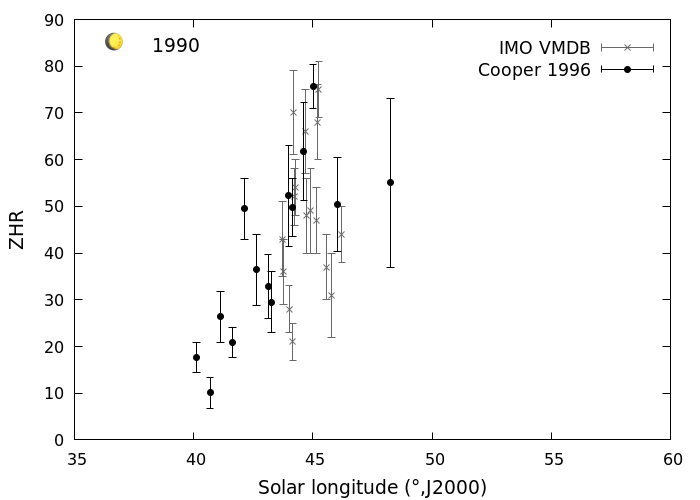}
      \includegraphics[width=.29\textwidth]{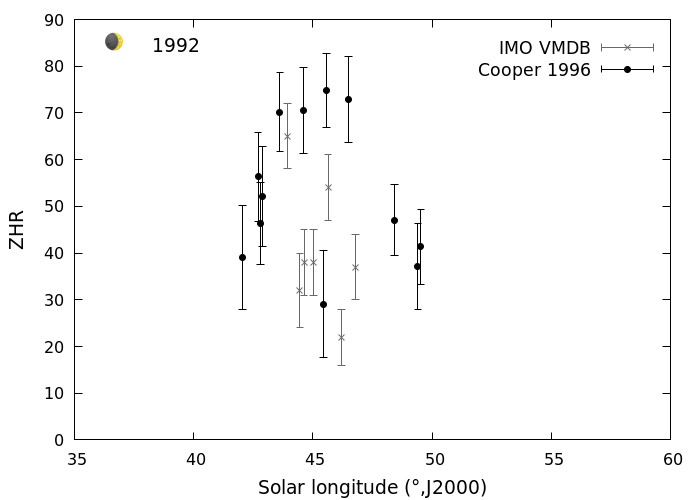}\\
      \includegraphics[width=.29\textwidth]{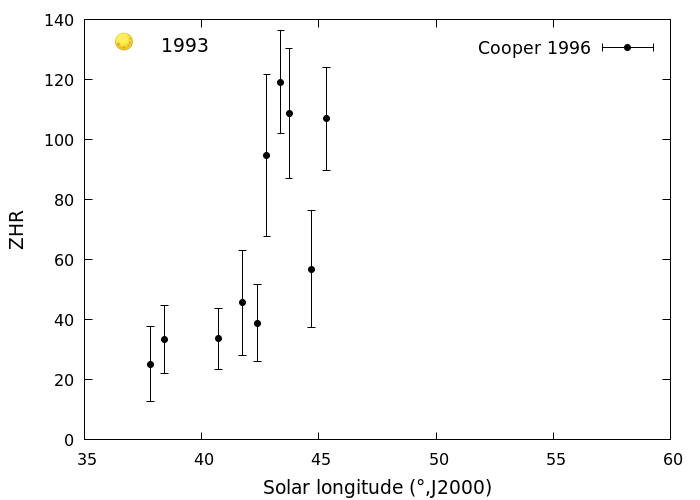}
      \includegraphics[width=.29\textwidth]{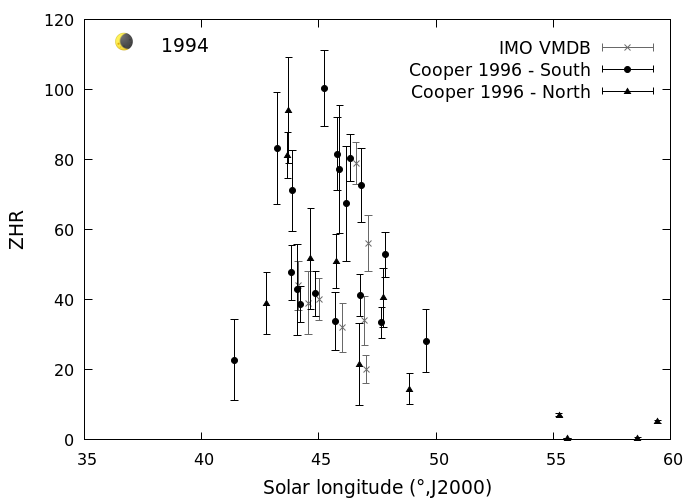}
      \includegraphics[width=.29\textwidth]{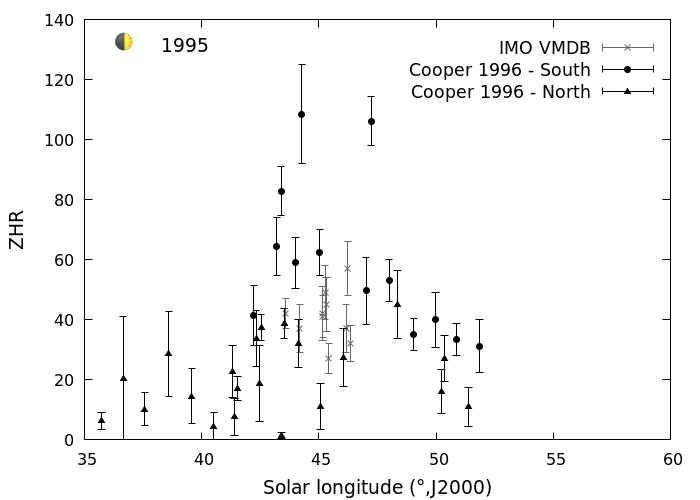}\\
      \includegraphics[width=.29\textwidth]{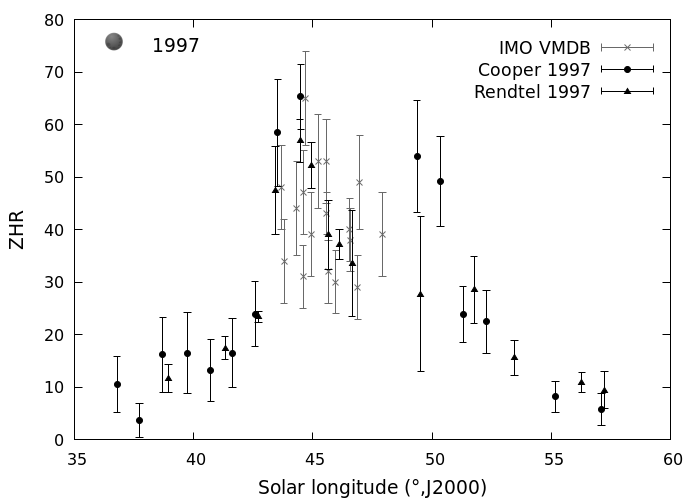}
      \includegraphics[width=.29\textwidth]{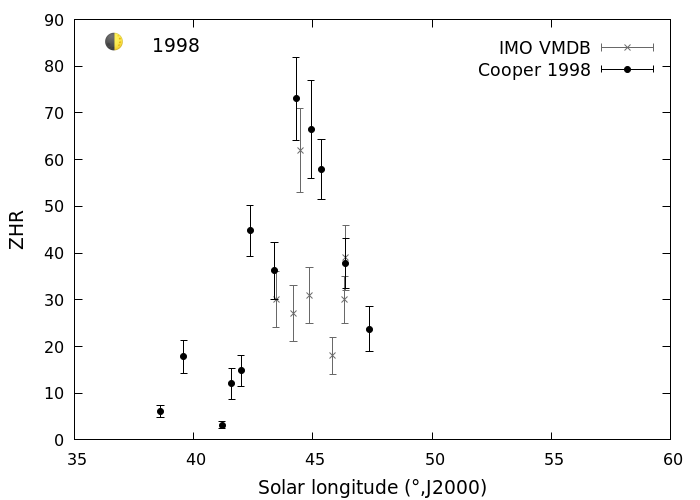}
      \includegraphics[width=.29\textwidth]{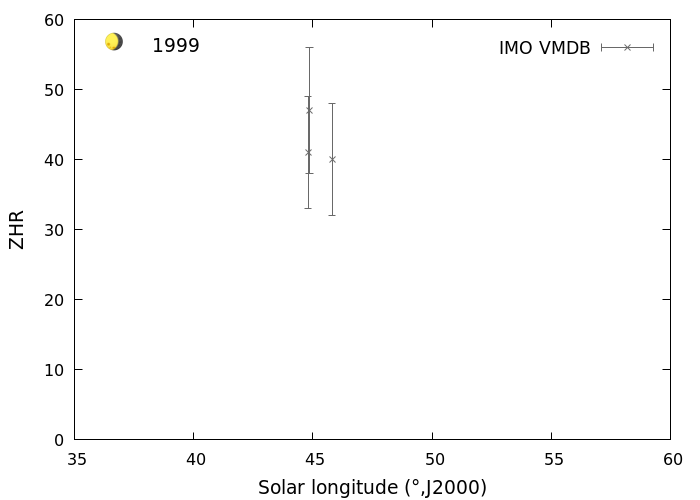}\\
      \includegraphics[width=.29\textwidth]{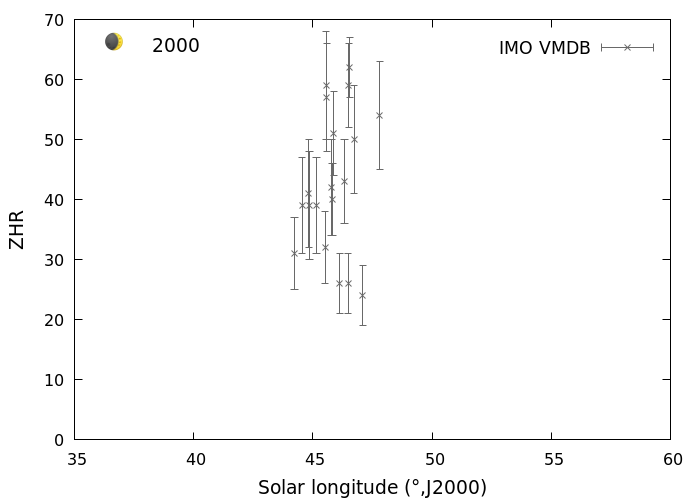}
      \includegraphics[width=.29\textwidth]{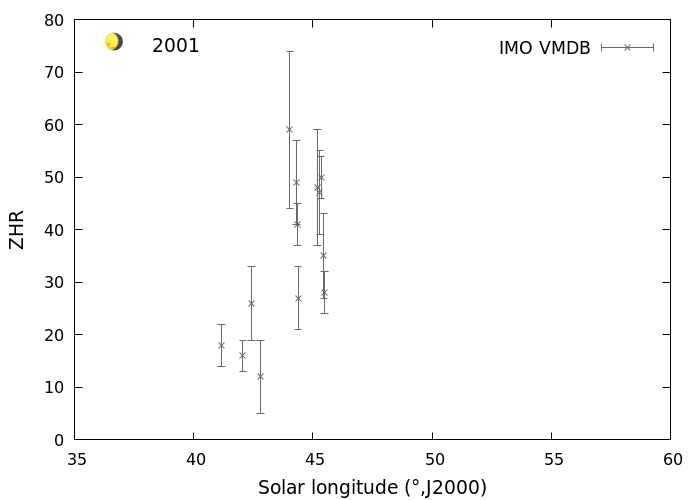}
      \caption{Available activity profiles of the $\eta$-Aquariids between 1985 and 2001. Original ZHR were rescaled to a common population index of 2.46. When possible, observations from the Northern and Southern Hemisphere are combined \citep[cf. ][]{Cooper1997}.}
      \label{fig:old_eta}
  \end{figure*}
  
  
   \clearpage
   \section{2002-2019 activity profiles} \label{new_obs}
   
   \mbox{ } \vspace{-0.5cm} \mbox{ }
     
    \begin{figure*}[!ht]
    \centering
      \resizebox{\dimexpr.9\textwidth-1em}{!}{
      \includegraphics[width=.32\textwidth]{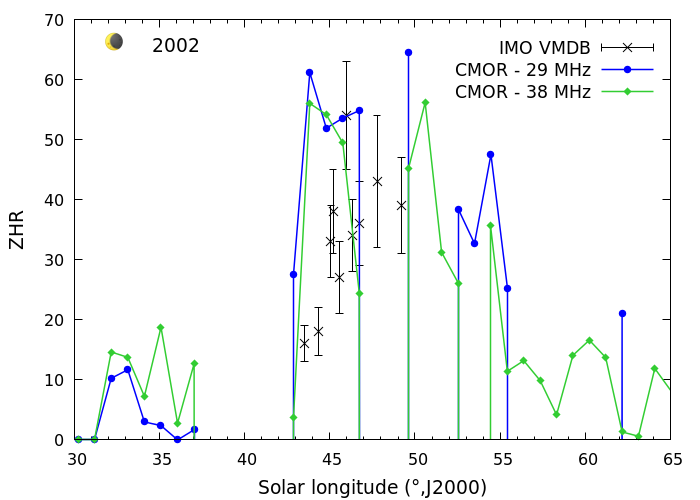}
      \includegraphics[width=.32\textwidth]{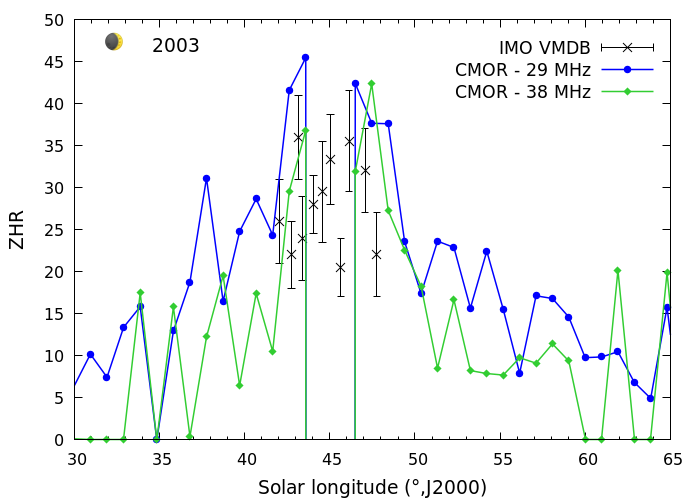}
      \includegraphics[width=.32\textwidth]{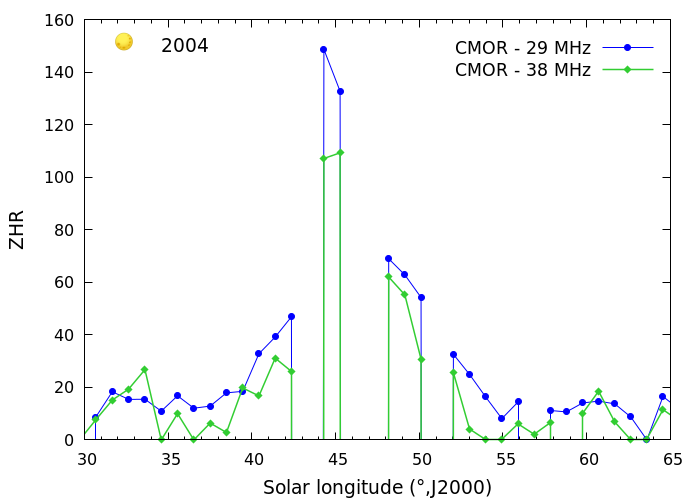}}\\
      \resizebox{\dimexpr.9\textwidth-1em}{!}{
      \includegraphics[width=.32\textwidth]{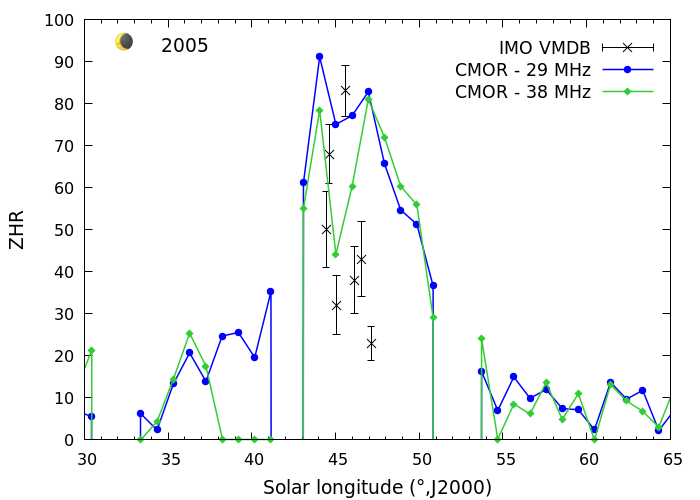}
      \includegraphics[width=.32\textwidth]{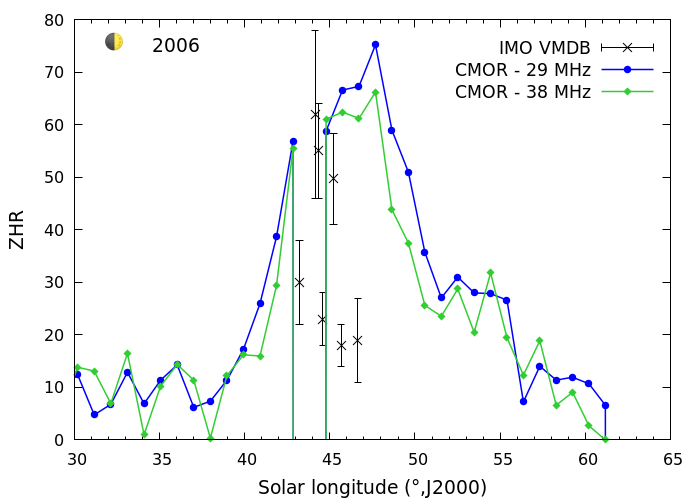}
      \includegraphics[width=.32\textwidth]{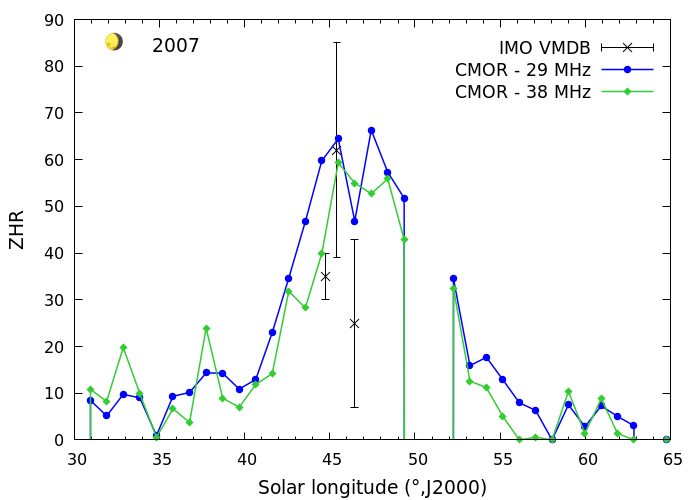}}\\
      \resizebox{\dimexpr.9\textwidth-1em}{!}{
      \includegraphics[width=.32\textwidth]{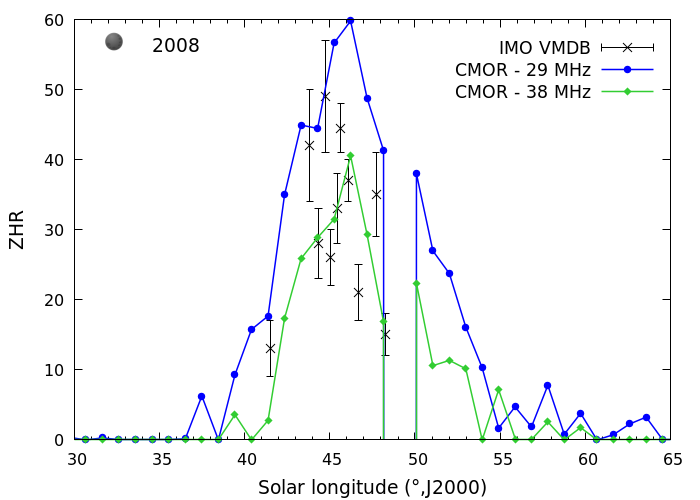}
      \includegraphics[width=.32\textwidth]{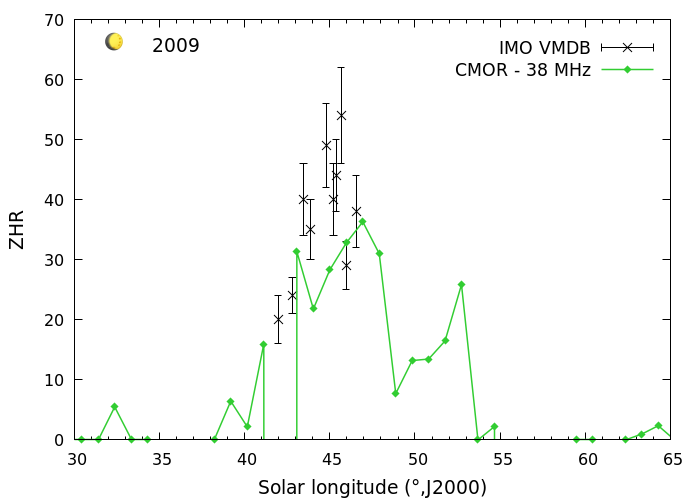}
      \includegraphics[width=.32\textwidth]{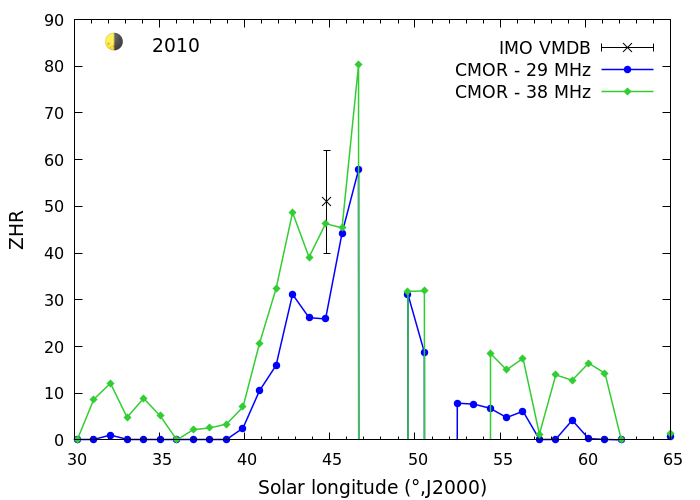}}\\
      \resizebox{\dimexpr.9\textwidth-1em}{!}{
      \includegraphics[width=.32\textwidth]{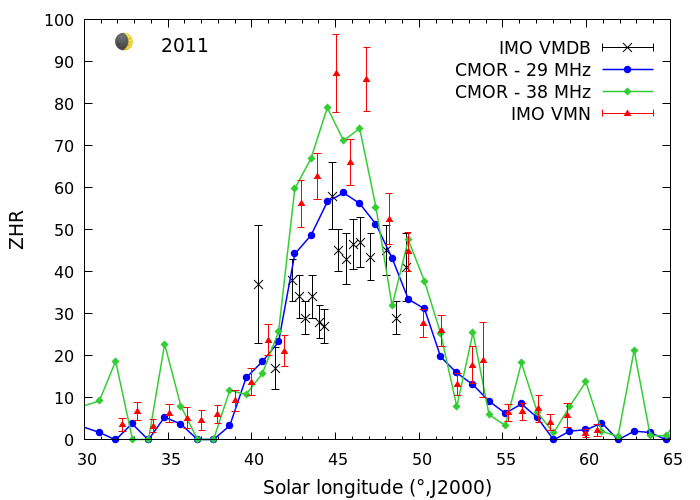}
      \includegraphics[width=.32\textwidth]{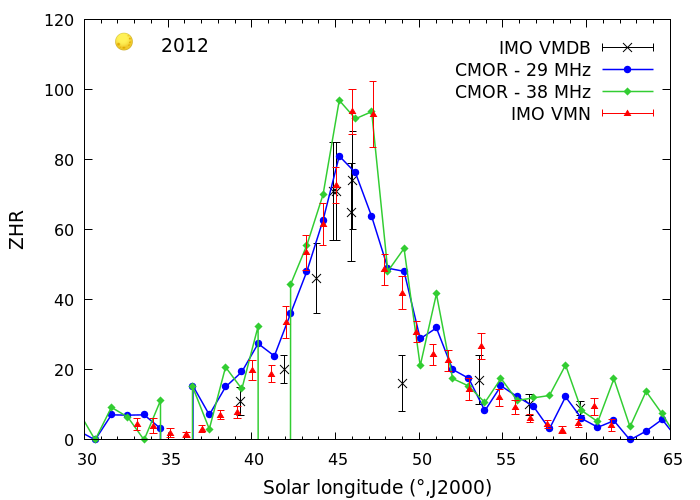}
      \includegraphics[width=.32\textwidth]{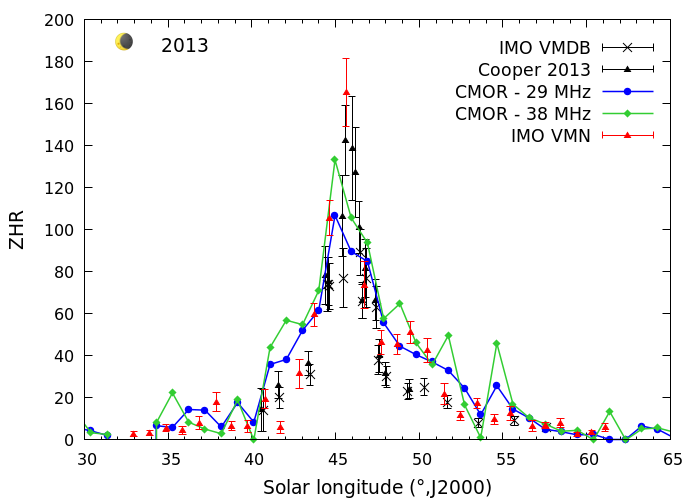}}\\
      \resizebox{\dimexpr.9\textwidth-1em}{!}{
      \includegraphics[width=.32\textwidth]{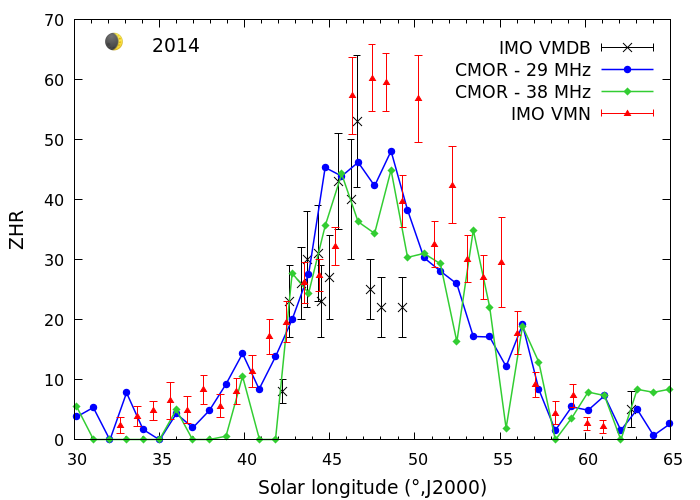}
      \includegraphics[width=.32\textwidth]{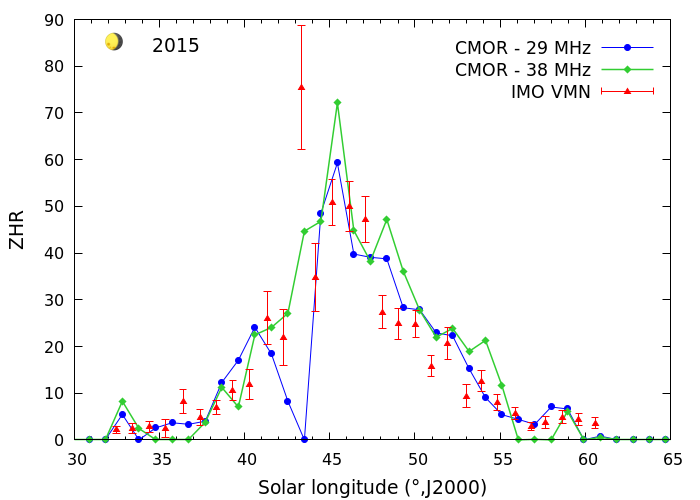}
      \includegraphics[width=.32\textwidth]{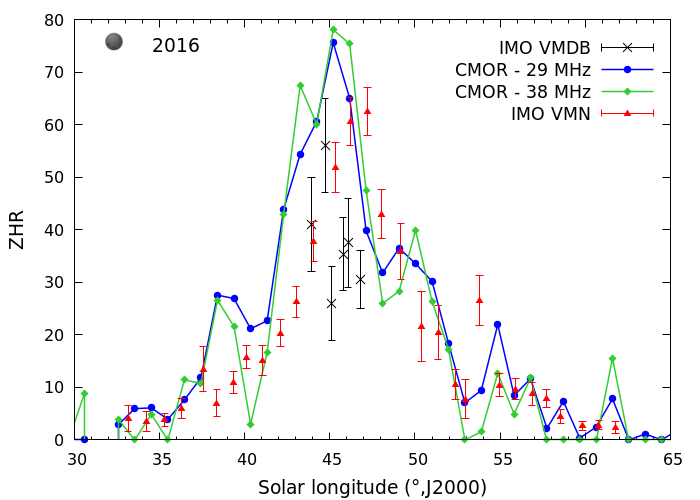}}\\
      \resizebox{\dimexpr.9\textwidth-1em}{!}{
      \includegraphics[width=.32\textwidth]{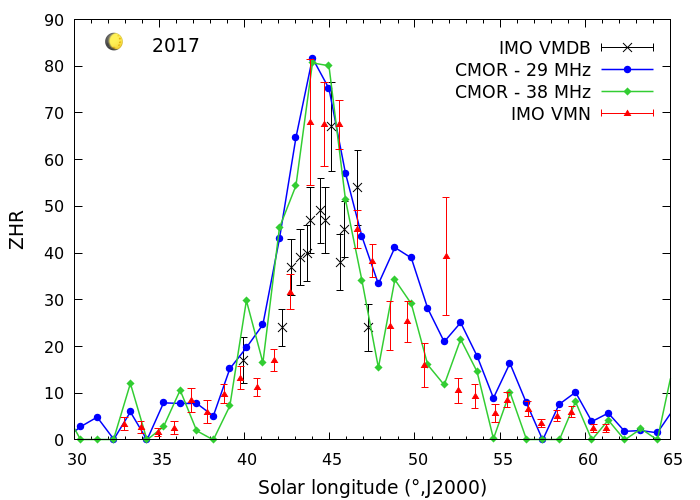}
      \includegraphics[width=.32\textwidth]{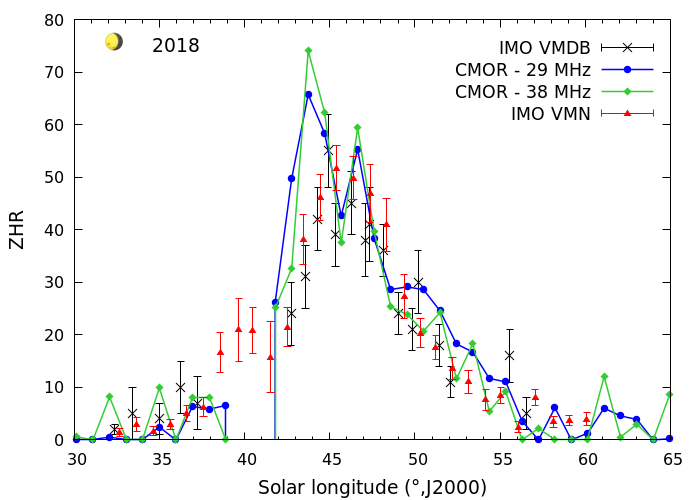}
      \includegraphics[width=.32\textwidth]{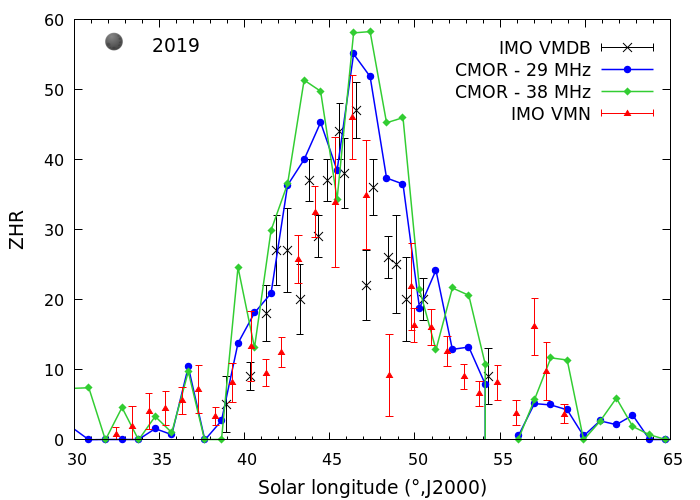}}
      \caption{Activity profiles of the $\eta$-Aquariids between 2002 and 2019.}
      \label{fig:new_etas}
  \end{figure*}
  
    \begin{figure*}[!ht]
      \centering
      \resizebox{\dimexpr.9\textwidth-1em}{!}{
      \includegraphics[width=.32\textwidth]{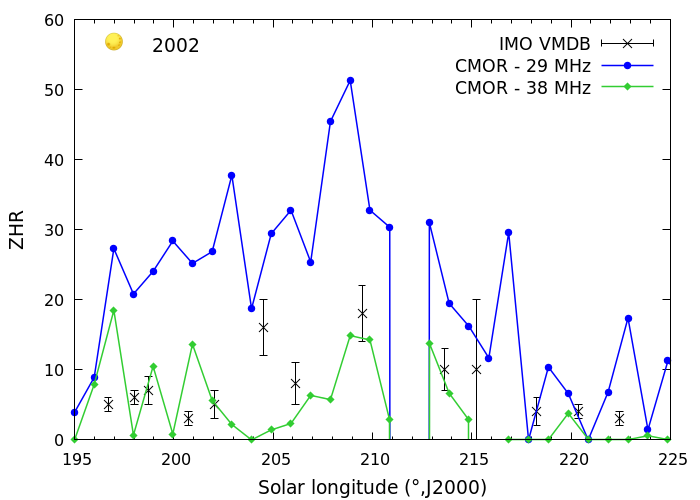}
      \includegraphics[width=.32\textwidth]{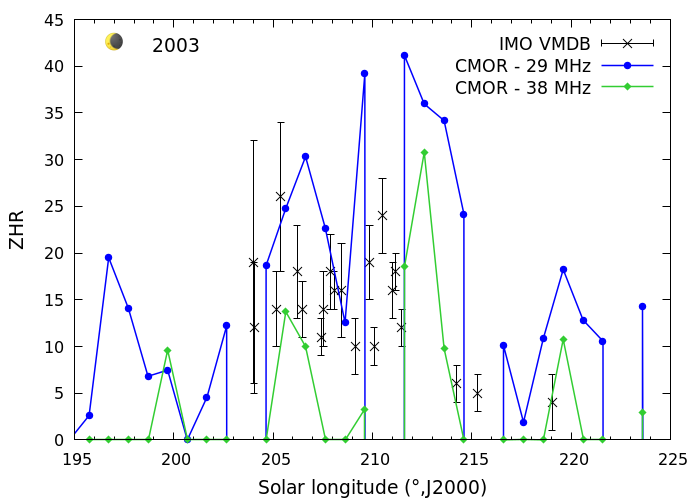}
      \includegraphics[width=.32\textwidth]{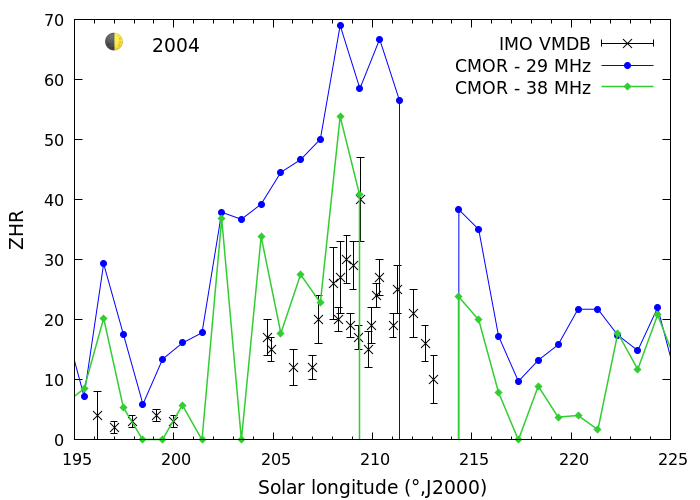}}\\
      \resizebox{\dimexpr.9\textwidth-1em}{!}{
      \includegraphics[width=.32\textwidth]{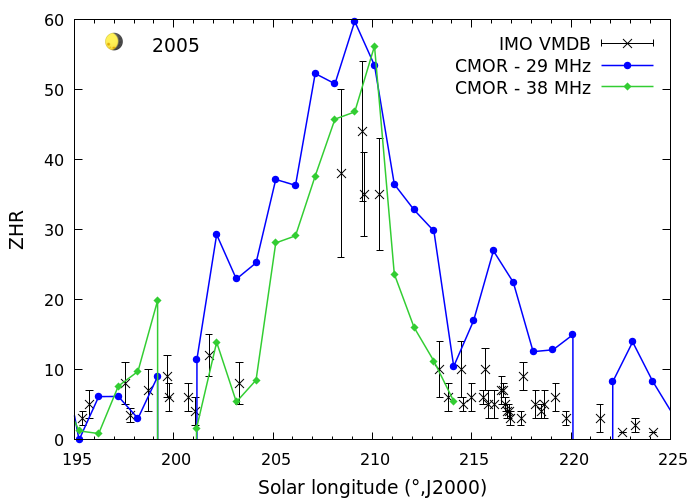}
      \includegraphics[width=.32\textwidth]{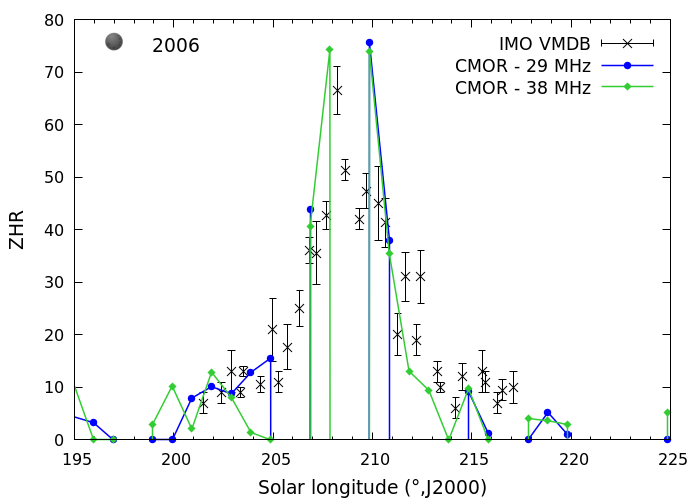}
      \includegraphics[width=.32\textwidth]{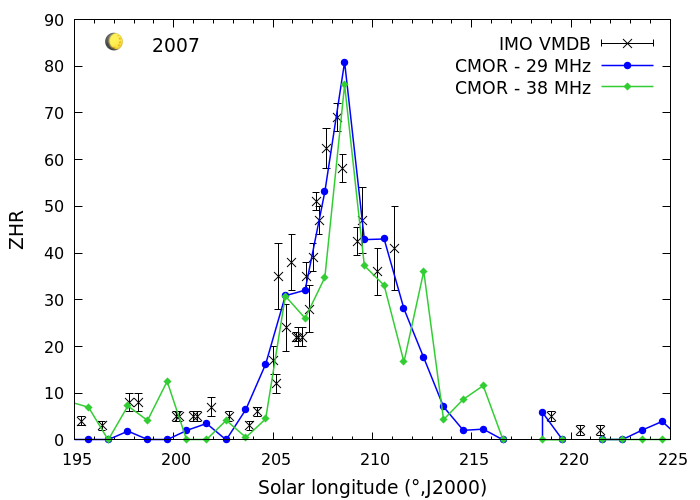}}\\
      \resizebox{\dimexpr.9\textwidth-1em}{!}{
      \includegraphics[width=.32\textwidth]{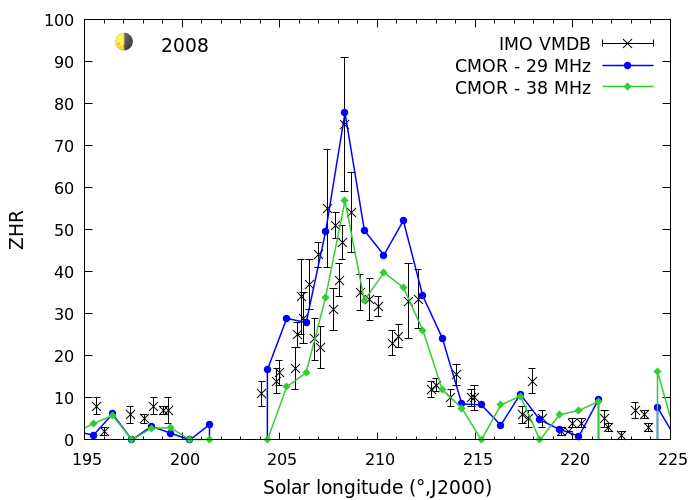}
      \includegraphics[width=.32\textwidth]{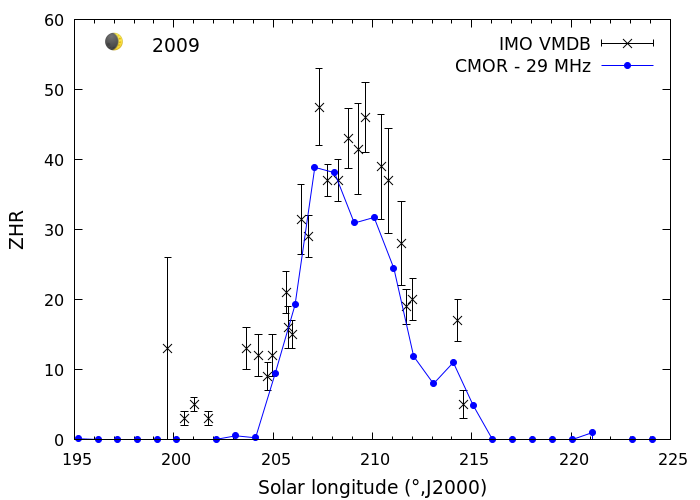}
      \includegraphics[width=.32\textwidth]{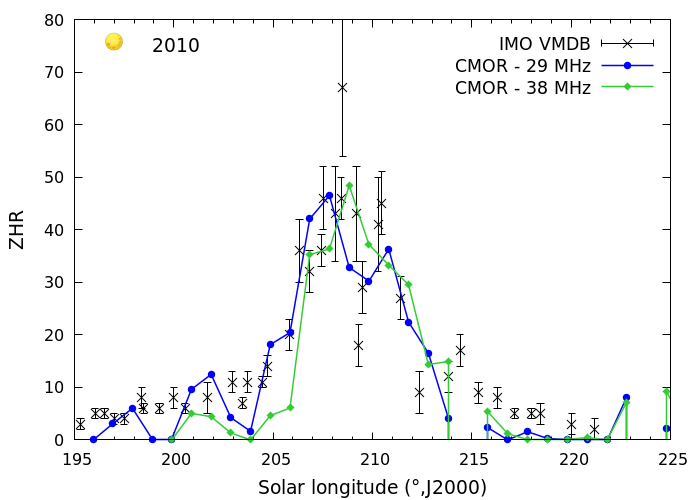}}\\
      \resizebox{\dimexpr.9\textwidth-1em}{!}{
      \includegraphics[width=.32\textwidth]{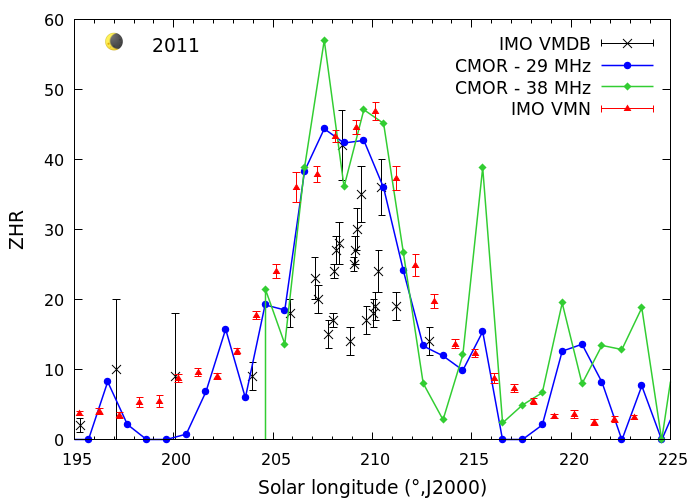}
      \includegraphics[width=.32\textwidth]{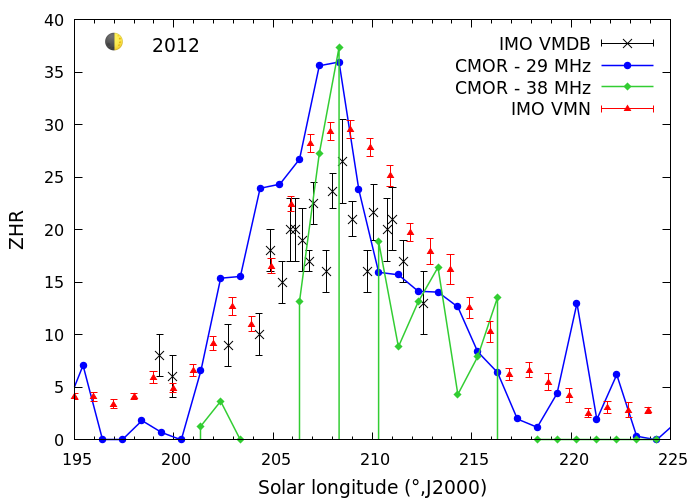}
      \includegraphics[width=.32\textwidth]{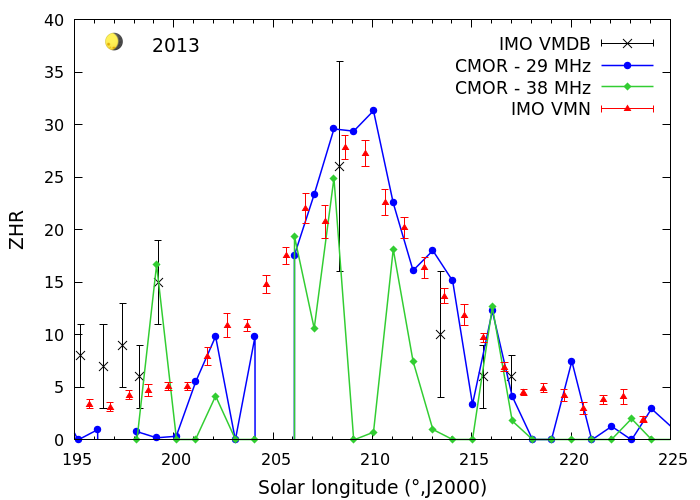}}\\
      \resizebox{\dimexpr.9\textwidth-1em}{!}{
      \includegraphics[width=.32\textwidth]{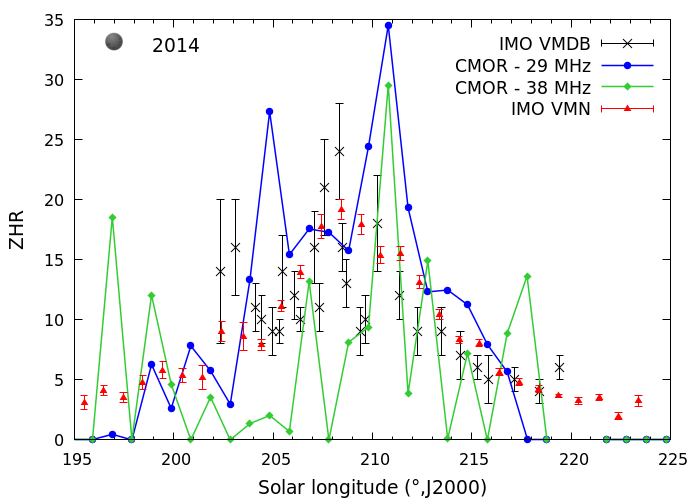}
      \includegraphics[width=.32\textwidth]{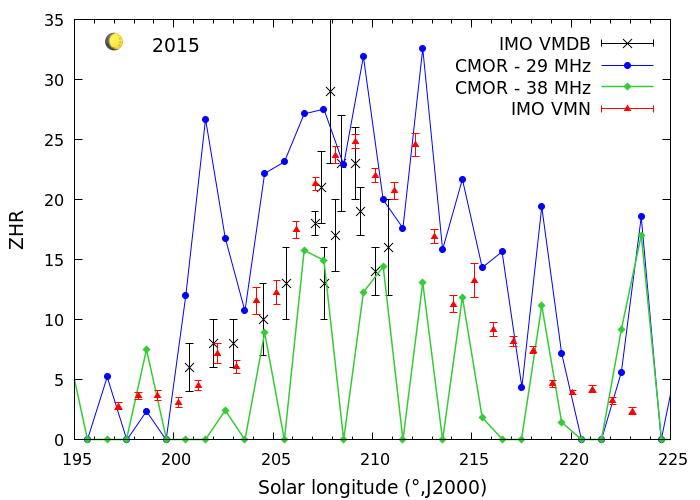}
      \includegraphics[width=.32\textwidth]{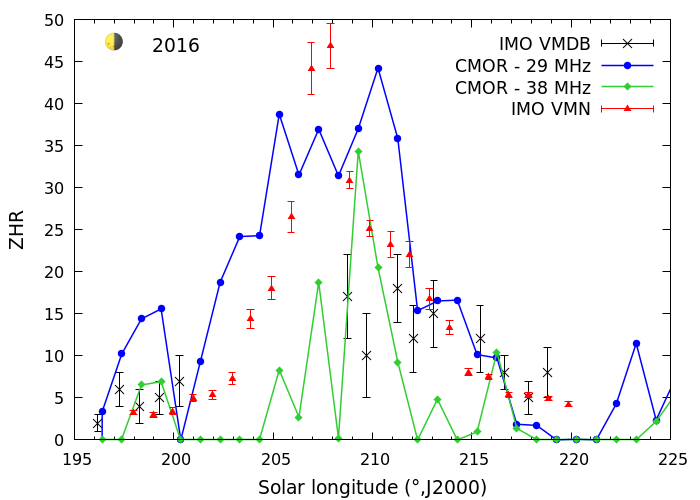}}\\
      \resizebox{\dimexpr.9\textwidth-1em}{!}{
      \includegraphics[width=.32\textwidth]{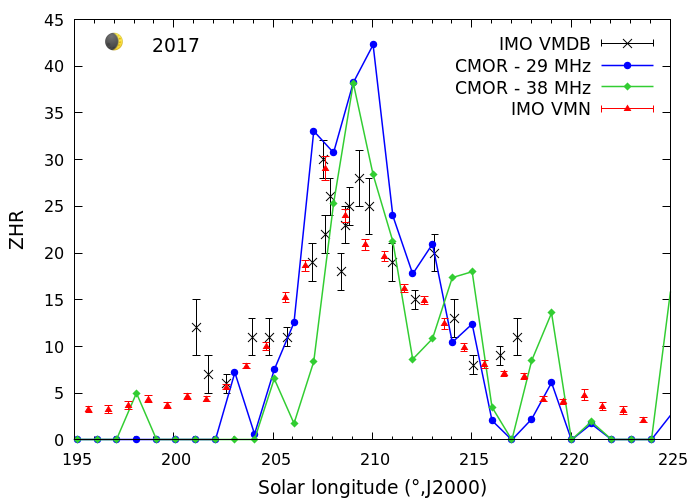}
      \includegraphics[width=.32\textwidth]{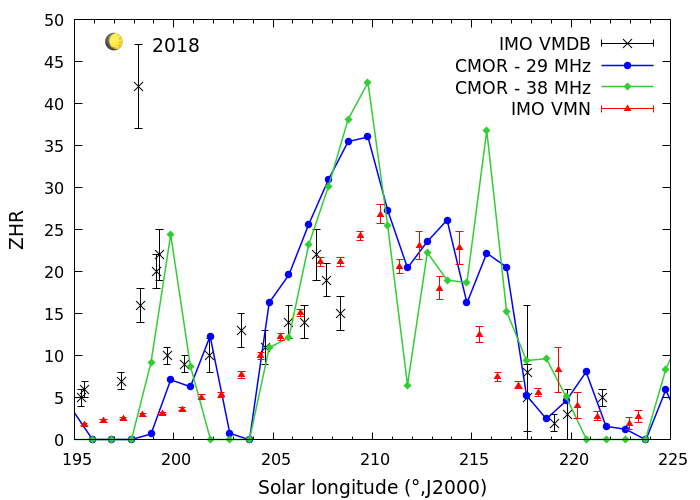}
      \includegraphics[width=.32\textwidth]{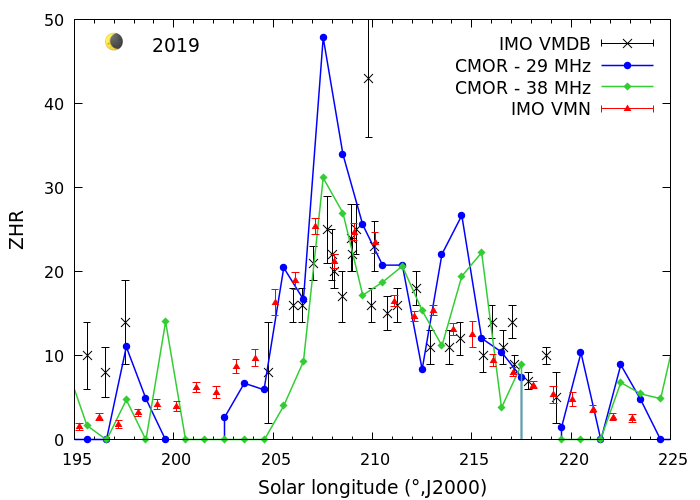}}
      \caption{Activity profiles of the Orionids between 2002 and 2019.}
      \label{fig:new_oris}
  \end{figure*}
  
  
  \clearpage
  \section{Additional intensity maps of the $\eta$-Aquariids and Orionids between 1985 and 2019} \label{maps}
  
     \mbox{ } \vspace{-0.5cm} \mbox{ }

    \begin{figure*}[!ht]
      \centering
      \resizebox{\dimexpr.92\textwidth-1em}{!}{
      \includegraphics[width=.49\textwidth]{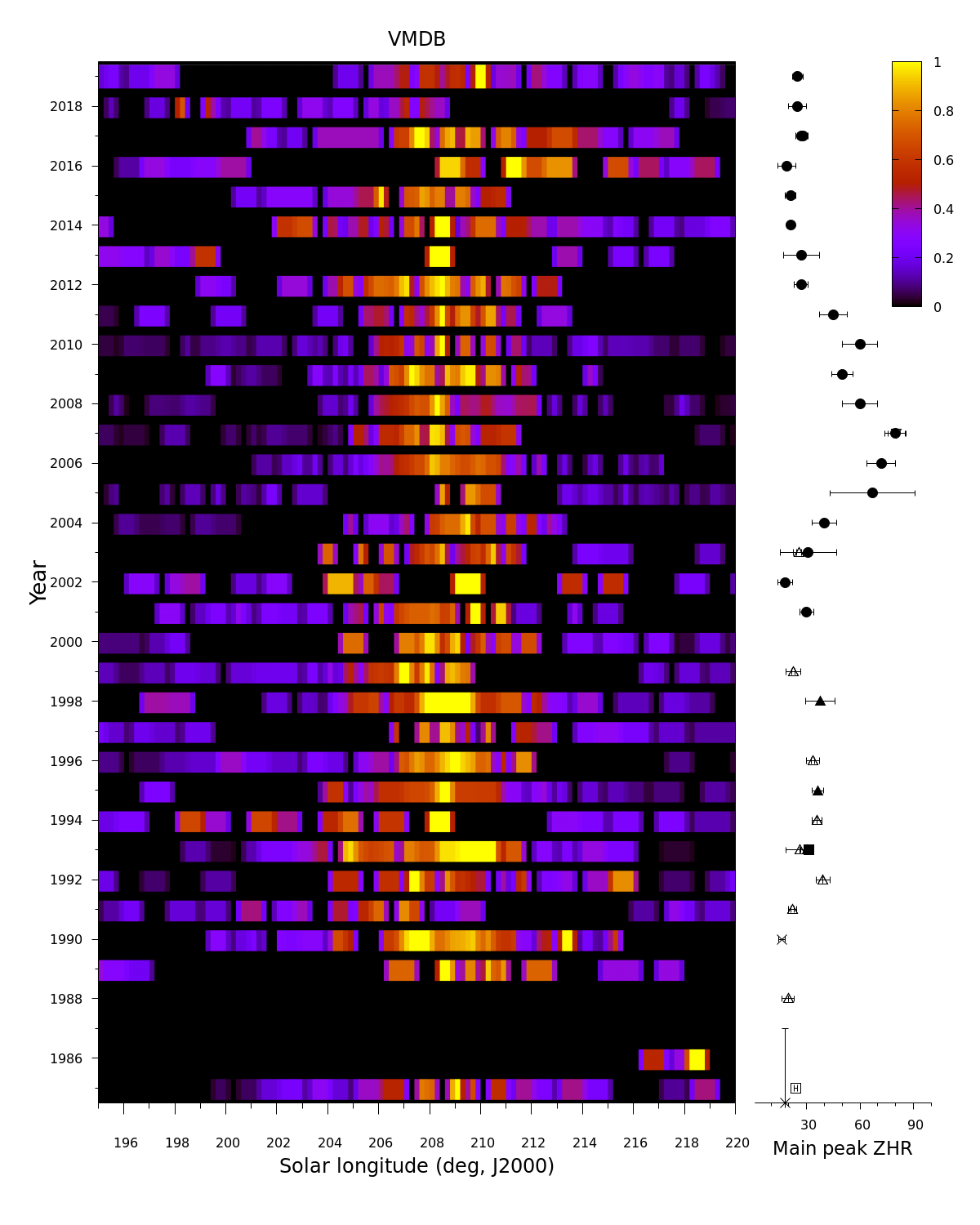}
      \includegraphics[width=.49\textwidth]{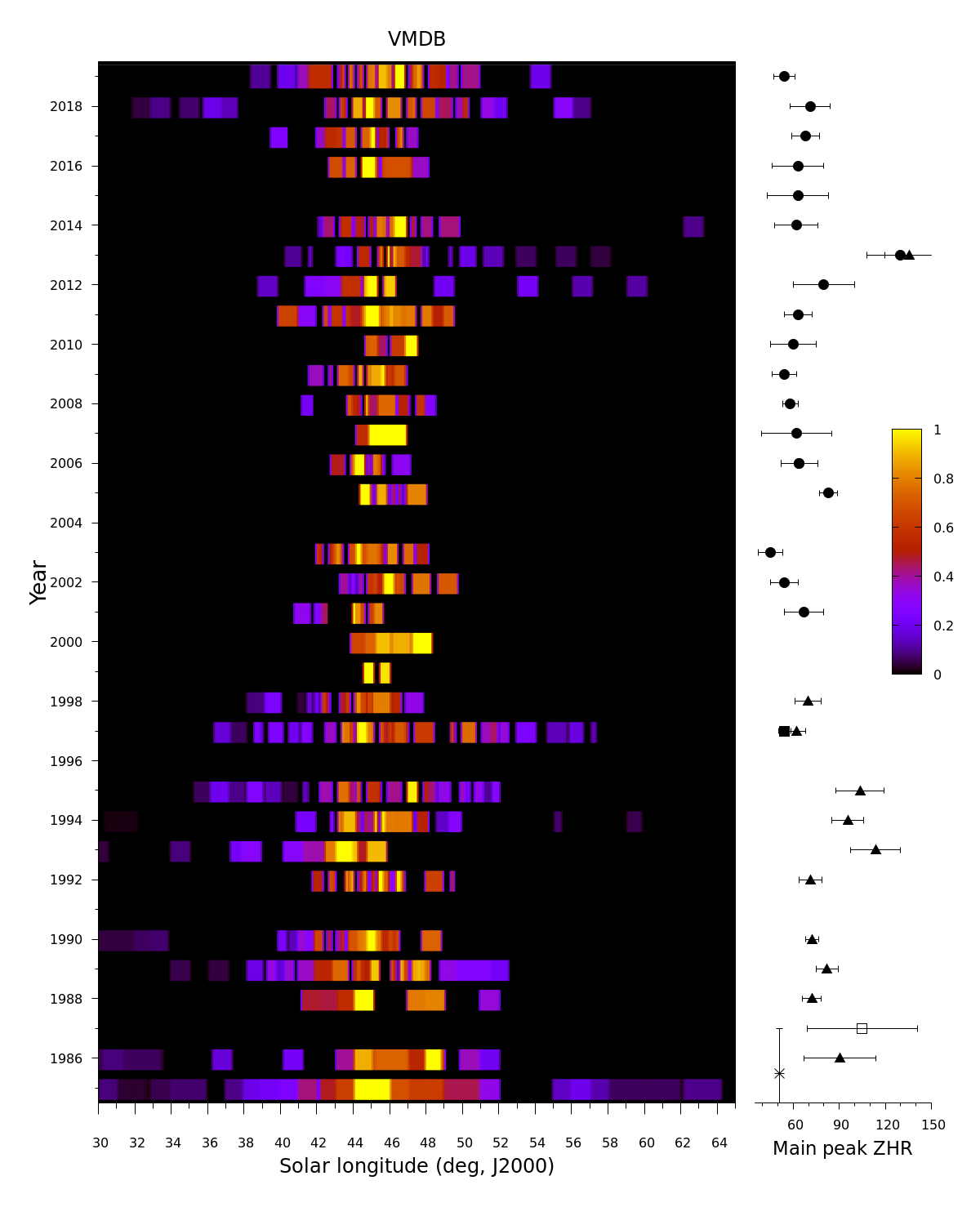}}\\[-0.5cm]
      \resizebox{\dimexpr.92\textwidth-1em}{!}{
      \includegraphics[width=.49\textwidth]{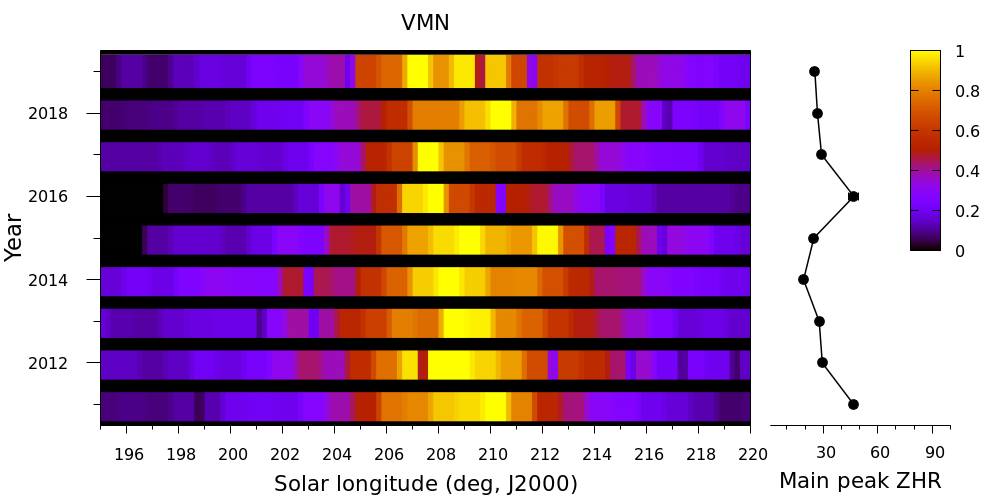}
      \includegraphics[width=.49\textwidth]{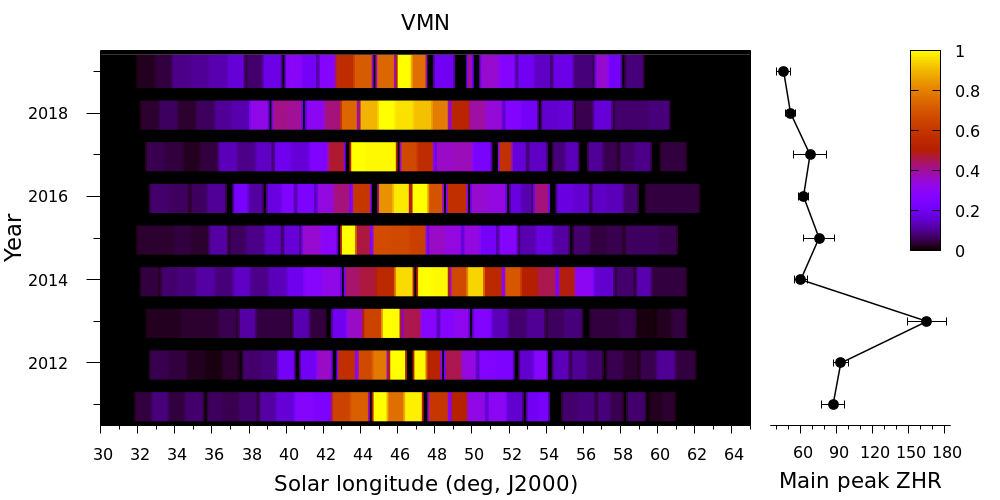}}\\[-0.5cm]
      \resizebox{\dimexpr.92\textwidth-1em}{!}{
      \includegraphics[width=.49\textwidth]{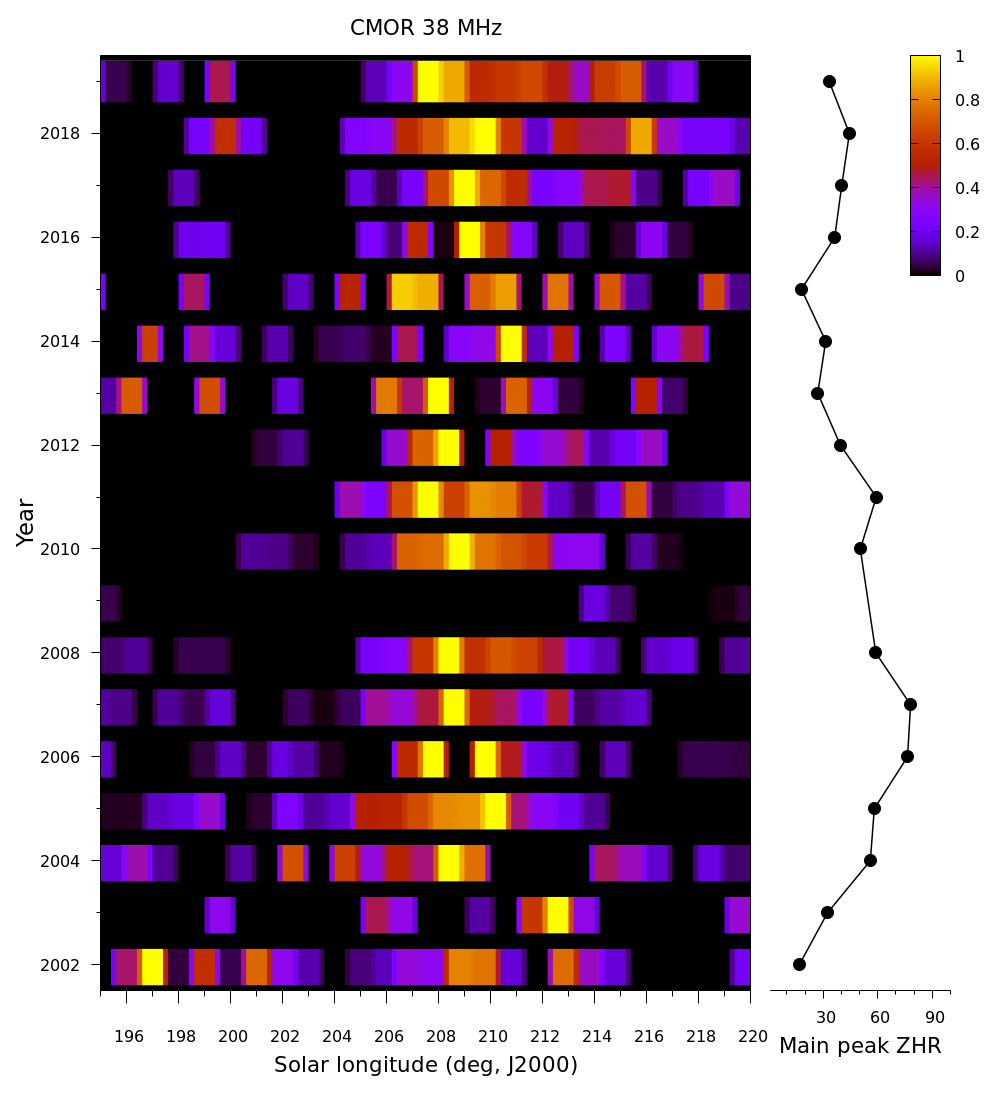}
      \includegraphics[width=.49\textwidth]{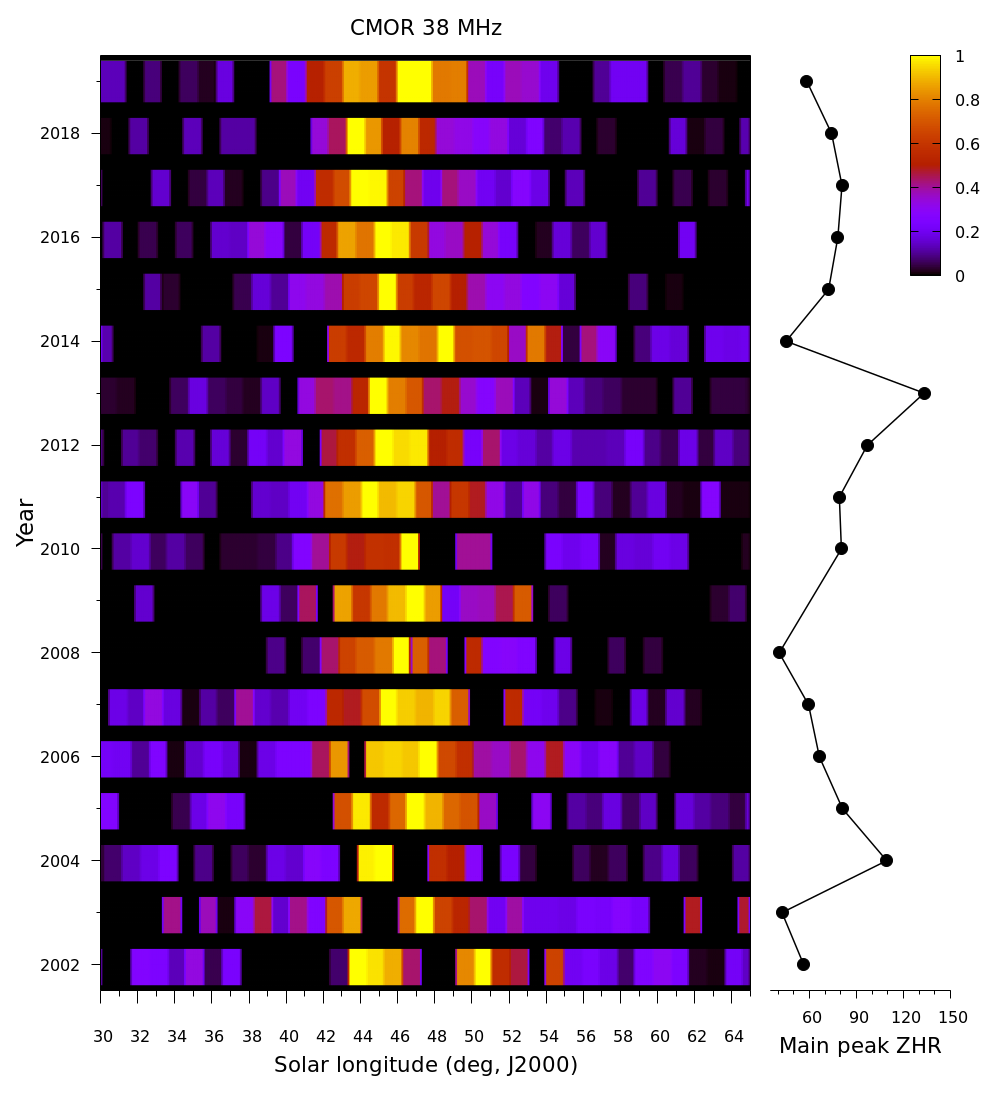}}
      \caption{VMDB (top), CMOR 38 MHz (middle), and VMN (bottom) intensity maps of the Orionids (left) and $\eta$-Aquariids (right).  For each apparition of the shower presented in Appendix \ref{old_obs} and \ref{new_obs}, the activity profile is normalized by the corresponding maximum ZHR determined in Section \ref{sec:ZHRmax} (reminded by the black curve at the right of the map).}
      \label{fig:Optical_maps}
  \end{figure*}
 

   \clearpage
   \section{Maximum visual rates of the $\eta$-Aquariids and Orionids meteor showers} \label{sec:maximum_rates}
   
  \begin{table*}[!ht]
      \centering
      \begin{tabular}{cccccl}
          \hline
          \hline
          Year & \Lsol main peak & ZHR main peak & FWHM &  $r$ at maximum & Source \\
          & (\si{\degree}, J2000) & & & & \\
          \hline
           1984-1987 & 45.28 & $\sim$51 & - &  2.5 & [1]$^\text{F}$ \\
          1986 & 48.55 & 90$\pm$24 & - & 2.3 & [2]$^\text{F}$ \\
          1987 & 44.2$\pm$0.9 & 105$\pm$36 & -  & 2.41$\pm$0.27 & [3] \\
          1988 & 44.66 & 72$\pm$7 & - & 2.3 & [2]$^\text{F}$\\
          1989 & 46.85 & 81$\pm$7 & - & 2.3 & [2]$^\text{F}$ \\
          1990 & 45.17 & 70$\pm$5 & - & 2.3 & [2]$^\text{F}$\\
          1992 & 45.66 & 71$\pm$34 & -- & 2.3 & [2]$^\text{F}$ \\
          1993 & $\sim$44 & 110 & -- & 2.3 & [2] \\
          1994 & 43.5 & 80 & -- & 2.3 & [2] \\
          1995 & $\sim$44 & 104$\pm$15 & -- & 2.3 & [2]$^\text{F}$ \\
          1997 & $\sim$44.5 & $\sim$62$\pm$6 & 8 days & 2.3 & [4] \\
               & 44.5$\pm$0.2 & 54$\pm$3 & 5\si{\degree} & 2.2 & [5] \\
          1998 & 44.36 & 69$\pm$3 & --  & 2.3 & [6] \\
          2001 & 45.5      & 67$\pm$13 & --         & 2.46 & [7]\\
          2002 & 46        & 54$\pm$9  & 45.5\si{\degree}- 49.5\si{\degree} \Lsol & 2.46 & [7] \\
          2003 & 44.1-44.4 & 45$\pm$8  & 43.0\si{\degree}- 47.3\si{\degree} \Lsol & 2.46 & [7]\\
          2005 & 45.6      & 83$\pm$6  & 44.3\si{\degree}- 46.8\si{\degree} \Lsol & 2.46 & [7]\\
          2006 & 44.4      & 64$\pm$12 & 43.3\si{\degree}- 45.6\si{\degree} \Lsol & 2.46 & [7]\\
               & 45.3      & 64$\pm$12 & 43.3\si{\degree}- 45.6\si{\degree} \Lsol & 2.46 & [7]\\
          2007 & 45.4      & 62$\pm$23 & 44.5\si{\degree}- 46.4\si{\degree} \Lsol & 2.46 & [7]\\
          2008 & 45.8      & 58$\pm$5  & 42.2\si{\degree}- 46.7\si{\degree} \Lsol & 2.46 & [7]\\
          2009 & 45.7      & 54$\pm$8  & 43.4\si{\degree}- 47.3\si{\degree} \Lsol & 2.46 & [7]\\
          2010 & 45.3      & 60$\pm$15 & --         & 2.46 & [7]\\
          2011 & 44.7-45.9 & 63$\pm$9   & 42.0\si{\degree}- 48.5\si{\degree} \Lsol & 2.46 & [7]\\
          2012 & 45        & 80$\pm$20 & 43.6\si{\degree}- 47.5\si{\degree} \Lsol & 2.46 & [7]\\
          2013 & 45.6       & 135$\pm$16 & -- & -- & [8] \\
          2013 & 45.75      & 130$\pm$22 & 44.0\si{\degree}- 44.6\si{\degree} \Lsol &  2.46 & [7]\\
          2014 & 45.65      & 62$\pm$14 & 44.0\si{\degree}- 46.3\si{\degree} \Lsol   &  2.46 & [7]\\
          2015 & 44.1       & 63$\pm$120 & -- & 2.46 & [7]\\
          2016 & 44.1       & 3$\pm$17 & 41.5\si{\degree}- 46.8\si{\degree} \Lsol & 2.46 & [7]\\
          2017 & 45.32      & 68$\pm$9 & 42.8\si{\degree}- 46.8\si{\degree} \Lsol & 2.46 & [7]\\
          2018 & 44.88      & 71$\pm$13 & 44.0\si{\degree}- 47.3\si{\degree} \Lsol & 2.46 & [7]\\
          2019 & 46.7       & 54$\pm$7 & 43.5\si{\degree}- 47.7\si{\degree} \Lsol & 2.46 & [7]\\
          \hline
      \end{tabular}
      \vspace{0.4cm}
      \caption{Compilation of main peak times and ZHRs of the $\eta$-Aquariids meteor shower between 1985 and 2019, deduced from visual observations. When available, the population index $r$ at the peak and the Full Half Width Maximum (FWHM) are also provided. Values from works labeled "F" were extracted from the published ZHR profiles and not specified explicitly in the source text. Estimates determined in this work are detailed in Section \ref{sec:ZHRmax}. References: [1] \cite{Porubcan1991}, [2] \cite{Cooper1996}, [3] \cite{Koseki1988}, [4] \cite{Cooper1997}, [5] \cite{Rendtel1997}, [6] \cite{Cooper1998}, [7] this work and [8] \cite{Cooper2013}.}
      \label{tab:visualZHR_eta}
  \end{table*}

  \begin{table*}[!ht]
      \centering
      \begin{tabular}{cccccl}
          \hline
          Year & SL main peak & ZHR main peak & duration / FWHM & $r$ at maximum & Source \\
          & (\si{\degree}, J2000) & & & & \\
          \hline
          \hline
           1982, 1984-1987 & 207.30 & $\sim$18 & -- &  2.5 & [1]$^\text{F}$ \\
          1985 & 207.69 & 23.8$\pm$0.9 & & 2.25 & [2] \\
          1990 & 207.35 & 15.6$\pm$0.8 & -- & 2.84$\pm$0.13 & [3] \\
          1990 & 209.7 & 16$\pm$1.9 & -- & 2.44$\pm$0.19 & [3] \\
          1993 & 204.82 & 31.1$\pm$0.09 & 204.74\si{\degree}- 206.14\si{\degree} \Lsol & 1.95$\pm$0.05 & [4] \\
          2001 & 207.9       & 30$\pm$4  & 206.9\si{\degree}- 210.8\si{\degree} \Lsol & 2.59 & [5]\\
          2002 & --           & 18$\pm$4  & --           &  2.59 & [5]\\
          2003 & 208.6       & 31$\pm$16 & --           &  2.59 & [5]\\
          2004 & 209.4       & 40$\pm$7  & 207.3\si{\degree}- 212.1\si{\degree} \Lsol &  2.59 & [5]\\
          2005 & 208.7       & 67$\pm$24 & --           &  2.59 & [5]\\
          2006 & 208.2       & 72$\pm$8  & 206.8\si{\degree}- 211.0\si{\degree} \Lsol &  2.59 & [5]\\
          2007 & 208.450 & 80.5$\pm$4.7 & 207.12\si{\degree}- 209.2\si{\degree} \Lsol & 2.11$\pm$0.07 & [6] \\
          2007 & 208.5       & 80$\pm$6  & 207.0\si{\degree}- 209.5\si{\degree} \Lsol &  2.59 & [5]\\
          2008 & 207.8-207.9 & 60$\pm$10 & 206.1\si{\degree}- 209.6\si{\degree} \Lsol & 2.59 & [5]\\
          2009 & 207.4-209.4 & 50$\pm$6  & 206.1\si{\degree}- 211.5\si{\degree} \Lsol &  2.59 & [5]\\
          2010 & 208.5       & 60$\pm$10 & 206.0\si{\degree}- 210.6\si{\degree} \Lsol &  2.59 & [5]\\
          2011 & 208.6       & 45$\pm$8  & 207.0\si{\degree}- 209.5\si{\degree} \Lsol &  2.59 & [5]\\
          2012 & 208.3       & 27$\pm$4  & 205.4\si{\degree}- 211.5\si{\degree} \Lsol & 2.59 & [5]\\
          2013  & 208.4      & 27$\pm$10 & -- & 2.59 & [5]\\
          2014  & 208.4      & 21$\pm$2  & 206.3\si{\degree}- 209\si{\degree} \Lsol & 2.59 & [5]\\
          2015  & 207.7      & 21$\pm$3   & 205.4\si{\degree}- 211\si{\degree} \Lsol & 2.59 & [5]\\
          2016 & 208.7      & 19$\pm$5   & --  & 2.59 & [5]\\
          2017  & 207.5     & 28$\pm$3   & 206.5\si{\degree}- 211.0\si{\degree} \Lsol & 2.59 & [5]\\
          2017  & 208.9     & 27$\pm$3   & 206.5\si{\degree}- 211.0\si{\degree} \Lsol & 2.59 & [5]\\
          2018  &  207.3    & 25$\pm$5   & -- & 2.59 & [5]\\
          2019  &  209.0    & 25$\pm$3   & 205.0\si{\degree}- 211.0\si{\degree} \Lsol & 2.59 & [5]\\
          \hline
      \end{tabular}
      \vspace{0.4cm}
      \caption{Compilation of main peak times and ZHRs of the Orionids meteor shower between 1982 and 2019, deduced from visual observations. When available, the population index $r$ at the peak and the Full Half Width Maximum (FWHM) are also provided. Values from works labeled "F" were extracted from the published ZHR profiles and not specified explicitly in the source text. Estimates determined in this work are detailed in Section \ref{sec:ZHRmax}. References: [1] \cite{Porubcan1991}, [2] \cite{Spalding1987}, [3] \cite{Koschack1991}, [4] \cite{Rendtel1993}, [5] this work and [6] \cite{Arlt2008}.}
      \label{tab:visualZHR_ori}
  \end{table*}

\end{appendix}
\end{document}